\documentclass[3p]{elsarticle}

\usepackage{todonotes}
\let\opentask\todo 
\renewcommand{\todo}[1]{\opentask[inline,color=red!40]{#1}}

\usepackage{amsmath,amssymb,amsfonts}
\usepackage{algorithm,algorithmic}
\usepackage{hyperref}
\usepackage{caption}
\usepackage{subcaption}
\usepackage{booktabs}
\usepackage{graphicx}
\usepackage{longtable}
\usepackage{eurosym}
\usepackage[normalem]{ulem}
\usepackage{footmisc}
\usepackage{makecell}
\usepackage{multirow}
\usepackage{array}
\useunder{\uline}{\ul}{}

\DeclareMathOperator*{\argmax}{argmax}

\usepackage[section]{placeins}

\journal{Transportation Research Part C, Special Issue on ISTTT 24}

\biboptions{authoryear}

\begin{document}

\begin{frontmatter}

\title{Regulating Mobility-on-Demand Services: Tri-level Model and Bayesian Optimization Solution Approach}

\author{Florian Dandl\textsuperscript{a,*}, Roman Engelhardt\textsuperscript{a}, Michael Hyland\textsuperscript{b}, Gabriel Tilg\textsuperscript{a}, \\ Klaus Bogenberger\textsuperscript{a}, Hani S. Mahmassani\textsuperscript{c}}
\address{\textsuperscript{a} Chair for Traffic Engineering and Control, Technical University of Munich, 80333 Munich, Germany}
\address{\textsuperscript{b} Department of Civil and Environmental Engineering, Institute of Transportation Studies \\ University of California, Irvine \\ 4000 Anteater Instruction and Research Bldg. (AIRB), Irvine, CA 92697-3600, United States}
\address{\textsuperscript{c} Department of Civil and Environmental Engineering, Northwestern University, Northwestern University Transportation Center, 600 Foster Street, Evanston, IL 60208, United States}

\cortext[corresponding]{Corresponding author: florian.dandl@tum.de}




\begin{abstract}
The goal of this paper is to develop a modeling framework that captures the inter-decision dynamics between mobility service providers (MSPs) and travelers that can be used to optimize and analyze policies/regulations related to MSPs. To meet this goal, the paper proposes a tri-level mathematical programming model with a public-sector decision maker (i.e. a policymaker/regulator) at the highest level, the MSP in the middle level, and travelers at the lowest level. The public-sector decision maker aims to maximize social welfare via implementing regulations, policies, plans, transit service designs, etc. The MSP aims to maximize profit by adjusting its service designs. Travelers aim to maximize utility by changing their modes and routes. The travelers' decisions depend on the regulator and MSP's decisions while the MSP decisions themselves depend on the regulator's decisions. To solve the tri-level mathematical program, the study employs Bayesian optimization (BO) within a simulation-optimization solution approach. At the lowest level, the solution approach includes an agent-based transportation system simulation model to capture travelers' behavior subject to specific decisions made by the regulator and MSP. At the middle and highest levels, the solution approach employs BO for the MSP to maximize profit and for the regulator to maximize social welfare. The agent-based transportation simulation model includes a mode choice model, a road network, a transit network, and an MSP providing automated mobility-on-demand (AMOD) service with shared rides. The modeling and solution approaches are applied to Munich, Germany in order to validate the model. The case study investigates the tolls and parking costs the city administration should set, as well as changes in the public transport budget and a limitation of the AMOD fleet size. Best policy settings are derived for two social welfare definitions, in both of which the AMOD fleet size is not regulated as the shared-ride AMOD service provides significant value to travelers in Munich.

\end{abstract}

\begin{keyword}
tri-level model; mobility services, AMOD, policy optimization, Bayesian optimization (BO)
\end{keyword}

\end{frontmatter}


\section{Introduction}
Historically, transportation systems (and transportation systems models) included two main sets of actors/agents/decision makers: public-sector entities and travelers. Public-sector entities directly or indirectly planned, designed, managed, operated, maintained, and controlled the physical and cyber transportation infrastructure, roadway networks, public transit (PT) systems, traffic signals, parking supply, etc., while travelers used this transportation system to access various opportunities. The goals of the public sector included, and still include, providing accessibility and mobility to travelers, subject to budgetary, environmental, political, and other constraints. The goal of each traveler is to maximize her utility, subject to the public sector's transportation system, as well as her own constraints (e.g. budgetary, social, professional, etc.). 

Transportation systems comprised of public-sector decision makers and travelers are commonly modeled using a bi-level modeling framework based on the Stackelberg game~\citep{Fisk1984}. In the bi-level modeling framework, the regulator is the upper level decision maker and the lower level decision makers are the individual travelers. The public-sector decision maker in the upper level aims to maximize some objective related to mobility, accessibility, revenue, or social welfare, subject to the aforementioned constraints. However, the public-sector decision maker is also subject to the decisions of the travelers in the lower level, which are typically modeled through a user-equilibrium constraint. The travelers in the lower level aim to maximize their own utilities given the public sector's decisions in the upper level. The bi-level modeling framework is common for road and transit network design problems, freeway ramp metering problems, congestion pricing problems, and various other transportation planning, design, and management problems~\citep{Xu2016, Yin2000}.

While this modeling framework effectively captures historical transportation systems that only include a regulator and travelers, the emergence and significant growth of private-sector mobility services providers (MSPs), such as ride-sourcing and ride-sharing companies, has altered the historical two-entity transportation system paradigm. Currently, MSPs use (i.e. they demand) the cyber and physical transportation infrastructure system provided by the public sector in order to provide (i.e. supply) transportation service to travelers. As such, the 
decisions made by public-sector entities impact MSPs directly and impact travelers indirectly and directly because travelers are subject to the decisions of the public sector but also the decisions of MSPs. These relationships between regulator, MSPs, and traveler suggest a tri-level, rather than a bi-level, modeling framework is appropriate for existing transportation systems that include MSPs. A tri-level framework is particularly valuable for analyzing and optimizing transportation policies, plans, designs, and management strategies that specifically impact MSPs.

\begin{figure}
    \centering
    \includegraphics[width=0.8\textwidth]{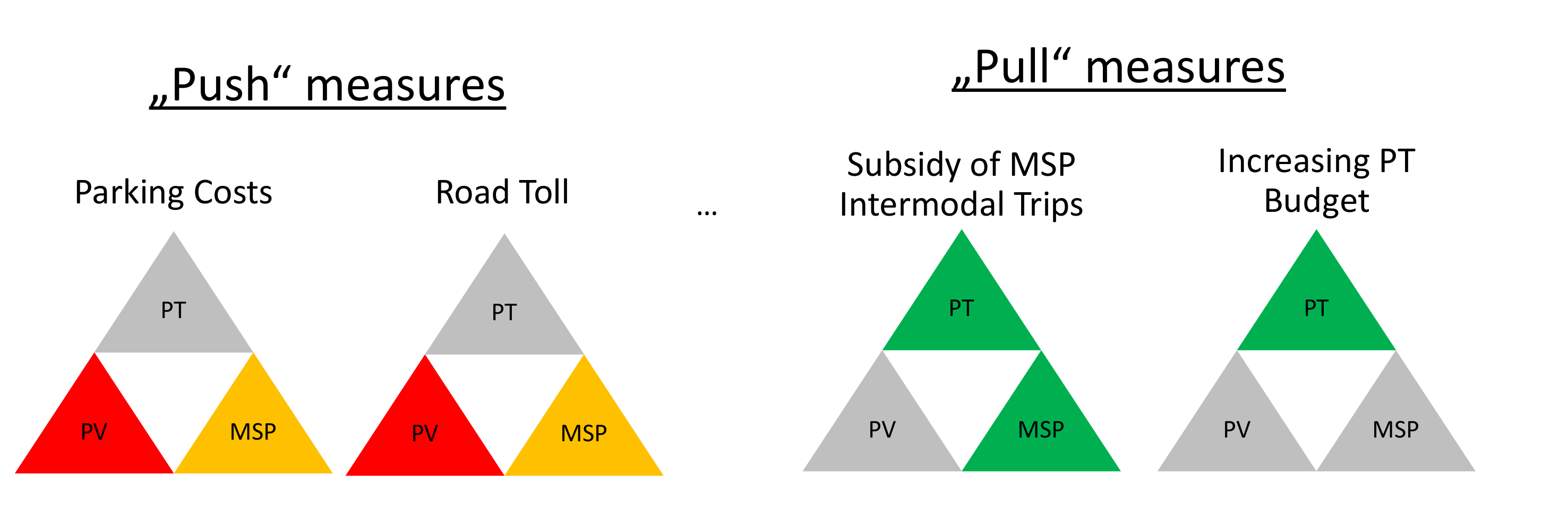}
    \caption{Possible regulatory push and pull measures in the context of MSPs affecting the traditional modes private vehicle (PV), PT and new mobility service providers (MSP).}
    \label{fig:push_pull}
\end{figure}

The significant presence of MSPs within transportation systems areas represents a substantial change, rather than an incremental change. Early empirical evidence suggests that while ride-sourcing and ride-sharing companies provide significant mobility and accessibility benefits to a wide range of travelers, they also increase vehicle miles traveled~\citep{Henao2018} and congestion~\citep{Erhardt2019}. Hence, transportation policymakers and regulators are considering different regulations that minimize the negative aspects of MSPs, while maximizing the societal benefits of MSPs. Typically, regulations can be classified into \textit{push} and \textit{pull} measures, depending on whether they increase the travel cost/dis-utility or they increase the utility/attractiveness of one or more modes (see Fig.~\ref{fig:push_pull}). In the context of a transportation system with MSPs, parking and/or road tolls can impact the input costs for MSPs, assuming the MSPs are not exempt. Moreover, public-sector decision makers can try to shape MSPs into a first/last mile service by incentivizing this particular service or increasing the public transport (PT) budget to offer a more attractive transportation service (see Fig.~\ref{fig:push_pull}). Given the uncertainty around MSPs and the potential impacts of various policies and regulations, there is a need for a modeling framework that can quantify the potential implications of specific policies and regulations before they are actually implemented in the field. Moreover, as sets of policies can have synergistic or conflicting impacts, it is important to analyze combinations of policies in the same model.

The goal of this study is to develop a modeling framework for current transportation systems with MSPs that can be used to evaluate relevant MSP policies/regulations. To meet this goal, this study presents a tri-level modeling framework that captures the decision contexts and interrelationships between the public-sector decision-maker (e.g. policymaker, regulator, transit planner, etc.) at the highest level, MSPs in the middle level, and travelers at the lowest level. Specifically, the study presents a tri-level mathematical programming formulation wherein the public sector aims to maximize social welfare subject to the MSP maximizing its profits and travelers maximizing their own utilities. To solve the tri-level problem, the study employs (1) an agent-based transportation system simulation model (henceforth, 'transportation model') that includes travelers, mobility services determined by the MSP, and the policies/regulations determined by the regulator to solve the user equilibrium problem at the lowest level, and (2) Bayesian optimization (BO) formulations and solution procedures to solve the middle-level MSP profit maximization problem and the upper level public-sector social welfare maximization problem. The study employs BO because the functional relationship between MSP decisions and travelers' responses as well as the functional relationship between regulator's decisions and MSP and travelers' responses are unknown and likely non-convex and non-smooth. 

The study focuses on future MSPs that own and operate their own fleet of fully automated vehicles (AVs) and provide automated mobility-on-demand (AMOD) service with shared rides. The study focuses on AMOD service rather than a mobility service without AVs because AVs are likely to significantly reduce the operational costs of existing MOD services thereby allowing AMOD services to have a significantly larger market share than current MOD services. 

This study makes several notable contributions to the academic literature. The study's main contribution is the development of a tri-level math programming modeling framework that includes a public-sector decision maker (henceforth, 'regulator' for simplicity), an MSP, and travelers in a clear hierarchy. As far as the authors are aware, no other studies capture this hierarchical nature of the three sets of decision-makers in current transportation systems that include MSPs. In related studies, \cite{Pinto2019} and \cite{Liu2019} study transportation systems with travelers and MSPs in bi-level modeling frameworks. The former study models an AMOD service as mainly a PT feeder service controlled by the PT agency; moreover, the former study does not analyze or evaluate regulations related to AMOD services. The latter study does evaluate an MSP regulation (i.e. a tax) but their bi-level model precludes policy/regulatory optimization. Hence, the tri-level modeling framework proposed in the current study provides significant conceptual, methodological, and practical value in terms of considering, modeling, evaluating and particularly optimizing transportation policies and regulations that impact MSPs. The generality of the proposed tri-level modeling framework in the current study also provides significant value.  The modeling framework can incorporate various objectives and decisions for the three sets of decision makers. Another notable contribution is the  simulation-optimization solution approach that includes a transportation model to simulate and evaluate the disaggregate (and aggregate) behaviors of travelers, PT networks, road networks,  mobility service fleets, etc. This simulation-optimization approach supports the generality of the proposed tri-level modeling framework as the transportation model can flexibly incorporate additional model components as well as increases or decreases in the resolution of model components depending on the regulations/policies under consideration. 

The remainder of the paper is structured as follows. The next section provides background information on regulating MSPs, multi-level transportation system models, and several modeling components employed in this study. Section 3 presents the general tri-level mathematical programming formulation along with the conceptual framework underlying the problem formulation. Section 4 presents the solution approach including the BO methods for optimizing social welfare and MSP profits, and a transportation model that solves for user-equilibrium. After choosing four explicit regulatory policies to consider, Section 5 presents the detailed model components within the transportation model, which were choosen/developed to be sensitive to the regulatory policies. Section 6 presents the results of a case study in which the modeling approach and solution approach are applied to Munich, Germany for validation purposes. Section 7 presents a discussion of the computational results as well as modeling limitations and areas for improvement. Finally, Section 8 concludes the paper with a summary of the study and discussion of future research directions.

\section{Background and Literature Review}
This section provides background information and reviews the academic literature related to regulating mobility services, multi-level transportation system models, and the modeling components within the tri-level modeling framework employed in this study.  

\subsection{Regulating Mobility Services}
Vehicle-based shared mobility services, particularly MOD services such as Uber, Lyft, and Didi, have grown rapidly over the past decade. Recent research indicates that MSPs are increasing overall VMT~\citep{Henao2018} and congestion~\citep{Erhardt2019} and subsequently fuel consumption and vehicle emissions. 
To avoid these negative societal outcomes, public-sector entities are considering and employing different regulations~\citep{Beer.2017,Cetin.2019}. For example, New York City currently caps the number of MSP vehicles in Manhattan~\citep{Bellon2019} and the city and state of New York have passed legislation to implement congestion pricing in Manhattan~\citep{Hu2019}. Additionally, in 2019, California  passed legislation that requires MSPs Uber and Lyft to hire drivers as employees rather than independent contractors~\citep{Campbell2019} but a ballot proposition was latter passed by Californians that exempted from classifying drivers as employees \citep{Conger.2020}.

A few studies in the academic literature have examined regulations related to MSPs. Using a queuing theoretic model that captures the stochasticity and dynamics of MOD service, \cite{Li2019} evaluate three different regulations: a driver minimum wage, a cap on the number of MOD vehicles/drivers, and a per trip congestion tax. The study concludes that a driver minimum wage could benefit both drivers and riders in the system. Conversely, they find a vehicle cap would hurt both drivers and riders as the MSP would capture the benefits of a supply limit. They find that a congestion tax would have the expected effect: fare increases, wage decreases, and MSP revenue decreases. Additionally, \cite{Zhang2019}, using an analytical market equilibrium model, find that a per-trip congestion tax policy that explicitly penalizes solo rides can harm social welfare even while promoting pooling/sharing rides. However, the same congestion tax can increase social welfare if implemented alongside a minimum wage policy for drivers. Notably, \cite{Zhang2019} assume a system where a single MSP has a monopoly. 

The expected advent of AVs and their inclusion in MOD service fleets is only increasing the urgency for the public sector to consider regulations related to MSPs. The reason for the added urgency is that AVs are expected to decrease the operational costs of MOD services via removing the main component of operational costs in existing MOD services (i.e. the driver)~\citep{Bosch2018, Dandl2019, Fagnant2015} thereby allowing MSPs to increase their market share. A recent study examines the impacts of various congestion pricing policies on a future transportation systems with AMOD~\citep{Simoni2019}. The study uses MATSim, an agent-based model, and finds that combinations of more complex congestion pricing strategies perform better in terms of social welfare than simple congestion pricing and road tolls.

\subsection{Multi-level Transportation System Modeling }
The usage of bi-level models is quite natural in transportation system models to evaluate policies, plans, designs, and managerial strategies as a system manager implements these changes to a transportation system, but their impact depends on how travelers respond to the changes. The key components of a bi-level model are the upper and lower level objective functions, the upper and lower level control/decision variables, and the response function that connects the upper and lower level decision variables. Since the lower level optimization problem is typically a non-linear constraint in bi-level programs, the bi-level problem is typically non-convex. This makes it difficult to obtain global optimal solutions to bi-level programming problems. Moreover, in cases such as the one in this paper, a functional relationship between the upper level and lower level decision variables is not easy to obtain. This makes the problem more difficult to solve using conventional optimization methods. 

Research that addresses the road network design problem commonly employs a bi-level modeling approach~\citep{Farvaresh2013, Gao2005, Xu2016, Yang1998}. The upper level system manager has a variety of control levers (e.g. links to add/remove from a road network) to optimize some objective (e.g. total network travel cost, total revenue, total flow, total network reserve capacity, etc.), while travelers change their route choices based on the network design~\citep{Yin2000}.

The two studies most related to the current paper both employ bi-level modeling frameworks~\citep{Liu2019, Pinto2019}. \cite{Pinto2019} develop a bi-level model wherein the lower level incorporates traveler mode choice, PT route choice, and AMOD fleet service and the upper level is the joint design of a PT network and AMOD fleet size. The study assumes that the PT agency operates their own AMOD fleet or contracts a specific service from an MSP. Similarly, \cite{Liu2019} develop a bi-level modeling framework that involves the design of an MOD service at the upper level (fleet size and fare) and traveler mode choice and traffic operations at the lower level. \cite{Liu2019} use BO to maximize MSP profit and the study evaluates the impact of a tax on MOD without shared rides. The current study also uses BO to optimize MSP profit; however, rather than evaluating policies, the current study adds a third level and once again uses BO to determine optimal policies in the highest level.

Notably, several studies in the transportation literature incorporate tri-level mathematical programming modeling approaches. In a recent study, \cite{Gu.2019} model the construction and pricing of new roadway links as a tri-level problem with a public-sector agency at the highest level, profit-maximizing private firms at the middle level, and travelers at the lowest level. Unlike the current study, the transportation model in \cite{Gu.2019} is static, deterministic, and only incorporates route choice.

\subsection{Modeling Components}

Embedded in the agent-based transportation simulation model proposed in this study is an AMOD fleet dispatching model that dynamically assigns AVs to open user requests. The importance of this problem from an operational efficiency perspective and transportation systems modeling perspective has motivated a sizable volume of research over the past five years. Several studies employ rule-based AV-traveler policies that mainly involve assigning new requests to the nearest idle AV~\citep{Bischoff2016, Chen2016, Fagnant2015, Fagnant2014, Gurumurthy2020}. These rule-based algorithms have been integrated into large-scale transportation network simulation models. Later research employs optimization-based AV-traveler assignment policies based on the assignment problem~\citep{Bertsimas2019, Dandl2019Mike, Hoerl2018, Hyland2019, Hyland2018, Maciejewski2016}. Other recent research involves reinforcement learning approaches to assign AV to user requests~\citep{Holler2019, Li2019M, Oda2018, Xu2018}. 

The studies mentioned in the previous paragraph address the operational problem in MOD/AMOD services without shared rides. However, the current study assumes only shared-ride AMOD services can operate in a given city. The shared-ride AMOD service operational problem has also received considerable attention in the literature; the service is also commonly referred to as dynamic ride-sharing with shared AVs in the literature~\citep{Fagnant2018, Gurumurthy2018}. Research on the shared-ride AMOD service problem includes rule-based policies~\citep{Fagnant2018, Gurumurthy2018}, and optimization-based strategies~\citep{AlonsoMora.2017, Engelhardt.2019, Bilali2020, Hyland2020, Simonetto2019}.

Computationally efficient path finding algorithms are crucial to simulate a large-scale shared-ride AMOD service. Typically, studies assume constant link and/or path travel times and store the travel time information offline based on pre-processed travel information tables~\citep{Santi.2014, AlonsoMora.2017}. \cite{Dandl.2020ITSC} develop a modeling framework in which routing is also based on pre-processed travel information, but travel times are dynamically adapted according to the current network state. The study incorporates network traffic performance using a network fundamental diagram (NFD)~\citep{daganzo2007urban, Geroliminis.2008}, which determines the adjustment of travel times. Since most urban networks are not homogeneous, introducing sub-clusters of the network improves the quality of these macroscopic models~\citep{Saedi.2020}. The current study extends the framework in \cite{Dandl.2020ITSC} by scaling travel times independently in different network clusters.

\section{Problem Description}
\subsection{Conceptual Framework}
\label{sec:concept_frame}
This study models a future transportation system with three key actors –-- the regulator, the AMOD service planner, and travelers/system users. The regulator sets the political, legal, and organizational framework; the AMOD service planner offers a profit-maximizing service depending on the existing regulatory environment; and each traveler chooses the best of her available mobility options. On the one hand, the choices of travelers affect the utilization of the road network and PT system as well as congestion levels, environmental impacts, and the profit/loss of both the AMOD business and the public agency. On the other hand, regulators and  AMOD service planners can influence the decisions of travelers. The regulator has direct and indirect methods to change travelers' mode decisions and to impact the  AMOD service planner's decisions. Increasing parking fees make PVs less attractive~\citep{Washbrook.2006}. Subsidies to PT or AMOD services represent indirect methods as these financial policies could motivate  AMOD service planners to offer cheaper trips. A road toll is an example that affects users both directly and indirectly. PV users directly pay the toll and AMOD companies -- if not exempted -- are likely to pass the additional costs on to their passengers. Furthermore, a zone-based distance-dependent toll can influence route choice decisions of AMOD operators and individual drivers.
\par
This study denotes (i) the decision variables of the regulator by $x^r$, where $r$ is the index for different regulations; (ii) the decision variables of the  AMOD service planner by $y^s$, where $s$ is the index for different service design parameters; and (iii) the mode-choice decision variable of traveler $i$ for mode $m$ by $z_i^m$ ($z_i^m=1$ for the chosen mode, $0$ for the others). For the sake of simplicity and clarity, this study limits the set of mobility options to PV, one AMOD service with shared/pooled rides and PT with walking access and egress.
\par
The decision-making processes of the three different actor classes represent a non-cooperative game. The study assumes there is a hierarchy of decision making: regulators effectively set the rules of the transportation game,  AMOD service planners define their service, and travelers make their mode decisions based on the regulator rules and the  AMOD planner's service. This approach reflects reality as policies and regulation changes generally require a long planning period; whereas, an  AMOD service planner can likely change/adapt their service characteristics relatively frequently (e.g. monthly or quarterly). Finally, a subset of travelers can update modal decisions on a weekly, if not daily, basis.
\par
\subsection{General Problem}
\label{sec:general_prob}
The goal of the tri-level decision problem is to find a set of regulator decisions $x^{r,*}$ that maximize social welfare $W$ subject to the conditions that (i) the AMOD business optimizes its service design to maximize profit $P$ for a given set of regulations $x^r$ and (ii) travelers optimize their travel behavior (combined mode and route choice) according to their personal utility $U_i = \sum_m z_i^m U_i^m(x^r,y^s)$, subject to regulations $x^r$ and the AMOD service design $y^s$.

\begin{align}
    x^{r,*} &= \argmax_{x^r} W \left(x^r, y^{s,*}(x^r), \{ z_i^{m,*}(x^r,y^{s,*}) \}_i \right) & ~ \label{eq:reg-opt}\\
    y^{s,*} &= \argmax_{y^s} P \left(y^s, \{ z_i^{m,*}(y^{s}) \}_i | x^r \right) & ~ \label{eq:op-opt}\\
    z_i^{m,*} &= \argmax_{z_i^m} U_i \left(z_i^m | x^r, y^s \right) & \forall i \label{eq:trav-opt}
\end{align}
Here, the $|$ symbol denotes that the right-hand variables are set by a higher hierarchy player.

\section{Solution Approach}
\label{sec:high_level_solution_approach}
Equations~(\ref{eq:reg-opt})-(\ref{eq:trav-opt}) represent a tri-level mathematical program. To solve the tri-level math programming problem, this study employs a simulation-optimization solution approach. The simulation-optimization approach includes the transportation model at the lowest level to capture the behavior of travelers given the middle and upper level decisions. The travelers are individually optimizing their own utility and the agent-based simulation solves for user-equilibrium. Additionally, the simulation-optimization solution approach employs BO to optimize the AMOD service planner's decisions in the middle level and to optimize the regulator's decisions at the highest level. 

The transportation model performs all traveler optimization processes and returns the aggregated quantities for social welfare $W$ and  AMOD service planner profit $P$ as functions of $x^r$ and $y^s$, which are inputs to the transportation model. Therefore, the transportation model must be sensitive to the regulations $x^r$ and the AMOD service design $y^s$. Section~\ref{sec:model} describes the details of the transportation simulation model and model components.

\begin{figure}
    \centering
    \includegraphics[width=1.0\textwidth]{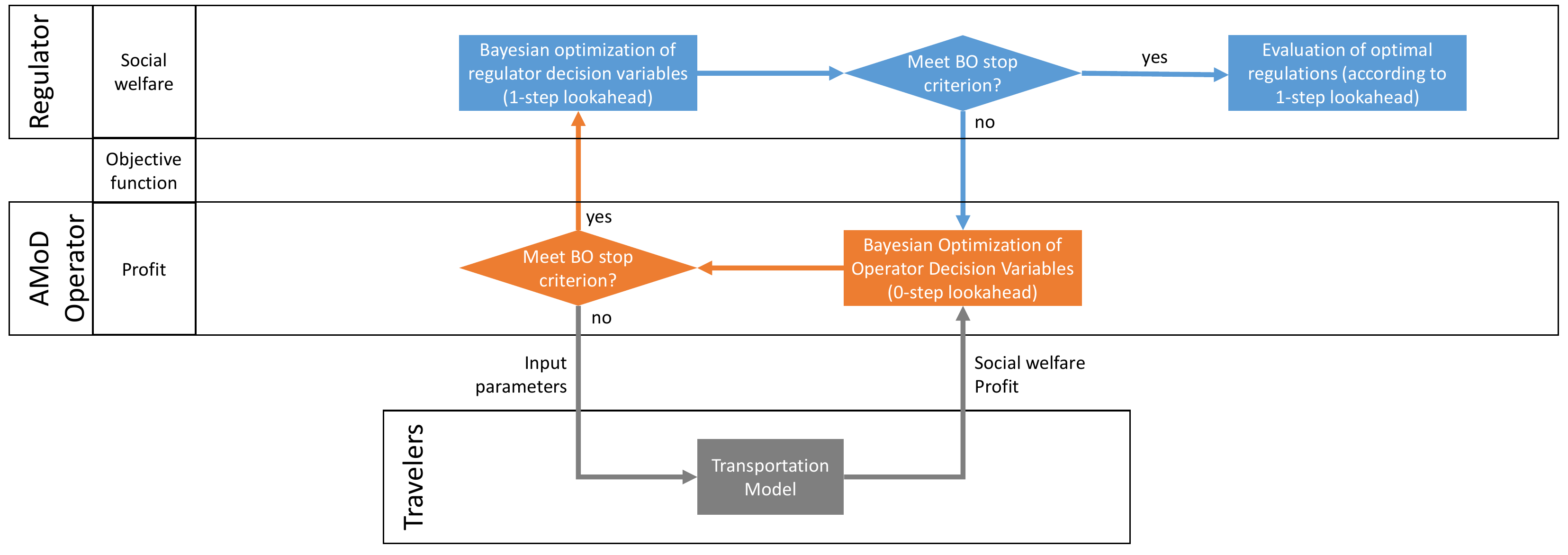}
    \caption{Flowchart describing the processes of the solution approach on a high level.}
    \label{fig:high_lvl_flow}
\end{figure}

The two-level optimization process is a two loop process. The  AMOD service planner optimization loop iterates until a solution to Eqn. ~(\ref{eq:op-opt}) is found. Then the process jumps to the regulator level and iterates at this level until the solution converges. Fig.~\ref{fig:high_lvl_flow} displays a flowchart describing the procedure for an arbitrary transportation model that returns $W$ and $P$ given $x^r$ and $y^s$.
\par
A naive approach would re-start the search for $y^{s,*}(x^r)$ from scratch for every hyper-plane in the solution space defined by a new set of regulator variables $x^r$. Conversely, BO can help reduce the number of lower level iterations and thereby overall computational effort.
\par
Both regulator and AMOD service planner level optimization are non-convex (multi-dimensional) problems. In these types of problems, an optimization algorithm needs to incorporate gradient search techniques to move closer to optima, but avoid getting stuck in local optima. Since a single function evaluation means running a full transportation model simulation in this study, the number of explicit evaluations should be as small as possible. BO in general is an iterative process for multi-dimensional non-convex optimization problems with two components: a \textit{surrogate model} representing the objective function and an \textit{acquisition function} determining the next point in the solution space for which the original function, in this case the transportation model, should be evaluated.
\par
The task of the surrogate model is to infer knowledge from prior iterations; the task of the acquisition function is to exploit this knowledge to search areas where good objective function values can be expected while also exploring the solution space to avoid ending up in a local optimum.
\par
Additional to this nice general feature, BO provides another key advantage in the regulator-AMOD service planner setting:  the surrogate function can infer approximations for the profit-optimal solution $y^{s,*}(x^r)$ from prior iterations in hyper-planes with different regulatory setting $x^{r,'}$. The surrogate function develops a sense of closeness in both $x^r$ and $y^s$ directions with regard to the underlying profit function $P(x^r, y^s)$. The 'closer' another setting $(x^{r,'}, y^{s,'})$ is, the larger its effect on the inference of unknown data points. This exploitation of prior knowledge reduces the number of required iterations.

\paragraph{Gaussian Process}
In this study, Gaussian Processes are used to model surrogate functions for AMOD profit $P(x^r, y^s)$ and social welfare. In order to anticipate an AMOD operator's reaction, the surrogate function for social welfare actually approximates $W(x^r, y^{s,*}(x^r))$. It considers only the social welfare data points from profit-optimized variables in the solution space $(x^r, y^{s,*}(x^r))$. Hence, this social welfare surrogate function is defined on the space of the regulator variables.
\par
Surrogate functions should be smooth representations of the true underlying function. They are estimated from prior data points. Like many other BO models, this study applies a Gaussian Process to infer the surrogate function value of any point in the solution space. Additionally, Gaussian Processes provide estimates of the uncertainty associated with an estimation. In general, this type of surrogate function is described through normal distributions $\mathcal{N}(\mu, \sigma)$ at each point of the solution space. For this section, let $f:X \to \mathbb{R}$ be an underlying function and $x_p, f_p=f(x_p)$ denote data points from prior explicit evaluations of $f$. Then a Gaussian process denotes a surrogate function
\begin{equation}
\label{eq:GP1}
f^S(x) \sim \mathcal{N}(\mu(x), \sigma(x))
\end{equation}
Both $\mu$ and $\sigma$ are inferred from the set of prior data points $(x_p, f_p)$ with the help of covariances. Let $X = \cup^{p}x_p$ be the set of prior solution space data points, $x, x_1, x_2 \in X$ be points in the solution space and $k(x_1, x_2)$ be ther covariance of two data points in the solution space. Together with Eqn.~\ref{eq:GP1}, the following equations describe a Gaussian process:
\begin{align}
\mu(x) &= \sum_{p \in X} \sum_{q \in X} k_p(x) \mathbf{K}^{-1}_{pq} f_q \\
\sigma^2(x) &= k(x,x) - \sum_{p \in X} \sum_{q \in X} k_p(x) \mathbf{K}^{-1}_{pq} k_q(x)
\end{align}
where $\mathbf{K}$ denotes the constant covariance matrix, where each entry $\mathbf{K}_{pq} = k(x_p,x_q)$ for all prior data points $x_p, x_q$. Furthermore, $k_p(x) = k(x_p,x)$ describes the covariance between a prior data point and the point in the solution space $x$, for which the Gaussian Process should be computed. $\mu(x)$ can be used as surrogate function and $\sigma(x)$ gives information about the uncertainty of the approximation. This study employs the commonly chosen Matern kernel:
\begin{align}
k(x_1, x_2) &= \frac{1}{2^{\zeta - 1}\Gamma(\zeta)}\left( 2\sqrt{\zeta} \left\| x_1 - x_2 \right\| \right)^\zeta H_\zeta \left( 2\sqrt{\zeta} \left\| x_1 - x_2 \right\| \right)
\end{align}
where $\Gamma$ is the Gamma function and $H$ the Bessel functions of order $\zeta$ and $\zeta = 5/2$ is set as in~\citep{Liu2019}. Essentially, the covariances determine the contribution of each prior observations $f(x_p)$ on the approximation of $f$ for an arbitrary point $x$ in the domain. 

\paragraph{Acquisition Function}
The iterative nature of the BO is controlled by the acquisition function. The point $x_{n+1} \in X$, for which the function $f$ should be evaluated explicitly in the next iteration is determined by the maximization problem:
\begin{equation}
x_{n+1}  = \argmax_{x} A_n(x)
\label{eq:bo_step}
\end{equation}
where $A_n: X \to \mathbb{R}$ is the $n$th-step acquisition function. This function in general has the same dimension as the data points and (in the best case) the unknown underlying function $f$. However, with the help of $\mu(x)$ (representing the surrogate function) and $\sigma(x)$ from the Gaussian Process, this optimization is performed on an analytical, smooth and differentiable function defined over the whole domain with cheap function evaluations. 

The surrogate function is an approximation of the underlying function $f$. If the surrogate function would be chosen as the acquisition function, the iterative process is likely to end up in a local rather than the global minimum. It might be possible that the approximation further away from the data points is rather bad and even though the surrogate function decreases, the underlying function increases. To avoid this phenomenon, the acquisition function generally contains a measure of the uncertainty as well. Various possible definitions exist in literature, see e.g.~\cite{Liu2019}). The Upper Confidence Bound method is applied in this study:
\begin{align}
A_n(x) &= \mu_n(x) + \kappa_n \sigma_n(x) \label{eq:acquitision_func}\\
\kappa_n &= \sqrt{2 \log \left(\frac{\pi^2 n^{2+d/2}}{3 \delta} \right)} \label{eq:kappa_def1}
\end{align}
where $d$ is the dimension of the solution space and $\delta=0.1$ is chosen as in~\cite{Srinivas.2012}. The inclusion of $n$ in the definition of $\kappa_n$ helps to adapt the weight of the uncertainty for higher iterations. However, it also causes the algorithm to keep searching in regions with low expectation value in case there are a larger number of data points. In order to exploit prior knowledge to a higher degree, $\kappa_n$ was limited to $1$ for the later stage of the optimization process.

\paragraph{Parallel Initialization of Two-Level BO Process}
In this study, BO is used for both optimization levels. The regulator optimization represents an ordinary BO with the variables $x^r$. However, the AMOD service planner optimization is only applied in a hyper-plane, namely the AMOD service design variables for a given set of regulatory variables. Hence, the optimization only takes place along $y^s$, but data points outside of the $x^r$ hyper-plane can help infer the surrogate function for the optimization $P(x^r, y^s)$.

The initialization process is divided in two stages: first, the operator variables $y^s$ are optimized for the case of no regulations. An initial set of simulations are generated by Sobol sequences in the operator solution space. This low-discrepancy sequence method creates a good spread of data points in high-dimensional spaces. Moreover, the corners of the variable space are added to the initial set of simulations. Finally, the BO method determines the profit-optimal solution in the operator space while keeping the regulator variables on the status quo.

After determining the profit-optimal solution in the operator space, an initial set of data points in the regulator space are created by Sobol sequences (with the corners of the regulator space). By initializing  processes for a multitude of regulatory settings $\{x^r\}$ in parallel and joining the information (data points) after each evaluation, each operator optimization process benefits from the results of 'close' regulator-variable hyper-planes, which can reduce the number of necessary iterations even further.

\section{Model}
\label{sec:model}

\begin{figure}
    \centering
    \includegraphics[width=1.0\textwidth]{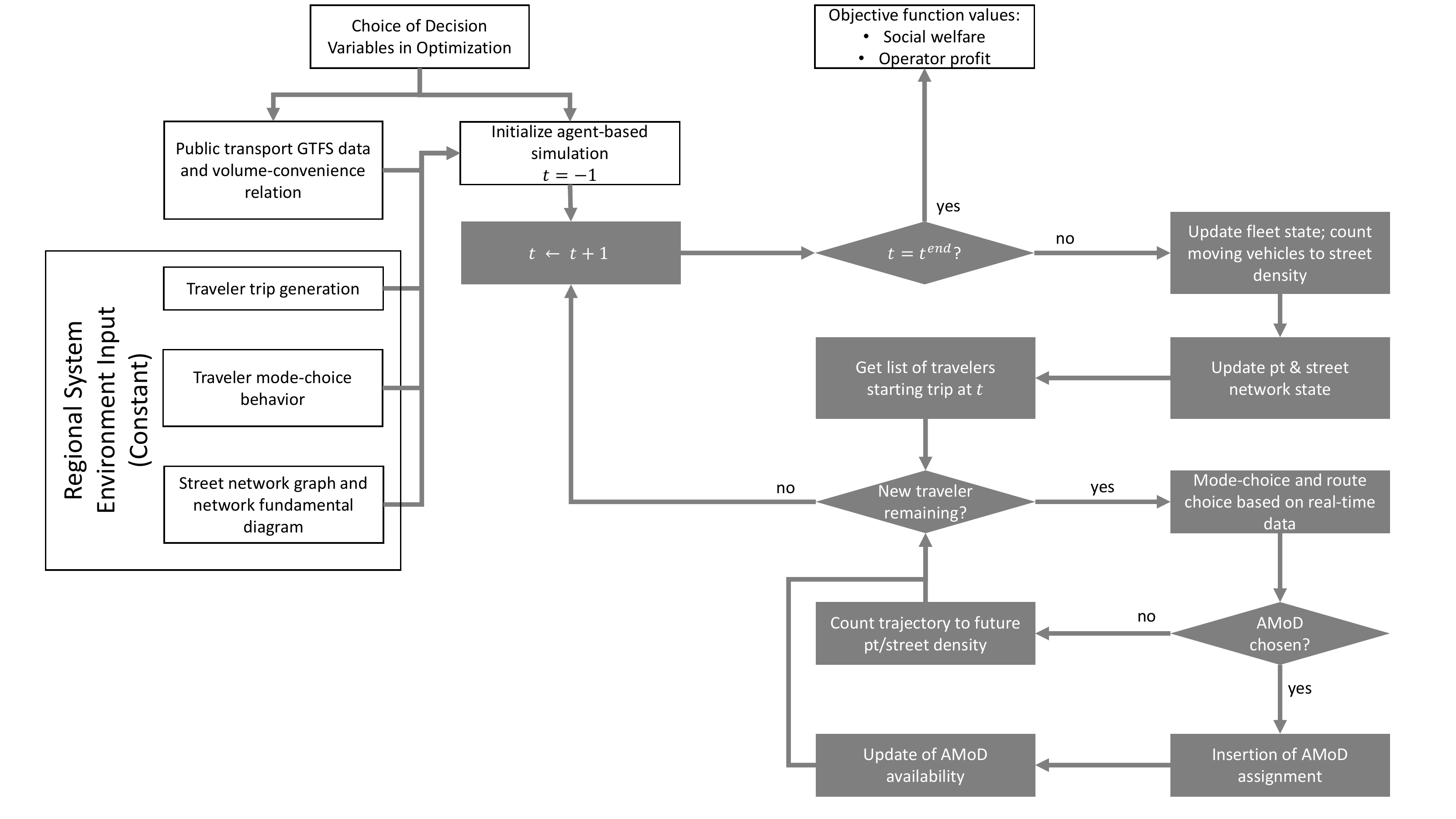}
    \caption{Flowchart describing the processes of the transportation model.}
    \label{fig:transportation_model_flow}
\end{figure}

The solution approach described in the previous section is general, and valid with any transportation model. However, this study proposes an agent-based transportation system simulation model at the lowest level, in order to model the experiences of individual travelers (particularly, AMOD users); the impacts of AMOD service decisions and operational strategies; and the impacts of specific regulation, in the appropriate detail to capture the experiences of travelers and the system impacts. This agent-based transportation model can evaluate social welfare $W$ and profit $P$ for a certain set of regulatory and AMOD parameters $x^r$ and $y^s$. 
The transportation model executes a loop (with index $i$) over time with time steps of size $\Delta t$ ($t \leftarrow t + \Delta t$); the time-step is one minute in this study. Figure~\ref{fig:transportation_model_flow} illustrates the model inputs and process flow at a high level. The following subsections describe each of the components in more detail.

Computation time is one of the key challenges of this tri-level optimization problem as the upper level requires multiple runs of the transportation model. Hence, the subsequently described model aggregates where possible while keeping the agent-based nature of travelers and AMOD vehicles. Moreover, cities are usually not spatially homogeneous regarding the density of inhabitants, work places, PT infrastructure, etc. To address the trade-off between computational efficiency and modeling accuracy, the proposed model separates the study area into two zones $Z_c$: an inner city zone $Z_I$ and an outer city zone $Z_O$. Note that the model is not limited to two zones and could be extended to more zones. More zones would require the specification of additional input parameters and decision variables.
\par
The transportation model must be sensitive to the upper level decision variable inputs. However, the model can remain relatively aggregate for portions of the transportation system that are not significantly impacted by potential AMOD regulations. Hence, the selection of decision variables influences the types and resolution of the transportation model components.

\subsection{Selecting a Set of Decision Variables}
There are many policies a city can test to improve the performance (i.e. social welfare) of a transportation system. Usually, these policies are studied independently of each other, thereby ignoring potential synergistic and opposing effects among multiple policies. This study and especially the general problem (section~\ref{sec:general_prob}) and high-level solution approach (section~\ref{sec:high_level_solution_approach}) aim to enable the optimization and analysis of a combination of policies. Nevertheless, the study focuses on a selected set of policies and AMOD service planner responses (i.e. decision variables) to keep the system computationally tractable.
\paragraph{Internalization of External Costs of Private Vehicles} Even though PVs wear the streets, cause congestion, and emit harmful pollutants as well as require valuable space for parking in urban areas, in most cities, PV owners do not pay (i.e. internalize) these cost. Higher costs for parking spaces and a toll for driving on urban roads could decrease the attractiveness of using the PV. AMOD services also utilize the street network and are not exempt from parking fees and road tolls in general. However, as they are likely to spend much less time parking~\citep{Fagnant2014} and might have their own depots, this study assumes that they do not need to pay parking fees.
\paragraph{PT budget} If the PT system wants to increase demand for PT, they should improve PT service quality. In most US cities to improve service quality, the regulator would need to provide a larger budget to the PT agency, who would subsequently determine how to improve their transit service. Increasing frequencies is necessary to accommodate more travelers and improve their comfort as the PT vehicles are less crowded.
The costs for increasing frequencies can serve as an input. However, it is not clear how many people will choose PT as their mode; the revenue and therefore the required PT budget are an output of a simulation run. Hence, an iterative process would be necessary with PT budget as regulator variable. For simplicity, the regulator can directly influence the PT frequencies in this model.
\paragraph{Fleet size limitation} City administrations could enforce an upper bound on the AMOD fleet size. This could be achieved by licensing of AMOD vehicles. Only licensed AMOD vehicles would be allowed to serve customers and the regulating authority controls the number of licenses that are handed to the AMOD provider.
\paragraph{AMOD service planner decision variables} The operation of AMOD services can vary in many details. The most influential service design parameter is fleet size. It determines the maximum number of travelers that can be served at one time and is one of the main factors for vehicle availability and traveler wait times. If a city enforces a fleet size limitation, the upper bound of this variable is set by the regulator. This study also looks into the price structure of an AMOD service planner as this influences the demand and thereby indirectly the vehicle movements and fleet distribution.

AMOD service planners or at least AMOD operators might also use a repositioning strategy to improve fleet efficiency. However, these strategies would add additional decision variables and require an estimation of future demand and supply, which is not available in this study because demand is endogenous and cannot be derived from historical data. For this reason, this study does not include repositioning.

Since substantial improvements to the transportation system by a pure ride-hailing service without shared rides are unlikely, this study assumes that the regulator constrains the AMOD operator to provide a pooling service. AMOD customers may be driven alone, but will share their ride if a match is possible.

\subsection{Traveler Mode Choice Model}
\label{sec:trav_model}
A traveler $i$ is modeled as an agent that wants to travel from an origin $o_i$ to a destination $d_i$ at a time $t_i$. Both $o_i$ and $d_i$ are within the boundaries of a pre-defined study area. For mode choice, the traveler considers following travel options $m$: PV, PT (with walking access/egress), and AMOD.

As stated in the problem formulation, each agent $i$ tries to maximize the utility related to a trip. In order to represent stochastic differences in travel behavior, mode-choice is determined by a logit model. The probability of traveler $i$ choosing mode $m$ is given by:
\begin{equation}
P_i(m) = \frac{\exp[U_i^m]}{\sum_m \exp[U_i^m]}
\end{equation}
where $U_m$ are the utilities of the modes. This study models the utility of each mode as the negative of the mode's traveling costs, i.e. $U_i^m = -C_i^m$. For the sake of simplicity, the case study uses one value of time $c^{VOT}$ for all time-related factors (e.g. in-vehicle time, wait time, walk time, etc.) in the utility/cost function.
\par
For PV trips, the fastest route is considered. The travel time $t_{od}$ is weighted by value of time $c^{VOT}$, the travel distance $d_{od}$ is weighted by the distance-dependent vehicle costs $c_{D}^{PV}$ and the utility is also impacted by the toll $C_{od}^{toll}(t_i)$, and parking fees $C^{park,PV}_{od}(t_i)$ for OD-relation $od$ of traveler $i$ at time $t_i$. The explicit toll and parking fees depend on the decisions of the regulators and are described in sections~\ref{sec:road toll} and~\ref{sec:park costs}, respectively. Additionally, the model includes an intercept term $u^{PV}$ to calibrate the modal split in the model in the absence of AMOD service and to reflect the convenience travelers typically associate with a PV, as well as unobserved attributes.
\begin{equation}
\label{eq:pv_mode_choide}
C_i^{PV} = c^{VOT} \cdot t^{PV}_{od} + c^{PV}_{D} \cdot d_{od} + C_{od}^{toll}(t_i) + C^{park,PV}_{od}(t_i) + u^{PV}
\end{equation}
\par
For PT, the model considers the fare $f_{od}^{PT}$, the access and egress distances $d_i^{walk}$ to and from PT stops with a walking speed $v^{walk}$, the number of transfers $N_T$, the transfer penalty $c^T$, crowding of the PT system $\eta$, and the travel time $t_{od}$. Each transfer represents a risk of additional travel time. Delays on one line can cause a traveler to miss the next connection; the resulting additional time for the traveler depends on the frequencies of the PT lines. Furthermore, crowding impacts utility through the value of travel time with the help of a monotonously increasing function $g^{VOT}(\eta)$ described in ~\ref{sec:appendix}. In other words, the perceived travel time increases with the level of crowding in vehicles.
\begin{equation}
\label{eq:mc_PT}
C_i^{PT} = f_{od}^{PT} + (g(\eta) \cdot c^{VOT}) t^{PT}_{od} + c^{VOT} \cdot v^{walk} \cdot d_i^{walk} + c^{VOT} \cdot t^{PT}_{wait} + c^T \cdot N_T 
\end{equation}
\par
For AMOD, the model assumes that the fare, the travel time, and the wait time affect the utility associated with AMOD and thus the mode choice of travelers. The AMOD service price/fare is one of the AMOD service planner's decision variables and will be described in more detail in section~\ref{sec:AMOD_fare}.
\begin{equation}
C_i^{AMOD} = f_{od}^{AMOD} + c^{VOT} \cdot t_{od} + c^{VOT} \cdot t^{AMOD}_{wait}
\end{equation}

\paragraph{Modeling Assumptions}
As this study aims to estimate the long-term effects rather than the short-term effect that AMOD can have on PV ownership and the transportation system, travelers do not have a PV ownership attribute. Instead, all travelers are given the choice to use PV, but at full costs (i.e. $c_D^{PV}$ includes depreciation, insurance, maintenance, fuel, etc.) instead of just operating costs. 
\par
For the sake of simplicity, the model does not incorporate impact of dynamic costs (toll, AMOD pricing) on the departure time choice of travelers.

\subsection{Network Model}
Route choices affect the costs of all modes. As modeling route choices is a computationally expensive process, the proposed framework pre-processes both the PT and the street network and makes some simplifying assumptions that enable computationally efficient network modeling. PT stations are matched and connected to the closest street network node with zero cost. Traveler origin and destination locations are matched to the nearest street network access/exit nodes.

\subsubsection{Street Network and Routing Model}
\label{sec:street_network}
PVs and AMOD fleet vehicles drive in the street network graph $G=(N,E)$. The attractiveness/utility of these two modes is severely influenced by route travel times in the network. The travel times, in turn, depend on the state of the network, which is affected by the number of vehicles on the streets. In order to be useful in the larger modeling framework, the street network model needs to reflect traffic dynamics. Furthermore, spatial granularity is necessary to model the spatio-temporal availability of AMOD vehicles and pooling of travelers for shared rides. Modeling AMOD in a microscopic traffic simulation~\citep{Dandl.2017} would satisfy both criteria, but is computationally very demanding and not suitable for the upper level optimizations.
\par
This study employs an approach that combines spatially granular but computationally efficient routing with macroscopic traffic dynamics. The approach involves pre-processing all vehicle paths based on free-flow velocities. The travel times on each path are scaled using factors $\psi^t_c$ derived from separate NFDs for the outer and inner city. Hence, vehicles always travel on the path that has the shortest travel time under free-flow conditions. However, the edge (i.e. link) travel times in the transportation model simulation are scaled according to the traffic state on each edge. The travel time on edge $e$ in cluster/zone $Z_c$ at simulation time $t$ is given by
\begin{equation}
t_e^t = \psi_c^t t_e
\end{equation}
where $t_e$ is the free-flow travel time on this edge. The cluster travel time factors $\psi_c^t$ are updated each time step based on the density in the respective part of the street network. The density comprises the number of PVs and AMOD vehicles driving at time $t$, as well as the vehicles starting or ending their trips outside of the study area, which only act as background traffic. The number of background traffic vehicles needs to be calibrated. The flow $q_c^t$ in cluster $Z_c$ at time $t$ is derived from the NFD $q_c = q_c(k_c)$.
\par
As in~\cite{Dandl.2020ITSC}, the travel time factors are given by the functional form
\begin{equation}
\psi^t_c \sim v_{1,c} \left(\frac{k_c^t}{q_c^t} + \frac{1}{v_{2,c}} \right)
\end{equation}
and the parameters $v_{1,c}$ and $v_{2,c}$ are fitted according to a travel time comparison with a dynamic traffic assignment simulation. In order to reduce oscillations, the moving average of the last 5 time steps is used to update the travel time factors.
\par
In order to reduce the computational burden, PVs are not rerouted and tracked in the model. Only AMoD vehicle states are updated each time step because they affect the offer the AMoD operator can make to new travelers and therefore their decisions.
\par
This NFD-based approach to scaling network travel times precludes meaningful evaluation of route choices. However, this approach effectively approximates the quantity affecting traveler's mode-choice and AMOD vehicle-passenger matching: the travel time between two points.

\subsubsection{PT Network Model}
A PT system includes the spatial network layout of all lines, the schedule for each each line, as well as the specification of vehicle types (e.g. buses and trams).

Such PT systems are specified in the widely used GTFS format\footnote[1]{https://developers.google.com/transit}. GTFS specifies the PT system including schedule, stops, routes, and vehicles. To process routing queries in such a system,  OpenTripPlanner\footnote[2]{https://www.opentripplanner.org/}, which is based on OpenStreetMap\footnote[3]{https://www.openstreetmap.org} is a useful tool. By specifying origin and destination, as well as a departure time, possible PT trips are found. Each trip includes access, egress, and transfer walking distances, total travel time and number of transfers. These parameters are required inputs for the traveler choice model. 

This study aims to investigate a large operating area and thus computational time is of essence. In order to increase efficiency, routing queries in the PT system between all stop-to-stop combinations are pre-processed. The result is a data table (i.e. skim matrix) in which the walking distance, travel time, and number of transfers for each stop-to-stop combination is stored. This procedure allows to reduce the OpenTripPlanner query to a simple look-up in the pre-processed data thereby substantially decreasing the computational cost of such queries. As the waiting time at the station depends on the schedule and whether the PT service operates on time, it is assumed that the traveler arrives just in time, i.e. with $t^{PT}_{wait} = 0$.
\par
Another PT attribute affecting a traveler's mode choice (Eqn.~(\ref{eq:mc_PT})) is the crowding within PT vehicles. This is modeled in an aggregated fashion similar to the street network. Whenever a traveler chooses PT, the system-wide count of PT travelers increases by one for every time step -- until the arrival. Crowding represents the equivalent of congestion of road-bound traffic. The literature suggests that the perceived travel time of (standing) passengers increases mostly linear with the number of passengers per vehicle. Therefore, the effects of crowding on the travel time can be represented as a (value of) travel time factor. Studies for various large cities suggest a maximum factor of 2.0~\citep{tirachini2017estimation}. Due to the lack of corresponding studies for the city of Munich, the average maximum factor to model the effects of travel time is applied in this case study. The crowding factor $\eta^t$ is defined as the ratio between the number of PT travelers $n_t^{PT}$ and the total PT capacity $\Omega_T^{PT}$.

\begin{align}
\eta^t &= \frac{n_t^{PT}}{\Omega_T^{Pt}} \label{eq:PT_crowding_factor} \\
\Omega_T^{PT} &= \frac{1}{T}\sum_{l} \sum_{\zeta_l^{T}} \Omega_l^{PT} =  \sum_l \frac{\vert\zeta_l^T\vert}{T} C_l^{PT} = \sum_l \nu^T_l C_l^{PT} \label{eq:PTcapacity}
\end{align}

where $\Omega_l^{PT}$ is the capacity of a PT vehicle on line $l$ (train, tram or bus) and $\vert\zeta_l^T\vert$ is the number of trips on this line during a time interval $T$ (chosen to be one hour) and $\nu_l^T$ the derived frequency. Similar to the street network, PT users traveling from or to areas outside of the study area need to be included in a calibration process.
\par
For the estimation of social welfare, operating costs and emissions have to be evaluated. The absolute value is actually not important to the framework, only the difference in measures across transit line frequencies. Therefore, operating cost $C_T^{PT}$ and emissions  $E_T^{PT}$ of the PT system during time interval $T$ are approximated based on the frequencies:
\begin{align}
C_T^{PT} &= T \sum_l \nu^T_l d_l c_l \\
\label{eq:PT_capacity}
E_T^{PT} &= T \sum_l \nu^T_l d_l e_l
\end{align}
where $d_l$, $c_l$ and $e_l$ are the mean length, cost, and emissions of a vehicle trip (from start to end station) of line $l$, respectively. The assumed cost values reflect typical per schedule-km costs beyond operating costs. Together with the revenue of PT travelers, they represent the PT budget.

\subsection{AMOD Fleet Control Model}
AMOD vehicles are explicitly represented as agents in the transportation model, which are controlled centrally by the AMOD operator. At the beginning of every time step, their state is updated. In this update phase, the vehicles move and customers board and disembark vehicles according to their respective currently assigned plan. The operator can update vehicle plans every time step according to the fleet control model, i.e. the operator can insert stops for the pick-up and drop-off of new customers into existing routes of vehicles regardless if they are currently idle, in the process of boarding, or moving with or without passengers on board. This section first elaborates on general considerations for the fleet control sub-problem in the context of this study's transportation model before introducing a mathematical description of the model.
\par
AMOD fleets can be utilized for autonomous ride-hailing and ride-pooling services. This study assumes that the regulator only allows mobility providers with pooling service in order to enable higer passenger flows on the street network. In a ride-pooling system, the number of possible combinations of request-to-request and vehicle-to-request matches grow exponentially~\citep{AlonsoMora.2017} with problem size (i.e. fleet size and demand rate). Even with advanced graph-based algorithms, parallelization on a large number of CPUs is necessary to optimize matching in city scale systems~\citep{AlonsoMora.2017,Engelhardt.2019}. Furthermore, these optimization procedures assume that users already made a binding request, i.e. the operator knows that the users will use the system~\citep{AlonsoMora.2017}, or assume that all customers accept the current solution and re-optimize the system once the information whether customers accepted the offer is available~\citep{Engelhardt.2019}. In the proposed model, all travelers make requests to the pooling system, but many travelers will not use the AMOD service. For these reasons, a fleet control approach with global ride-pooling optimization is computationally not feasible. Instead local optimization with an insertion heuristic are employed for each traveler sequentially (see Fig.~\ref{fig:transportation_model_flow}). When a traveler makes a request, the AMOD operator returns an offer based on the result of the insertion heuristic. Then, the traveler chooses his/her immediately given the AMOD offer and the attributes of the other possible modes. If the traveler chooses AMOD, the model assigns the traveler to the vehicle and inserts the request inside the vehicle's route. The state of this vehicle is updated, and the algorithm moves to the next traveler request.
\par
The strategy of a private-sector AMOD provider is likely to be profit-driven. Generated profits and consequently, the control strategy, in a ride-pooling system are depending critically on the fare-sharing strategy. For simplicity, the approach includes a simple strategy, where the fare for a trip is only dependent on the distance of the fastest direct route between a traveler's origin and destination. Assuming distance-based vehicle costs, this fare-sharing strategy aligns city and AMOD goals rather well: both want to maximize the difference between passenger mileage and vehicle mileage. Moreover, customers of an AMOD system value low waiting and detour times. Therefore, the AMOD control function objective is defined as
\begin{equation}
\label{eq:fc_obj}
C = c^D_v  \sum_{v \in V} d_v + c^{VOT} \sum_{i \in R^s} \left( t_i^{do} - t_i \right)
\end{equation}
where $t^{do}_i$ and $t_i$ correspond to the expected drop-off and request time of traveler $i$, respectively. $R^s$ is the set of all served customers and $d_v$ is the complete driven distance of vehicle $v$.
\par
The second term in the control function objective in Eqn.~(\ref{eq:fc_obj}) minimizes user detours and wait times. As such, in general, travelers do not have to make very long detours just to share part of their trip with somebody else. Since this is no guarantee for an individual passenger, the model includes a hard constraint for passenger detour to be smaller than $t^{d,rel}_{max}$ \% of the fastest direct route travel time between the traveler's origin and destination. Additionally, hard constraints on the maximal customer wait time $t_w^{max}$ are imposed to increase computational performance. Taking these constraints into account, many possible vehicle-traveler combinations as well as traveler-traveler route permutations can be excluded. Therefore, the first step in the local optimization is to find all vehicles $v$ that are able to pick-up the new traveler $i$ within $t_w^{max}$. For a further speed-up, the insertion of a new traveler request $i$ into the existing route is only performed for $N^{i}_{v,max} = 10$ vehicles in this study: 5 vehicles, which are idle or will become idle next, and 5 vehicles, for which the direction, i.e. the angle to x-axis in a x-y plane, between current location and last stop is closest to the direction between customer origin and destination. The origin and destination of the new traveler are inserted into the currently assigned route $\xi_v(R_v)$ serving the set of customers $R_v$ in all possible positions in the currently assigned stop list (considering that the traveler has to be picked up first). These stop permutations are checked for feasibility next. A sequence of stops is feasible, if all customers in $R_v^i = R_v \cup \{i\}$ can be picked up before their maximum waiting time expires and they are dropped off within their detour time limit. A constant boarding time of $B$ is considered whenever a customer boards or disembarks an AMOD vehicle at a stop. Furthermore, the vehicle cannot accommodate more than $\rho$ passengers at the same time.

If no vehicle can provide a feasible route, no offer is made to the traveler. If there are more than one feasible options, the fleet control objective is computed for all of them. The operator selects the vehicle and tour $\xi_v(R_v^r)$ with the minimal control objective value and creates an offer for the traveler containing the relevant values for mode choice, i.e. fare, wait time and travel time (according to the currently planned route). This study treats pricing as one of the AMOD service planner's decision variables. The pricing model will be described in section~\ref{sec:AMOD_fare}.
 
Both routes $\xi_v(R_v^r)$ and $\xi_v(R_v)$ are kept in memory until traveler $r$ decides which mode will be used. If AMOD is chosen, $\xi_v(R_v^r)$ is assigned to $v$, else $\xi_v(R_v)$ remains the currently assigned route for vehicle $v$. Due to dynamic travel times, it is possible that 
wait and detour time constraints become invalid from one time step to the next. In these cases, the currently assigned solution is still treated as feasible, but the addition of a new request to any tour requires the validity of all time constraints.

\subsection{Modeling the Impacts of Decision Variables}
\subsubsection{Parking Fee Model}
\label{sec:park costs}
Parking fees for PVs are generally dependent on the location and the amount of time a vehicle is parked. In an activity and agent-based travel demand model, the parking time can be approximated very well by the time of a given activity. For trip-based models, the parking costs become much more difficult to approximate. Furthermore, there is no way of keeping track of vehicle consistency, i.e. a traveler who chooses a PV to get to work is very likely to choose the car back on the way home. Unfortunately, only trip-based data is available in the investigated case study (see section~\ref{sec:case_study}). For the sake of simplicity, it is assumed that PV trips in the morning start from a traveler's home and trips in the afternoon end there as well. Additionally, it is assumed that parking at home is free and only parking fees for a constant duration in the intermediary location have to be paid. Parking fees depend on the zone where a PV is parked, whereas for the sake of simplicity it is assumed that parking fees are only introduced in the inner city zones $Z_I$, while there are none in the outer part $Z_O$). For the model of this study, parking fees are equally split across both trips; therefore, the destination of morning trips and the origin of afternoon trips determine the amount:
\begin{equation}
C^{park,PV}_{od}(t_i) = \left\{
\begin{matrix}
x^{P} & t_i < 12:00 ~ \& ~ d \in Z_I \\
0 & t_i < 12:00 ~ \& ~ d \in Z_O \\
x^{P} & t_i \geq 12:00 ~ \& ~ o \in Z_I \\
0 & t_i \geq 12:00 ~ \& ~ o \in Z_O \\
\end{matrix}
\right.
\end{equation}
$t_i$ reflects the time a traveler wants to start her trip and $x^{P}$ the regulator decision variable for parking regulations.\\

\subsubsection{Road Toll Model}
\label{sec:road toll}
Similar to the work of~\cite{Bracher.2017}, a dynamic toll controlled by the NFD is applied in this study. A distance-based toll is used for this study, where only the distance within the toll area in the inner city $\left.d_{od}\right|_{Z_I}$ is relevant. The coefficient is $0$ up until a certain threshold $k_0$. Above this threshold, the toll coefficient increases linearly with the vehicle density in the inner city zone and is scaled by the regulator decision variable $x^{RT}$:
\begin{equation}
C_{od}^{toll}(t) = max \left( \frac{k^t_{Z_I} - k_0}{k_0}, 0 \right) \cdot x^{RT} \cdot \left.d_{od}\right|_{Z_I}
\end{equation}

\subsubsection{PT Frequency Model}
\label{sec:pt freq}
The frequency of PT lines impacts the transportation system in multiple ways. In this model, it affects the total capacity of the PT system and thereby the crowding $\eta^t$ of PT vehicles. Additionally, a higher frequency improves traveler's acceptance of transfers $c^T$ because the average transfer and wait times decrease. For simplicity, the model includes a linear relationship between frequency and the coefficient in the traveler logit model. The effect of scaling the PT frequency by a factor of $x^{PT}$ is modeled as follows:
\begin{align}
\eta^t &= \frac{\eta^t_0}{x^{PT}} \\
c^T &= \frac{c^T_0}{x^{PT}}
\end{align}
where the "0" lower-case index refers to the value from the original data.
\par
Of course, higher PT frequencies require operating more PT vehicles and generate higher costs and emissions. For simplicity, a linear relation between number of trips $n_l^t$ on line $l$ per time period and costs are assumed:
\begin{align}
C_T^{PT} &= x^{PT} \cdot C_{T,0}^{PT} \\
E_T^{PT} &= x^{PT} \cdot E_{T,0}^{PT}
\end{align}

\subsubsection{AMOD Fare Model}
\label{sec:AMOD_fare}
This study incorporates a distance-based fare $y^{PD} d_{od}$, where $d_{od}$ is the direct route distance. Additionally, the model includes a minimum base fare of $f^{AMOD}_{min}$ in order to reduce demand for very short trips. To balance demand and supply,  AMOD service planners use utilization-dependent scale factors to optimize their profit. A simple model is used within this framework: the  AMOD service planner scales offered fares by $y^{PU}$ if at least 75\% of their vehicles are utilized. 

\subsection{Social Welfare and Profit Model}
\label{sec:soc_welfare_profits}
The objectives of regulator and AMOD service planner are social welfare and profit, respectively. The profit can be defined straightforwardly as the difference between revenues from fares of served travelers and operating costs:
\begin{equation}
P = \sum_{i \in R^{s}} f_i^{AMOD} - \sum_{v=1}^{y^F} \left(c_v^F + c_v^D d_v\right) - \sum_{t} C_t^{toll,AMOD}
\end{equation}
where $R^s$ is the set of served requests, $f_i^{AMOD}$ the fare paid by request $i$, $c_v^F$ the fixed costs of a vehicle (e.g. for leasing and insurance),  $c_v^D$ distance-dependent operating costs,  $d_v$ the total driven distance of vehicle $v$ and $C_t^{toll}$ is the total toll of all fleet vehicles moving in the toll area ($\vert V \vert_{Z_I}$) during times with high traffic demand:
\begin{equation}
C_t^{toll,AMOD} = max \left(\mathbf{k}^t_{Z_I} - k_0, 0 \right) \cdot x^{RT} \cdot \vert V \vert_{Z_I}
\end{equation}
\par
The definition of social welfare is more complex and not unique. Every city could weigh different objectives differently or have additional objectives. The objective function in this study includes i) the sum of travelers' utilities of their chosen mode $m*$, ii) revenues and costs for PT, iii) revenues from parking and tolls and iv) emissions of the transportation system.
\begin{align}
W &= \sum_{i} U_i^{m*} + \sum_{i:m_i=PT} f_i^{PT}- C^{PT}\sum_{l,t} n_l^t d_l \nonumber\\
~ &+ \sum_{i:m_i = PV} C^{park,PV}_{od} + \sum_{i:m_i = PV} C^{toll,PV}_{od} + \sum_{t} C_t^{toll,AMOD} \nonumber\\
~ &- C^{CO2} \left( E^{PV}\sum_{i:m_i = PV} d_i^{od} + E^{AMOD}\sum_{v=1}^{y^F} d_v + E^{PT}\sum_{l,t} n_l^t d_l \right)
\label{eq:social_welfare}
\end{align}

\section{Case Study}
\label{sec:case_study}
\subsection{Case Study Description}
\begin{figure}
     \centering
     \begin{subfigure}[b]{0.6\textwidth}
         \centering
         \includegraphics[width=\textwidth]{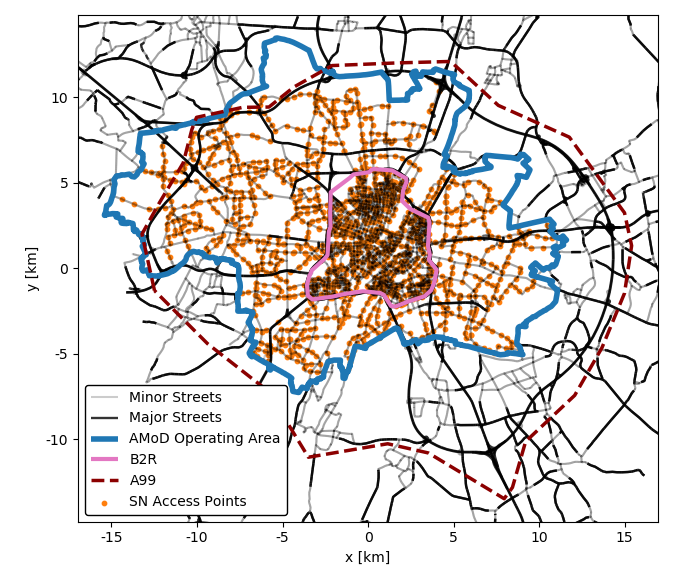}
         \caption{Street Network}
         \label{fig:PVnetwork}
     \end{subfigure}
     \hfill
     \begin{subfigure}[b]{0.6\textwidth}
         \centering
         \includegraphics[width=\textwidth]{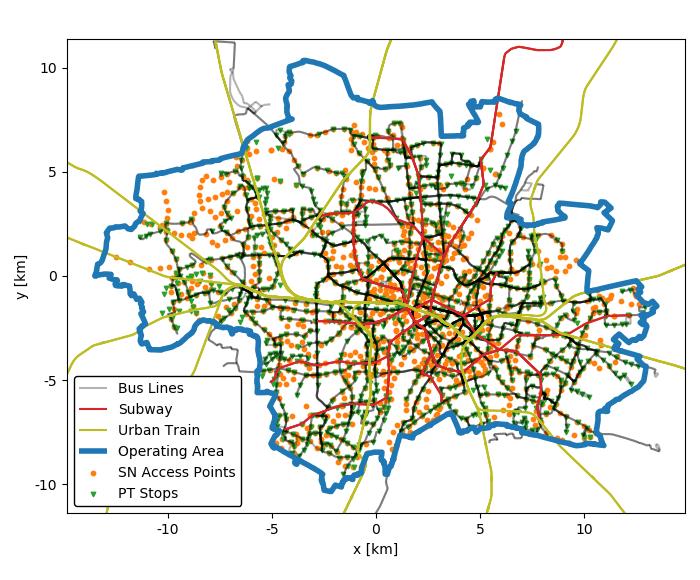}
         \caption{PT Network}
         \label{fig:PTnetwork}
     \end{subfigure}
    \caption{Street network (a) and PT network (b) of the city of Munich used in this case study. Blue shows the AMOD operating area and orange points correspond to street network (SN) access points, where travelers enter and leave the simulation. Road tolls and parking fee regulations are applied in the "iB2R" area with the "B2R" as outer boundary (a). The area "iB2R" and the remaining study area denoted by "LHM-iB2R" also resemble the two network clusters for calculating the NFDs. In (b) PT stops area shown in green. To connect the PT network and the street network, each PT stop is connected to the nearest street network node.}
    \label{fig:munichnetworks}
\end{figure}

The city of Munich, Germany, is the focus of this case study. The transportation model includes the street network and PT system up to the outer highway belt, as illustrated in Fig.~\ref{fig:munichnetworks}. This highway belt called A99 with a speed limit of predominantly 120 km/h, the inner highway belt denoted as B2R with speed limit of predominantly 60 km/h and highways leading towards the city center with speed limits between 60 and 120 km/h are highlighted in the street network as major streets (see Fig.~\ref{fig:PVnetwork}). Most other streets have a speed limit of 50 km/h (general) or 30 km/h (residential areas). The PT network (see Fig.~\ref{fig:PTnetwork}) consists of two high-capacity rail-based systems (Urban Train and Subway), a few tram-lines, and several bus lines. The train lines are mostly directed to the center and are much sparser in the outer part of the city (outside of B2R).
\par 
The PT mode share in Munich is $40\%$ within the study area (operating area of AMOD). However, the high-capacity trains and corresponding stations in the city center are often overly crowded resulting in stations which act as bottlenecks in the system. Additionally, trips starting or ending in the outskirts of the city require time-consuming access and egress trip legs to/from transit which might reduce the attractiveness of high-capacity train lines. Moreover, lower real-estate prices in the outer regions of Munich increases the attractiveness of owning a PV in such regions. As a consequence, the modal share of PVs in Munich is large leading to one of the most congested street networks in Germany\footnote{https://inrix.com/press-releases/2019-traffic-scorecard-german/}.

\subsection{Network and Demand Model}
\begin{figure}[th]
\centering
\includegraphics[scale=0.5]{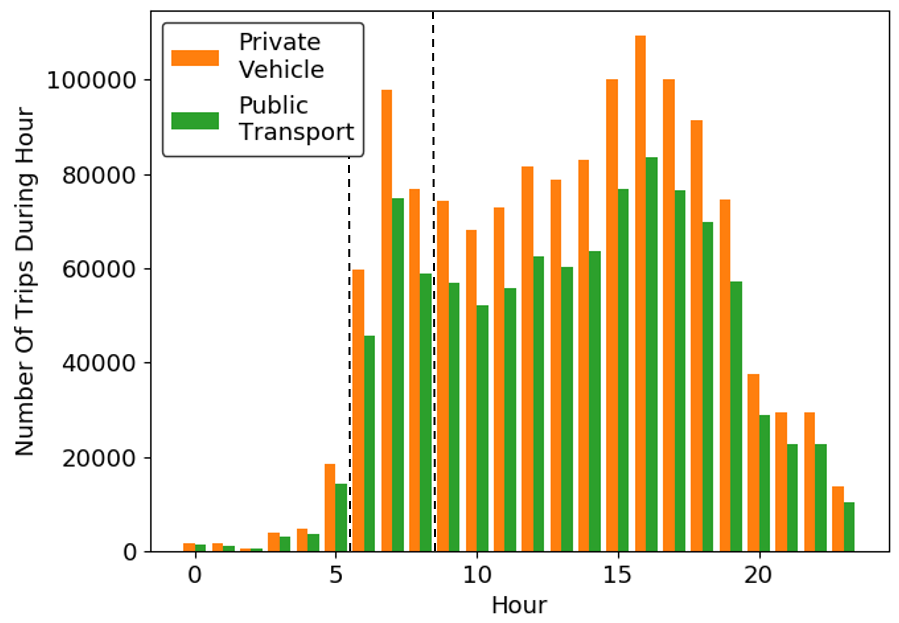}
\caption{Hourly trips within AMOD operating area from PT and PV trip data. Overall the data covers 2.3 million trips used as travelers in this study with a mode share of $56.6\%$ PV and $43.4\%$ PT. Because of computational complexity, the analysis only simulates the morning peak from 6am to 9am -- indicated by the dashed lines. The morning peak contains $18\%$ of all trips.}
\label{fig:hourlydemand}
\end{figure}

The street network graph $G = (N,E)$ was exported from a microscopic traffic simulation model~\citep{Dandl.2017}. Travelers can access or leave the simulated network at 1305 access points marked in orange in Fig.~\ref{fig:munichnetworks}. These points are generated at intersections with the following characteristics: 
\begin{itemize}
    \item Omit intersections connected to major roads (speed limit higher than $60 km/h$).
    \item Prioritize intersections close to PT stops.
    \item Distribute the access points homogeneously such that at least one access point is reachable within $300 m$ walking distance.
\end{itemize}
\par
PT stops, lines, and schedules were extracted from publicly available GTFS-data\footnote{https://gtfs.de/}. Travelers can access or leave the PT network at the PT stops marked in green. Each PT stop is connected to the nearest street node. Additionally, each street network access point connects to the nearest PT network stop.
\par
The AMOD operating area is shown in blue in Fig. \ref{fig:munichnetworks}. This area corresponds to the municipal area of the city of Munich. Only trips starting and ending within this operating area qualify as possible AMOD trips. In this study, the  AMOD service planner employs vehicles with a capacity of $\rho = 4$.
\par
To create trips, OD matrices referring to PV and PT trips are defined. Hourly PV trip OD-matrices are extracted from the microscopic traffic simulation model~\citep{Dandl.2017} containing an hourly time-dependent OD-matrix for a working day. Daily PT OD-matrices are extracted from a macroscopic traffic model for the region of Bavaria~\citep{Maget.2019}. Only OD pair trips with starting and ending locations within the AMOD operating area are considered as travelers within the simulation. Since the macroscopic traffic model for the region of Bavaria only contains information on trip counts for one whole day, the hourly aggregated PV trip counts within the operating area are utilized to disaggregate the PT OD-matrices for different hours of the day. 
\par
Overall, the data includes 2.3 million daily trips within the operating area with a modal split of $56.5 \%$ PV and $43.4 \%$ PT. The resulting time-dependent trip counts are shown in Fig.~\ref{fig:hourlydemand}. The transportation simulation model generates travelers by applying Poisson processes with rates corresponding to the respective OD-matrix entries. Street network access points within a particular origin and destination area are chosen randomly as traveler origins and destinations, respectively.
\par 
NFDs for two separate regions are estimated based on data from the microscopic traffic simulation model. One refers to the city center (zone inside B2R in Fig.~\ref{fig:PVnetwork}), and the other one to the outer city area including the highway belt (B2R to A99 in Fig.~\ref{fig:PVnetwork}). The microscopic traffic simulation model also includes PV trips passing, entering, or leaving the A99 area as background traffic as these trips are not represented in the traveler list. To calculate background traffic density the model creates two additional traveler sets. The first one includes all PV trips available in the data, while the second one only includes PV trips with origins and destinations within the AMOD operating area. The simulation is run with both traveler sets while enforcing the travelers to always choose PV. The difference between the measured hourly averaged network densities for each NFD-area is considered the background traffic density.
\par 
The time-dependent PT capacity is calculated using the GTFS data by evaluating Eq.~\ref{eq:PT_capacity}. The PT background travelers (leaving or entering the study area) are determined the same way as previously described for PVs. Two traveler sets are created: one includes all PT trips in the data and the other one only includes travelers with both origin and destination within the AMOD operating area. A simulation for each traveler set is run while enforcing the travelers to chose PT. The hourly average difference in PT travelers is used in the model as background value for calculating Eqn.~\ref{eq:PT_crowding_factor}.
\par 
By treating travelers from outside the A99-area as background traffic these travelers are effectively exempted from mode choice decisions. Nevertheless, only a minority of travelers, who also have no access to the AMOD service, are affected by this assumption, while computational complexity is reduced.
\par
Simulations for calibration showed that the crowding factor in the PT system reached a level of approximately 0.2 for realistic modal splits between PV and PT. In the real PT system, central stations are highly crowded while many buses, especially in the outer regions, are rarely crowded. In order to model this inhomogeneity in Munich's PT system, where a large share of PT travelers experience crowding, the total capacity computed according to Eqn.~\ref{eq:PTcapacity} is reduced by a factor of 3.
\par
Finally, the mode choice model needs to be calibrated to resemble the mode share between PV and PT in the data. By applying the simulation model without an AMOD service, the PV intercept parameter $u^{PV}$ defined in Eqn.~\ref{eq:pv_mode_choide} is adjusted until a mode share of $43.4 \%$ PT travelers is observed over the course of whole day. The calibrated model uses the value $u^{PV} = -4.70$\euro~. During the morning peak, the PT share amounts to $45.2\%$.

\subsection{Summary of Input Parameters}
\paragraph{Constant Parameters} ~\ref{sec:appendix} includes a table of all input parameters for the transportation model simulations. The results of each set of variables were created from 3 morning-peak (6am-9am) simulations with different random seeds (also in the generation of traveler data sets from the hourly OD matrices).
\paragraph{Decision Variables}
In order to improve the livability in the city, the administration and regulator might apply various policies to regulate traffic in a future with AMOD fleets, especially in the inner districts. The case study incorporates the policies described in section \ref{sec:model}. Table \ref{tab:dec-vars} shows upper and lower bounds for the corresponding regulator and  AMOD service planner decision variables based on discussions with experts from the city administration. The scenario "No Regulation" denotes the set of regulator variables $x^P=2.50$~\euro, $x^{RT}=0$~\euro, $x^{PT}=1$, and $x^F=50,000$ vehicles. Test simulations showed that this number is sufficiently high to not constrain the profit-optimal operator fleet sizes.
\par 
For  application in the Gaussian Process and framework, the variable ranges are linear transformed to [0,1] in order to have comparable scales. These limits act as boundaries in the optimization process of the acquisition function in Eqn.~\ref{eq:acquitision_func}.

\begin{table}[h]
    \centering
    \begin{tabular}{p{15mm}cccp{80mm}}
\toprule
Decision Variable & Player & Range & Unit & Short Description \\
\midrule
$x^{P}$ & R & 2.50 - 5.00 & \euro & Parking fees for PVs ending/starting within toll area (morning/afternoon)  \\
$x^{RT}$ & R & 0 - 1.00 & \euro/km & Dynamic road toll per driven km for AMOD and PVs within toll area \\
$x^{PT}$ & R & 0.25 - 2.0 & ~ & Scale factor to decrease/increase PT frequencies \\
$x^F$ & R & 1-50,000 & vehicle & Number of AMOD vehicle licenses \\
\midrule
$y^F$ & FO & 0-$x^F$ & vehicle & Number of AMOD vehicles \\
$y^{PD}$ & FO & 0.25 - 2.00 & \euro/km & Distance-based AMOD fare \\
$y^{PU}$ & FO & 1.0 - 10.0 & ~ & Scale factor to increase fares in times of high vehicle utilization \\
\bottomrule
    \end{tabular}
    \caption{Decision variables of regulator (R) and AMOD fleet operator (FO) in this study.}
    \label{tab:dec-vars}
\end{table}

\subsection{Second Social Welfare Definition: Pro-PT Scenario}
A second definition of social welfare will be used in the case study to show impacts of other weights of this multi-objective problem. This scenario will be denoted by \textit{Pro-PT} scenario. Following adaptions are assumed:
\begin{align}
C^{CO2} &\rightarrow 25 \cdot C^{CO2} \label{eq:sw_re_def1}\\
C^{park} &\rightarrow C^{park} / 4 \\
C^{toll} &\rightarrow C^{toll} / 4 \\
C^{PT} &\rightarrow C^{PT} / 10 \\
E^{PT} &\rightarrow E^{PT} / 10 \\
E^{AMOD} &\rightarrow 3 \cdot E^{AMOD} \label{eq:sw_re_def2}
\end{align}
The modified social welfare definition puts a much higher weight on the emissions of the transportation system. Moreover, the standard parameter definition assumes that all parking revenues are paid to the city, thereby ignoring private parking garages. The valuation of parking revenue can also be lower to incorporate the space that has to be reserved for parking instead of put to other productive uses. Similarly, the prior definition ignored operating costs of the toll, a valuation that is given for the required space and infrastructure costs for roads. Therefore, the full parking costs (PV users) and toll costs (PV users and AMOD operator) have to be paid, but only one forth is added as revenue for the city administration. The costs for PT could be lowered with automation and the valuation of its costs could also be lower as PT has the obligation to serve travelers, which naturally will lead to higher costs, but could also be valued. A full electrification, right-sizing, and adaption of the schedule could bring significant reductions in PT emissions. Finally, the AMoD operator might use internal combustion engine vehicles and have an increased $CO_2$ footprint.
\par
A first impression to sensitivities of different social welfare parameters can be checked very quickly as these coefficients do not affect the input of simulations. Therefore, all simulations are still valid and the result database from simulations with other SW coefficients can be utilized as long as the database contains values for the respective single social welfare components.

\subsection{Case Study Results}

\subsubsection{Evaluation After Initial Simulations}
\begin{figure}[ht]
    \centering
    \includegraphics[width=0.9\textwidth]{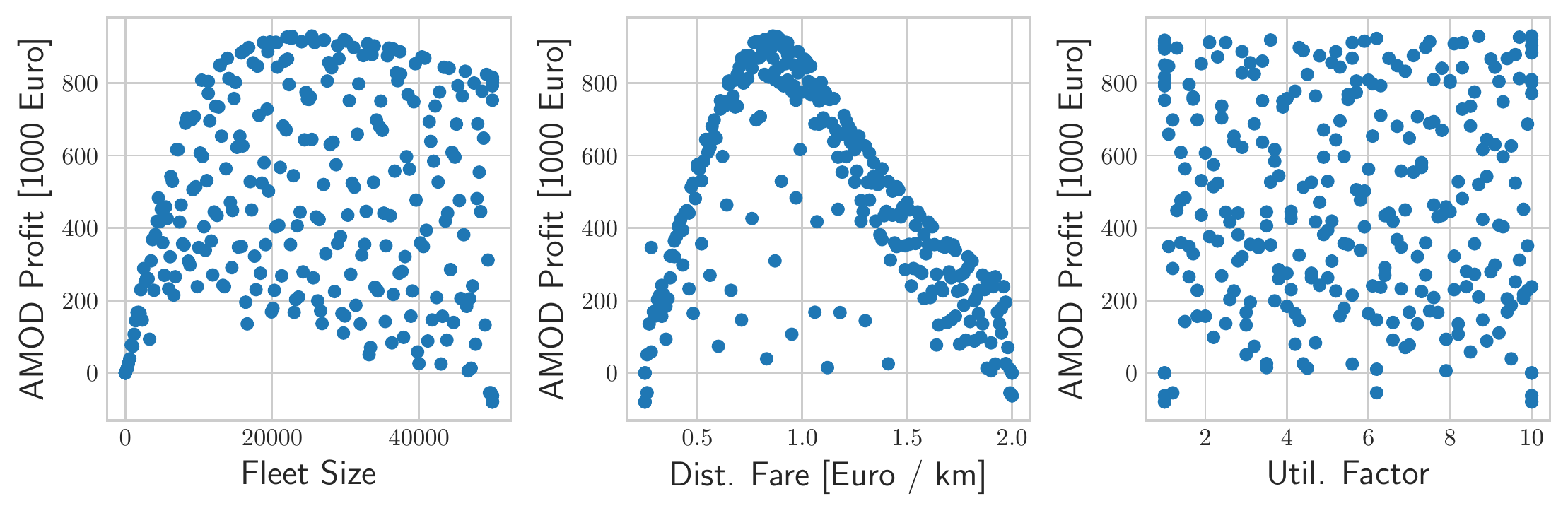}
    \caption{Operator profit over operator variables after the initial set of operator simulations for the case without regulation.}
    \label{fig:init_operator_rev}
\end{figure}
An initial set of 272 operator settings for the no-regulation case revealed that the two operator variables fleet size $y^F$ and distance-based fare $y^{PD}$ are more critical than the utilization-based pricing factor $y^{PU}$.
\par
Fig.~\ref{fig:init_operator_rev} illustrates a sharp increase in operator profit for fleet sizes below 10,000 vehicles. In this region, each addition of an AMOD vehicle generates much more \textit{variable profit}, i.e. revenue minus variable distance-dependent costs, than its fixed vehicle costs. The marginal benefit of adding vehicles decreases up to fleet sizes of 20,000 to 25,000 vehicles before the additional fixed costs of adding vehicles finally become larger than the extra variable profit. Interestingly, the curve does not show a steep decline of profit for larger fleet sizes. This means that vehicles do not have to be utilized the whole time to be profitable. It can be expected that full-day simulations would create a steeper curve since fewer vehicles would be able to serve the demand in off-peak hours.

\begin{figure}[th]
    \centering
    \includegraphics[width=0.8\textwidth]{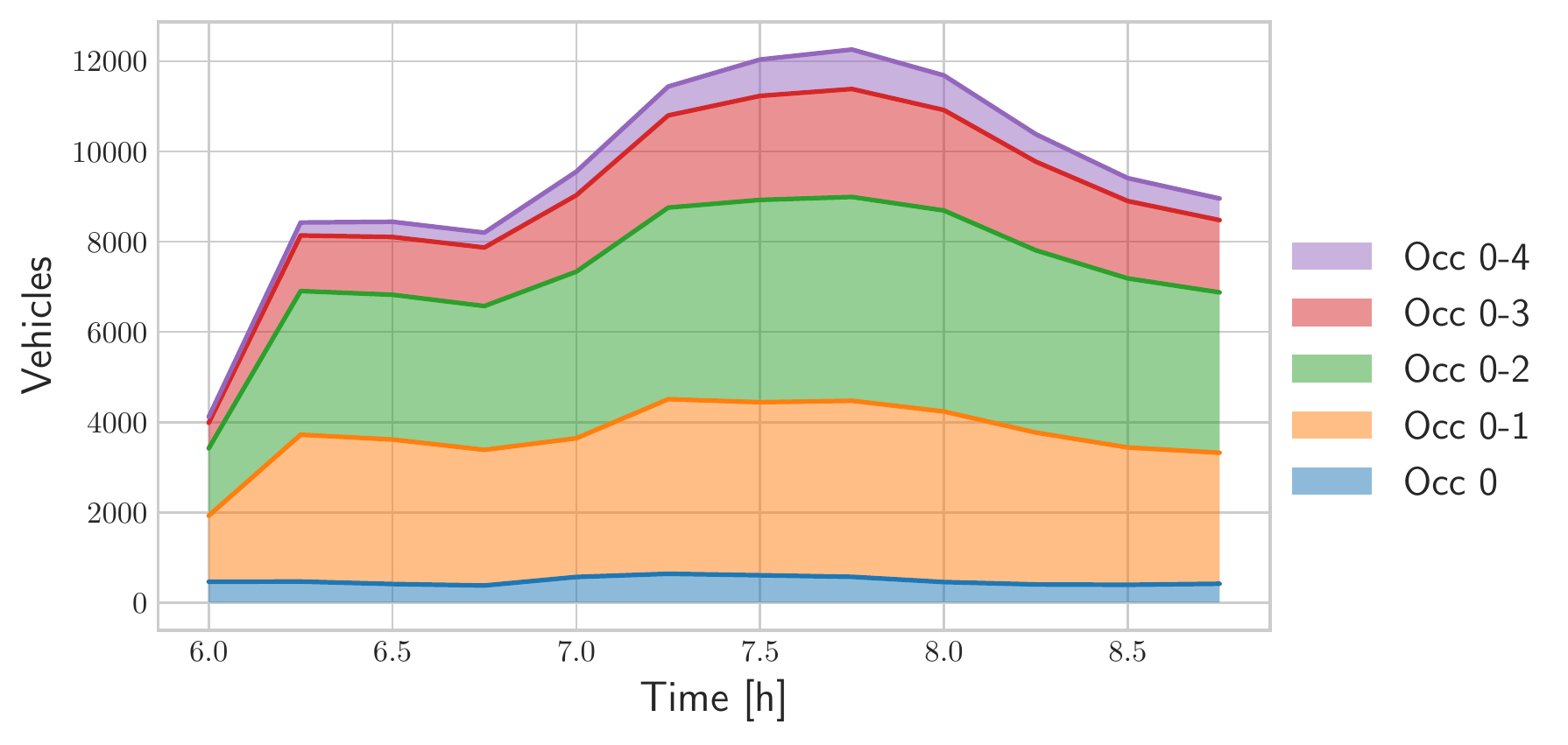}
    \caption{AMOD vehicle occupancy in the best operator scenario with 25,400 vehicles and a distance fare of 0.86~\euro~per km.}
    \label{fig:stack_plot_rev}
\end{figure}
\par
The distance fare shows a clear maximum at about 0.85~\euro~per km (of direct customer route distance). In the used mode choice model, fares below that hardly affect the attractiveness of AMOD compared to PT and PV. Hence, higher fares generate larger AMOD revenues. Larger costs reduce the number of AMOD users more than additional revenue per user can bring the operator. Likely, reduced pooling efficiency for a lower number of customers also plays a role. It should be noted again that this study looks at long-term impacts and computes the PV mode with full costs rather than operating costs. It is probable that AMOD operators have to offer lower fares in order to convince travelers to give up their PVs.
\par
As illustrated in Fig.~\ref{fig:stack_plot_rev}, no more than 12,000 of the more than 25,000 AMOD vehicles are in use at any given time, of which only a few are driving without any passengers, most with either one or two passengers, a considerable share with three passengers, and a small share with the maximum of four passengers. This results in a km-averaged occupancy of 1.9 passengers/vehicle. The average travel time and fare of customers are approximately 11 minutes and approximately 3.3~\euro, respectively.
\par
A simple estimation for the lower bound of profitable vehicle utilization can be conducted with these quantities. With 1.9 passengers on board, a vehicle produces revenues and variable costs of 3.8 and 0.25~\euro~per km, respectively. Therefore, 93\% of revenues are variable profit. Three trips with this variable profit are sufficient to compensate for the fixed vehicle costs. Hence, it can be concluded that the fixed vehicle costs are compensated with approximately half an hour of activity.
\par
In reality, operators are likely to spend more computational resources and utilize demand estimations to improve the fleet control by more advanced customer-vehicle assignments and repositioning. This will reduce the fleet size needed to serve the same amount of demand thereby reducing fixed vehicle costs. Empty travel and variable costs might increase, but to a lesser degree. For the purpose of this study, the over-supply of vehicles is acceptable since for the choice of travelers it is critical that vehicles are available; a customer does not consider whether the AMOD vehicle was idle or had to be repositioned beforehand.
\par
The utilization-based pricing factor $y^{PU}$ only makes a difference for fleets in under-supply or near-to-under-supply conditions. Unless the regulator sets a low boundary on fleet size, this will not be the case in the subsequent scenarios.

\begin{figure}[ht]
\centering
\begin{subfigure}[t]{.35\textwidth}
  \vspace{0pt}
  \centering
  \includegraphics[width=1.0\textwidth]{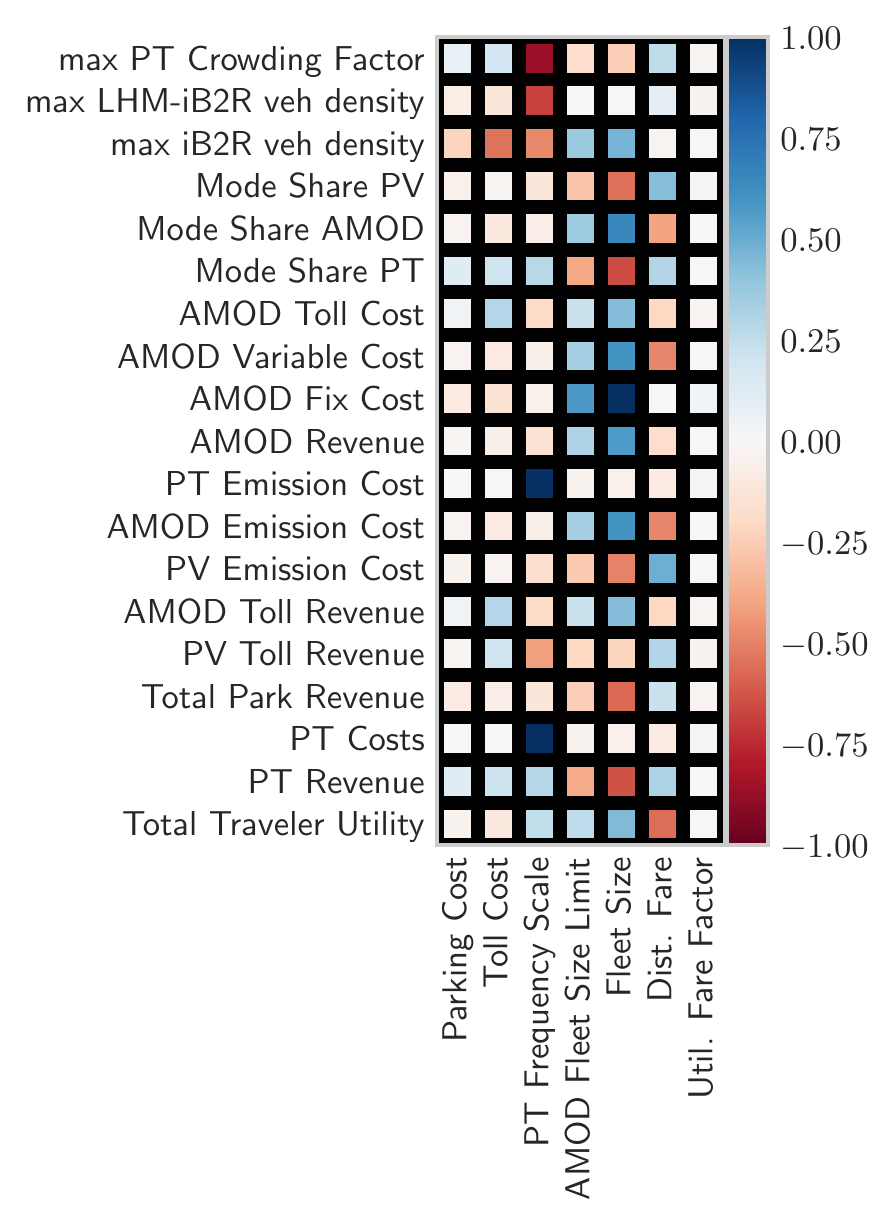}
\end{subfigure}%
\begin{subfigure}[t]{.6\textwidth}
  \vspace{0pt}
  \centering
  \includegraphics[width=1.0\textwidth]{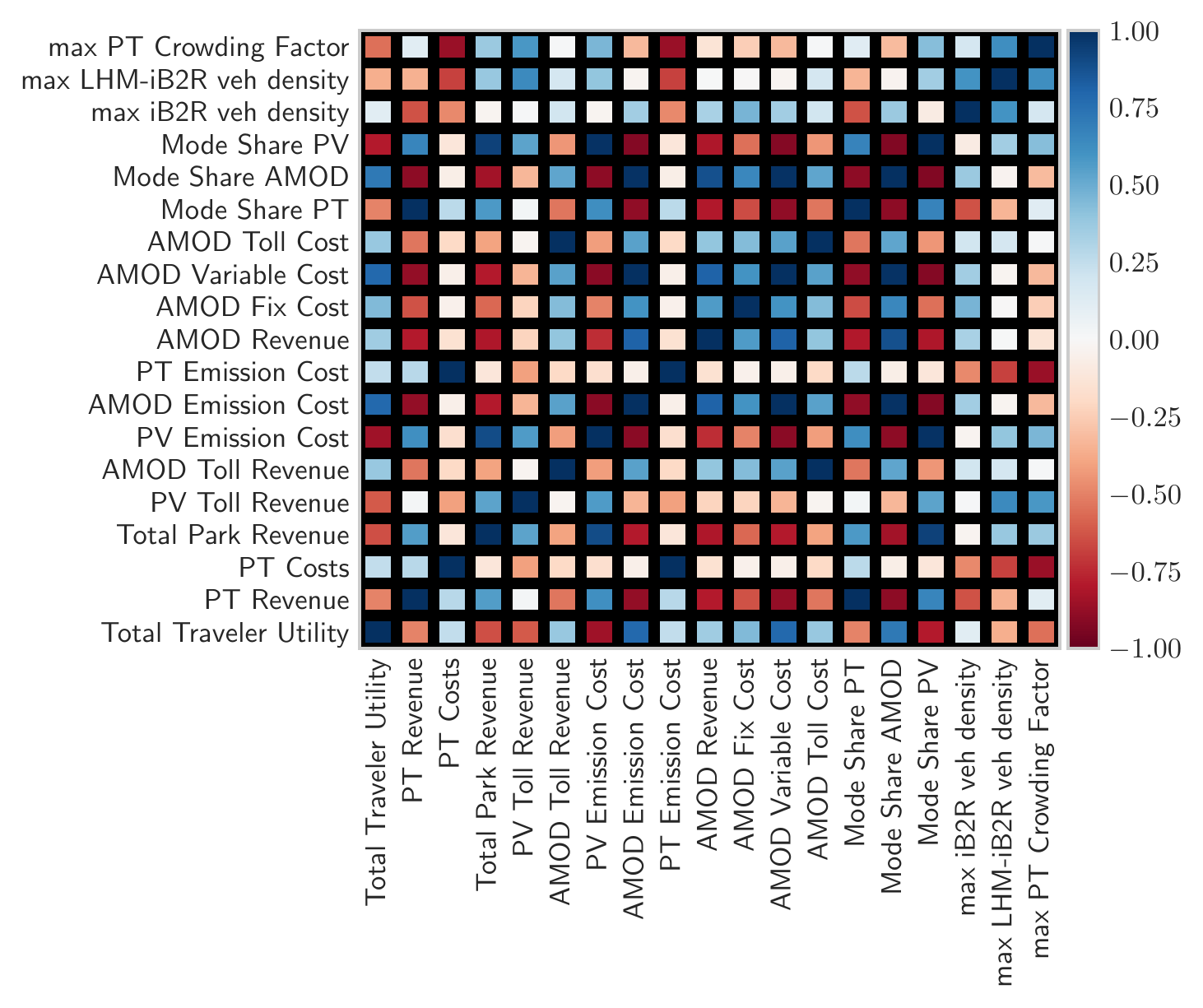}
\end{subfigure}
\caption{Pearson-Correlations of various KPIs with the decision variables (left) and other KPIs (right).}
\label{fig:correl_rev}
\end{figure}

After the simulations of the initial set of regulator settings, correlations were used as a sanity check for the sensitivity of (regulator) variables in the transportation model. Fig.~\ref{fig:correl_rev} displays the Pearson correlation coefficients between several KPIs and specific variables as well as between KPI pairs.
\par
As expected, the modal split of the AMOD service correlates positively with the fleet size and negatively with the base fare. The total traveler utility increases with higher AMOD availability and lower fares. A better AMOD offer also decreases crowding in the PT system and the PV mode share and thereby parking revenues. Obviously, vehicle fixed costs completely correlate with fleet size. Since a fleet size limitation affects the fleet size for low values, its correlations follow the fleet size, but are weaker.
\par
An increased PT frequency effectively decreases crowding, and therefore increases the total utility of all travelers and the modal split of PT. As expected, an increase in PT mode share also reduces the density in both clusters inside (iB2R) and outside (LHM - iB2R) of the inner highway belt (B2R) of the street network. 
\par
Parking costs and toll costs generally seem to have smaller impacts than changing the PT frequency and the available AMOD offer (fleet size and fare). Nevertheless, they both serve their purpose increasing the modal share of PT and decreasing the density within the inner city street network.
\par
Most KPI correlations are clear. For instance, low PT emissions are the direct consequence of low PT frequency. Similarly, AMOD fixed costs are a direct consequence of the fleet size variable and therefore, show the same correlations.
\par
Interesting correlations include a negative correlation of PT revenue and total traveler utility. Push measures, which intend to increase the PT mode share to make PT relatively more attractive, increase the general travel costs and therefore decrease the total travel utility. Moreover, high PT mode share is related to scenarios with low AMOD revenue (either due to low fleet size or high fares), which also decrease the total travel utility. All KPIs indicating a high mode share of AMOD (including a low PV mode share) seem to have a positive impact on travel utility.
\par
While correlations are meaningful to derive first insights and perform qualitative model checks, they cannot give quantitative answers to questions, such as how high a regulator should set the toll. Moreover, the correlations do not account for the structure of the tri-level optimization problem; rather the correlations consider every simulation equally no matter whether the AMOD service planner variables are optimal or not.

\subsubsection{Convergence to the Optimal Solution}
\begin{figure}[htb]
\centering
\includegraphics[width=0.7\textwidth]{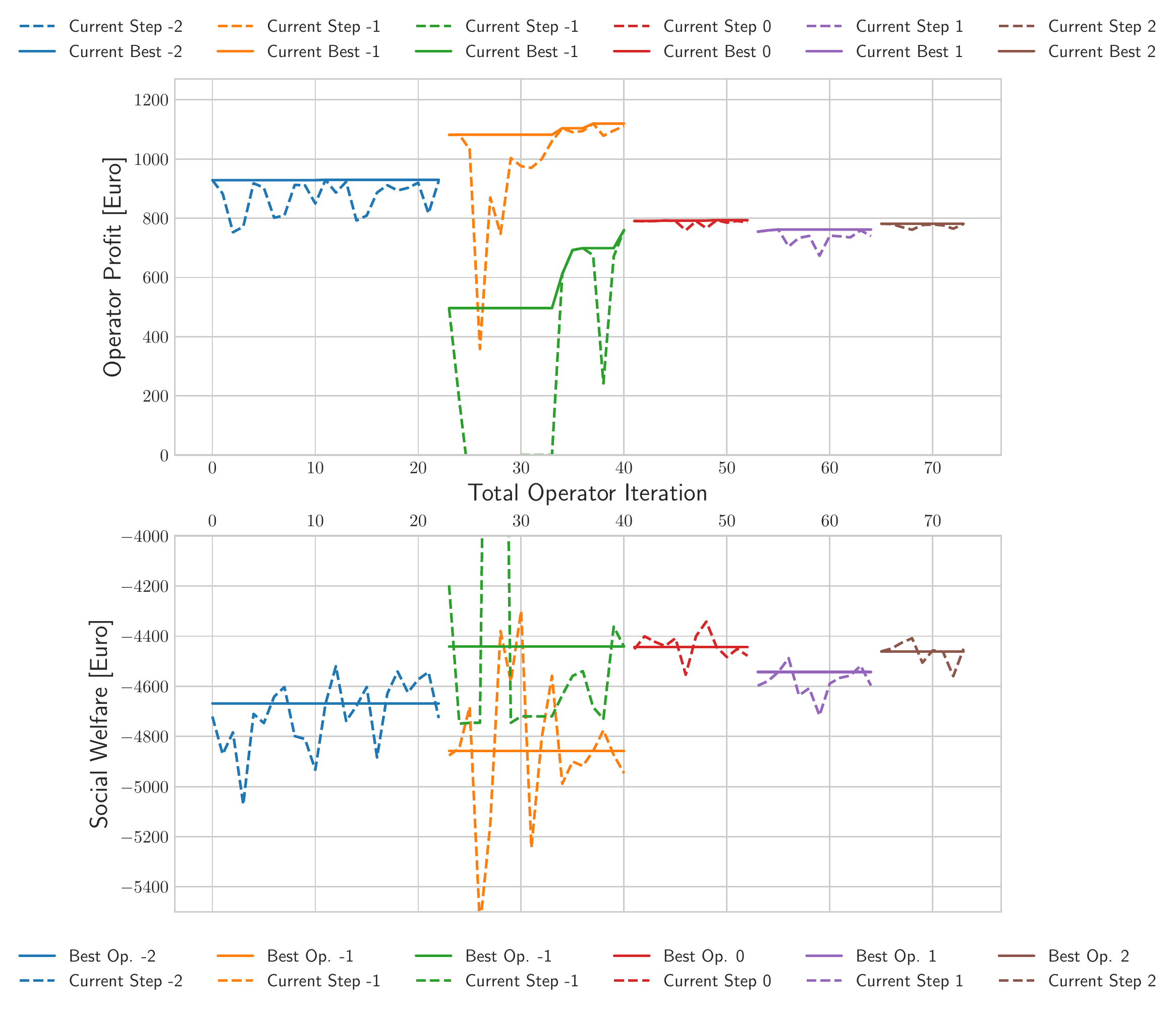}
\caption{Convergence behavior of two-level BO framework. Different colors represent different regulator settings, for which an operator variable optimization is performed. The "-2" iteration represents the initial no regulation setting and the "-1" iteration contains the initial set of regulator settings. The two colors for this iteration show the regulator setting with the best operator profit (orange) and best social welfare (green) result.}
\label{fig:convergence_rev}
\end{figure}
The solution approach contains multiple convergence processes as the lower level operator optimization is supposed to converge for each regulator step. Fig.~\ref{fig:convergence_rev} illustrates the convergence behavior of the two-level framework plotted on a single axis representing the continuously counted operator iteration.
\par
The iteration labeled "-2" denotes the no-regulation case. Its first operator iterations contains 272 different operator settings. After that, each operator step created one new operator setting according to the acquisition function optimization in Eqn.~\ref{eq:bo_step}. A minor improvement over the best initial setting was achieved.
\par
After that, an initial 272 regulator settings were set up, where the respective operator variables were derived via BO. If the fleet size limit was above the optimal fleet size, the best operator variables could be used, for smaller fleet sizes, other operator variable settings were derived. As shown in Fig.~\ref{fig:convergence_rev}, the first operator iterations of the initial regulator iteration ("-1") often produced really low profit values because the algorithm explored low AMOD fleet sizes or too high or low distance-fare values. The search hardly exploited the prior knowledge from the no-regulation case with $\kappa$ according to Eqn.~\ref{eq:kappa_def1}, which motivated the limitation of $\kappa$ to 1 after 34 operator iterations. With this change, the exploitation of prior simulation results helped to quickly derive near-optimal operator settings.
\par
The large and well distributed set of initial regulator settings in the "-1" iteration ensures that a good solution in the regulator variable space is available.

\subsubsection{Comparison of Scenario Results}
There are different mindsets when it comes to regulating a transportation system. In the end, the definition of the social welfare function will determine what the optimal scenario will be. In the following, the results from several scenarios are compared:
\begin{itemize}
    \item \textit{No AMOD}: status quo without an AMOD system
    \item \textit{No Regulation}: transportation system with AMOD system without changes in regulation
    \item \textit{Default Social Welfare Definition}: operator and regulator variables defined by result of BO process
    \item \textit{Pro-PT Social Welfare Definition}: operator and regulator variables defined by result of BO process, where the weight of social welfare components are changed according to Eqns.~\ref{eq:sw_re_def1}-\ref{eq:sw_re_def2}
\end{itemize}

\begin{figure}[th]
    \centering
    \includegraphics[width=0.9\textwidth]{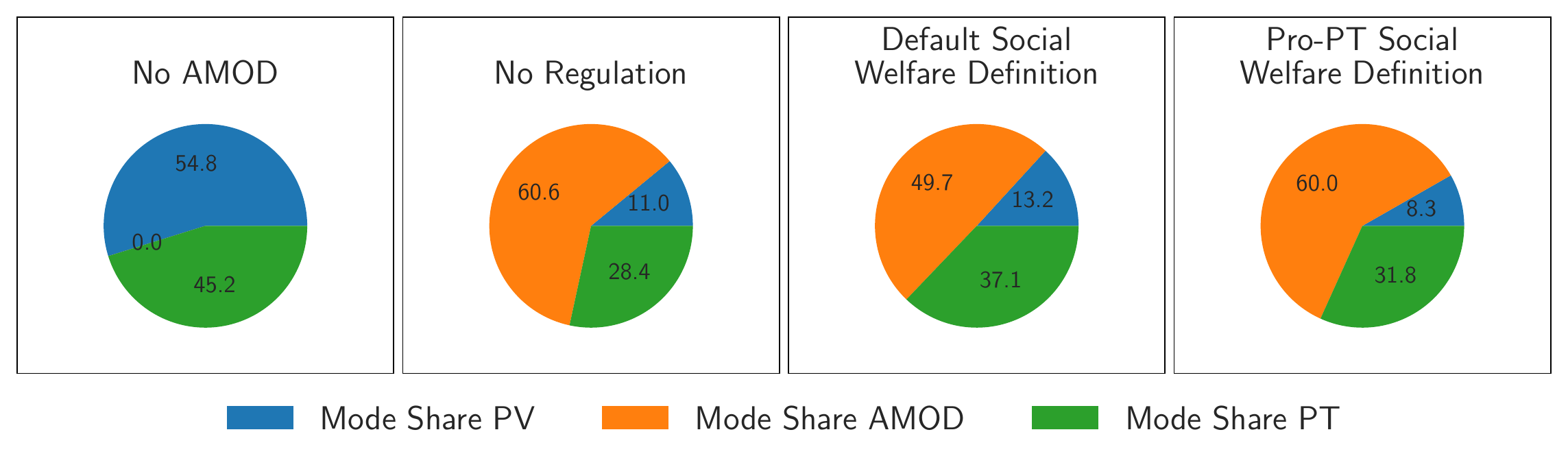}
    \caption{Mode split of motorized trips in the described scenarios.}
    \label{fig:mode_split_rev}
\end{figure}

As mentioned in the introduction, AMOD has the potential to disrupt urban mobility. Figure~\ref{fig:mode_split_rev} illustrates that AMOD will take shares from both PV and PT. In all AMOD scenarios, AMOD was chosen by approximately 50-60\% of travelers, whereas a larger share of AMOD travelers were prior PV than PT users. Regulations can impact whether travelers switch from PT to AMOD or PV to AMOD. Interestingly, the two SW definitions introduced measures with quite different outcomes: the standard definition limited the AMOD share at around 50\% thereby increasing both PV and PT share; the Pro-PT definition left the AMOD share unchanged, but further limited the PV share from 11\% in the unregulated scenario to 8\%.
\par
\begin{figure}
    \centering
    \includegraphics[width=0.75\textwidth]{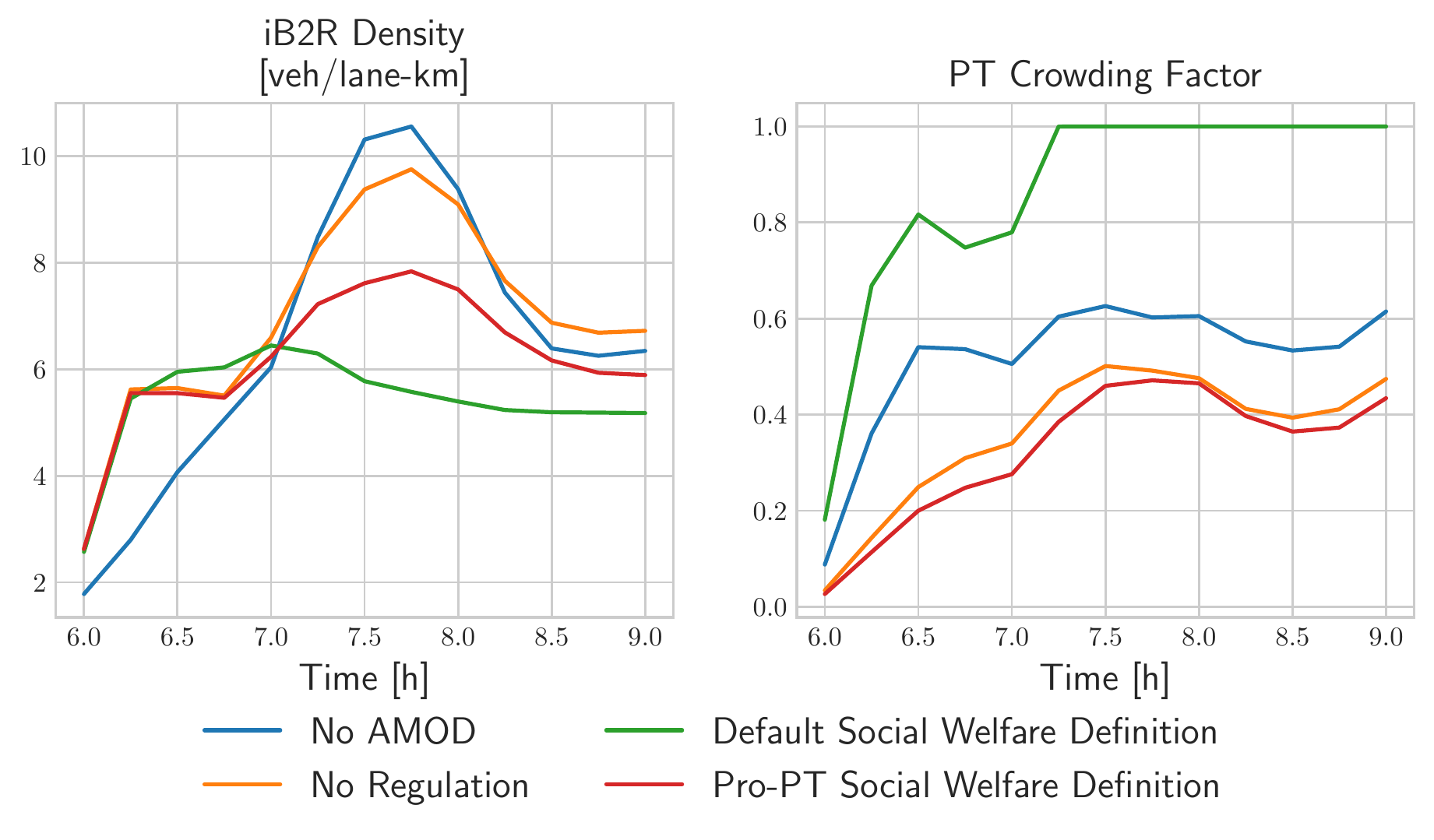}
    \caption{Evolution of street and PT network utilization over time in the described scenarios.}
    \label{fig:network_temporal_rev}
\end{figure}

Table~\ref{tab:comp_zero_no_with_regulation_rev} summarizes the operator and regulator variables and components for all scenarios. The first observation is that adding AMOD as an additional mode increases the total travel utility considerably, which is also reflected in the large AMOD mode shares. These require large fleet sizes of more than 10,000 vehicles. The fleet size is not constrained by the best regulatory solution in either regulation scenario, i.e. $x^F > y^F$. The fixed costs for the AMOD operator are lower than the variable costs and only a fraction of the revenue resulting in a high profitability of an AMOD service. The AMOD service performance is similar in all three scenarios (with AMOD system). The average waiting time is 130 seconds in the No Regulation and the Pro-PT SW scenario and 140 seconds in the Default SW scenario, which can be attributed to the smaller fleet size. For the same reason, the share of travelers with an AMOD offer, which was made if there is a feasible vehicle tour keeping both waiting time and detour time constraints, is slightly lower for this scenario. The average relative detour is between 17 and 18\% in all cases, which in absolute values is in the same range as the waiting time. Since repositioning was excluded in this study, the share of empty VKT is less than 4\% in all AMOD scenarios, which is a very low value. On the downside, large vehicle numbers are necessary to cover the area thereby resulting in rather low fleet utilization values between 40 and 55 \% in the three AMOD scenarios. Therefore, the utilization surge pricing factor, which would increase fares from 75\% utilization is not relevant. The km-weighed average occupancy is 1.9 in all scenarios; this values represents a considerable improvement over the average occupancy of 1.0 for PVs. With a large share of AMOD users originating from the PV sector, the densities in the street network are reduced even without regulation (see Fig.~\ref{fig:network_temporal_rev}). Moreover, parking revenues are drastically decreased compared to the No AMOD scenario, which also can be viewed as positive since parking space is freed for other use.

\begin{table}[ht]
\centering
\footnotesize{
\begin{tabular}{llllll}
\toprule
                            &                & No AMOD & No Regulation & \makecell[l]{Default Social \\  Welfare Definition} & \makecell[l]{Pro-PT Social \\  Welfare Definition} \\
 &  &         &               &                                                    &                                                   \\
\midrule
\multirow{4}{*}{\makecell[l]{Regulator\\Variable}} & Parking Cost [Euro / h] &     2.5 &           2.5 &                                                2.5 &                                              4.84 \\
                            & Toll Cost [Euro / km] &       0 &             0 &                                                  1 &                                              0.06 \\
                            & PT Frequency Scale &       1 &             1 &                                               0.25 &                                              1.23 \\
                            & AMOD Fleet Size Limit &   50000 &         50000 &                                              50000 &                                             46900 \\
\cline{1-6}
\multirow{3}{*}{\makecell[l]{Operator\\Variable}} & Fleet Size &       0 &         25400 &                                              14600 &                                             22900 \\
                            & Dist. Fare [Euro / km] &    0.25 &          0.86 &                                               1.01 &                                              0.84 \\
                            & Util. Fare Factor &       1 &            10 &                                                4.3 &                                                10 \\
\cline{1-6}
\multirow{11}{*}{\makecell[l]{Social\\Welfare\\KPIs}} & Social Welfare &   -5400 &         -4669 &                                              -4442 &                                     -4666 / -4727 \\
                            & Total Traveler Utility &   -5331 &         -4402 &                                              -4751 &                                             -4359 \\
                            & $\Delta$Total Traveler Utility &       0 &           929 &                                                580 &                                               972 \\
                            & PT Revenue &     178 &           113 &                                                149 &                                               127 \\
                            & PT Costs &     334 &           334 &                                                 76 &                                          405 / 40 \\
                            & Total Parking Revenue &     173 &            18 &                                                 15 &                                             7 / 2 \\
                            & AMOD Toll Revenue &       0 &             0 &                                                211 &                                            31 / 8 \\
                            & PV Toll Revenue &       0 &             0 &                                                 33 &                                             6 / 1 \\
                            & PV Emission Cost &      32 &             8 &                                                  9 &                                           6 / 142 \\
                            & AMOD Emission Cost &       0 &             2 &                                                  2 &                                           2 / 159 \\
                            & PT Emission Cost &      54 &            54 &                                                 12 &                                          66 / 165 \\
\cline{1-6}
\multirow{5}{*}{\makecell[l]{AMOD\\Profit\\KPIs}} & AMOD Profit &       0 &           930 &                                                760 &                                               899 \\
                            & AMOD Revenue &       0 &          1365 &                                               1275 &                                              1348 \\
                            & AMOD Fixed Costs &       0 &           193 &                                                111 &                                               174 \\
                            & AMOD Variable Costs &       0 &           242 &                                                193 &                                               243 \\
                            & AMOD Toll Costs &       0 &             0 &                                                211 &                                                31 \\
\cline{1-6}
\multirow{2}{*}{\makecell[l]{Traveler\\KPIs}} & Avg. Travel Utility &  -12.86 &        -10.63 &                                             -11.47 &                                            -10.53 \\
                            & Avg. AMOD Fare &       0 &           3.3 &                                               3.08 &                                              3.25 \\
\bottomrule
\end{tabular}
}
\caption{Regulator and operator variables (section 1 and 2), objective function components (section 3 and 4) and per traveler average values for the described scenarios. The left values in the Pro-PT scenario reflect the original coefficients while the right use the definitions from Eqn.~\ref{eq:sw_re_def1}-\ref{eq:sw_re_def2}. The units of the per traveler average values are \euro~and thousands of \euro~for all social welfare and profit component rows are thousands of \euro.}
\label{tab:comp_zero_no_with_regulation_rev}
\end{table}

The standard definition of social welfare employs an emission cost coefficient from the German Plan for Federal Traffic Routes. The contributions with this definition are an order of magnitude too small to effectively steer the regulatory policy.
Assuming a fully electric fleet and considering only the emissions for the production of the current (instead of life-cycle considerations for simplicity), the introduction of an AMOD system is clearly beneficial over the No AMOD scenario representing the status quo. The combined emissions of PV and AMOD are reduced by approximately 69\%. On the contrary, the PT system in Munich in its current state is neither economical nor ecological. Besides the city center, there are many low occupancy trips, which produce both high operating and emission costs. Especially the operating costs are a substantial contribution to the social welfare. Hence, the regulator in the default social welfare definition reduces the PT frequency to the minimal allowed value and increases the toll to its maximal value while leaving the parking costs unchanged. The dynamic toll sets in for densities over 5 veh/lane-km and increases with increasing density and has to be paid by PV and AMOD. Therefore, the high toll in the Default SW scenario is most effective in keeping the density of vehicles low in the inner city.  The AMOD operator reacts by decreasing the fleet size and increasing the fare. Interestingly, the fare is only increased by a fraction of the toll, namely by $0.15$ \euro~per km. As illustrated in Fig.~\ref{fig:network_temporal_rev}, the toll is high enough to keep the density in the inner city rather low, but the crowding in the PT system is really high due to the frequency reduction. Both effects are reflected in an decreased value of the traveler utility compared to the unregulated scenario. Since the AMOD price model assumes an increased per-km price also in the outer region, where no toll is applied to PV and additionally trip lengths and AMOD fares are higher, the ratio of PV to AMOD changes towards PV for trips with origin and destination outside of the city center (see last section in Table~\ref{tab:zone_od_analysis}).

\begin{table}[ht]
\tiny{
\centering
\setlength\extrarowheight{2pt}
\begin{tabular*}{0.955\textwidth}{lllrrrrrr}
\toprule
         &          &                                  &  \makecell[r]{Mode Share \\ PV [\%]} &  \makecell[r]{Mode Share \\ PT [\%]} &  \makecell[r]{Mode Share \\ AMOD [\%]} &  \makecell[r]{Avg. PV \\ Park Cost [\euro]} &  \makecell[r]{Avg. PV \\ Toll Cost [\euro]} &  \makecell[r]{Avg. Offered \\AMOD Fare [\euro]} \\
Origin & Destination & Scenario &                    &                    &                      &                            &                            &                                 \\
\midrule
\multirow{4}{*}{iB2R} & \multirow{4}{*}{LHM-iB2R} & No AMOD &               66.0 &               34.0 &                  0.0 &                        0.0 &                        0.0 &                             0.0 \\
         &          & No Regulation &               16.9 &               30.7 &                 52.3 &                        0.0 &                        0.0 &                             6.3 \\
         &          & Default SW &               14.3 &               41.8 &                 43.9 &                        0.0 &                        2.6 &                             7.1 \\
         &          & Pro-PT SW &               13.8 &               33.2 &                 53.0 &                        0.0 &                        0.4 &                             6.1 \\
\cline{1-9}
\cline{2-9}
\multirow{4}{*}{LHM-iB2R} & \multirow{4}{*}{iB2R} & No AMOD &               39.2 &               60.8 &                  0.0 &                        2.5 &                        0.0 &                             0.0 \\
         &          & No Regulation &                3.2 &               33.0 &                 63.8 &                        2.5 &                        0.0 &                             6.5 \\
         &          & Default SW &                3.2 &               42.1 &                 54.7 &                        2.5 &                        2.4 &                             7.5 \\
         &          & Pro-PT SW &                0.6 &               36.2 &                 63.1 &                        4.8 &                        0.3 &                             6.3 \\
\cline{1-9}
\cline{2-9}
\multirow{4}{*}{iB2R} & \multirow{4}{*}{iB2R} & No AMOD &               34.0 &               66.0 &                  0.0 &                        2.5 &                        0.0 &                             0.0 \\
         &          & No Regulation &                4.6 &               26.3 &                 69.1 &                        2.5 &                        0.0 &                             3.1 \\
         &          & Default SW &                3.0 &               39.2 &                 57.9 &                        2.5 &                        2.6 &                             3.4 \\
         &          & Pro-PT SW &                0.9 &               33.9 &                 65.2 &                        4.8 &                        0.4 &                             2.9 \\
\cline{1-9}
\cline{2-9}
\multirow{4}{*}{LHM-iB2R} & \multirow{4}{*}{LHM-iB2R} & No AMOD &               72.2 &               27.8 &                  0.0 &                        0.0 &                        0.0 &                             0.0 \\
         &          & No Regulation &               16.8 &               24.3 &                 58.9 &                        0.0 &                        0.0 &                             7.3 \\
         &          & Default SW &               26.9 &               28.4 &                 44.7 &                        0.0 &                        0.7 &                             8.5 \\
         &          & Pro-PT SW &               14.9 &               25.7 &                 59.4 &                        0.0 &                        0.1 &                             7.1 \\
\bottomrule
\end{tabular*}
}
\caption{Analysis of zone-to-zone travel relations in the described scenarios.}
\label{tab:zone_od_analysis}
\end{table}

\par
The Pro-PT scenario assumes a more efficient operation of the PT system, both ecologically and economically. Moreover, the regulator is assumed to put a much higher weight on emissions. In order to optimize social welfare in this scenario, the regulator asks the the PT operator to increase the frequency and only sets a minor toll while drastically increasing the parking costs. This further improves the space gained by the introduction of an AMOD system. There are hardly PV trips ending in the iB2R region, where PVs have to pay parking fees (see zone-to-zone travel relations in Table~\ref{tab:zone_od_analysis}). The AMOD system is operated with similar parameters as in the unregulated case. The densities in the street are lower than in the No AMOD and No Regulation scenario and the PT system shows much lower levels of crowding (Fig.~\ref{fig:network_temporal_rev}).

\subsubsection{Sensitivity of AMOD Operator and Regulator Variables}

\begin{figure}[th]
    \centering
    \includegraphics[width=0.4\textwidth]{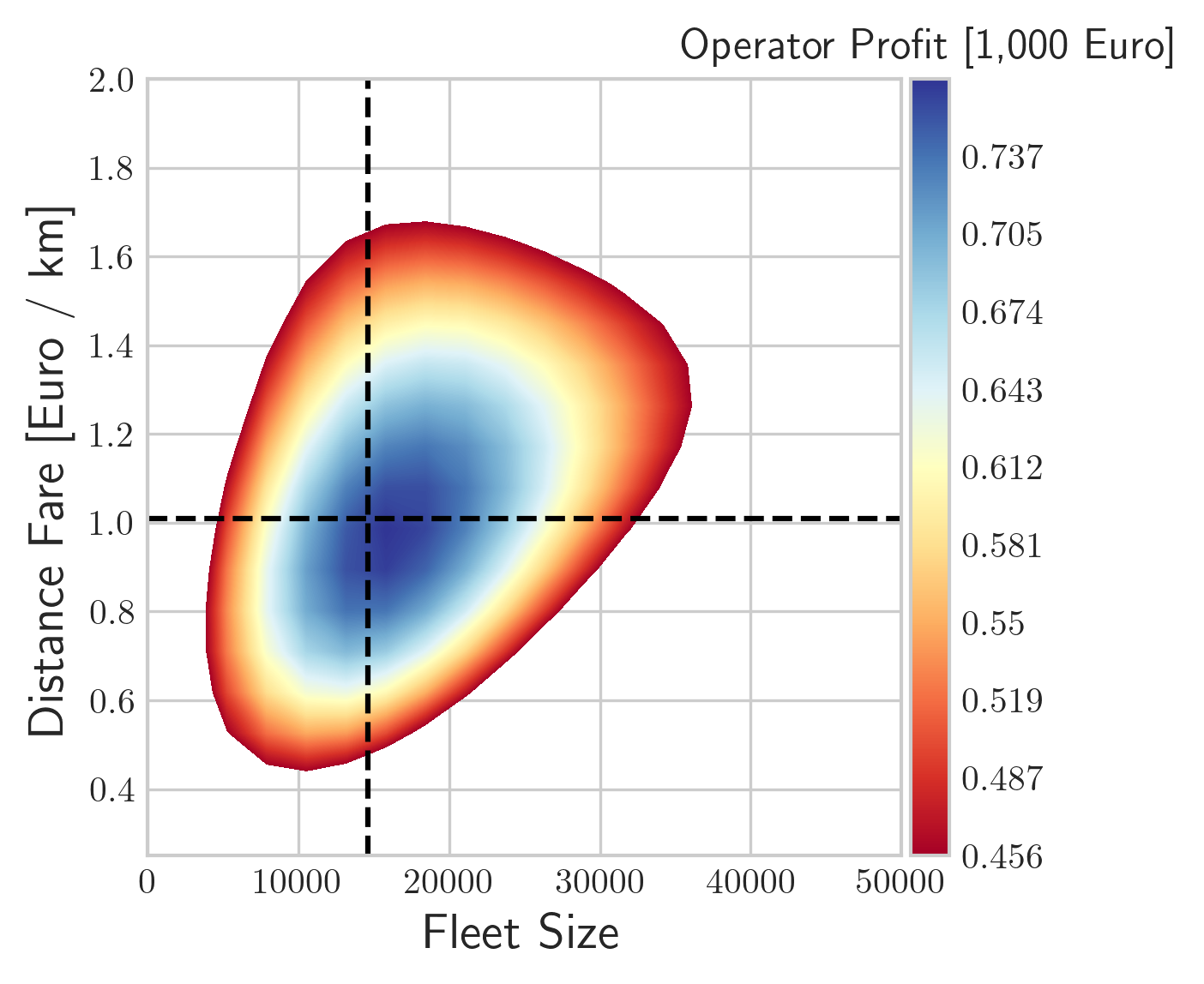}
    \includegraphics[width=0.4\textwidth]{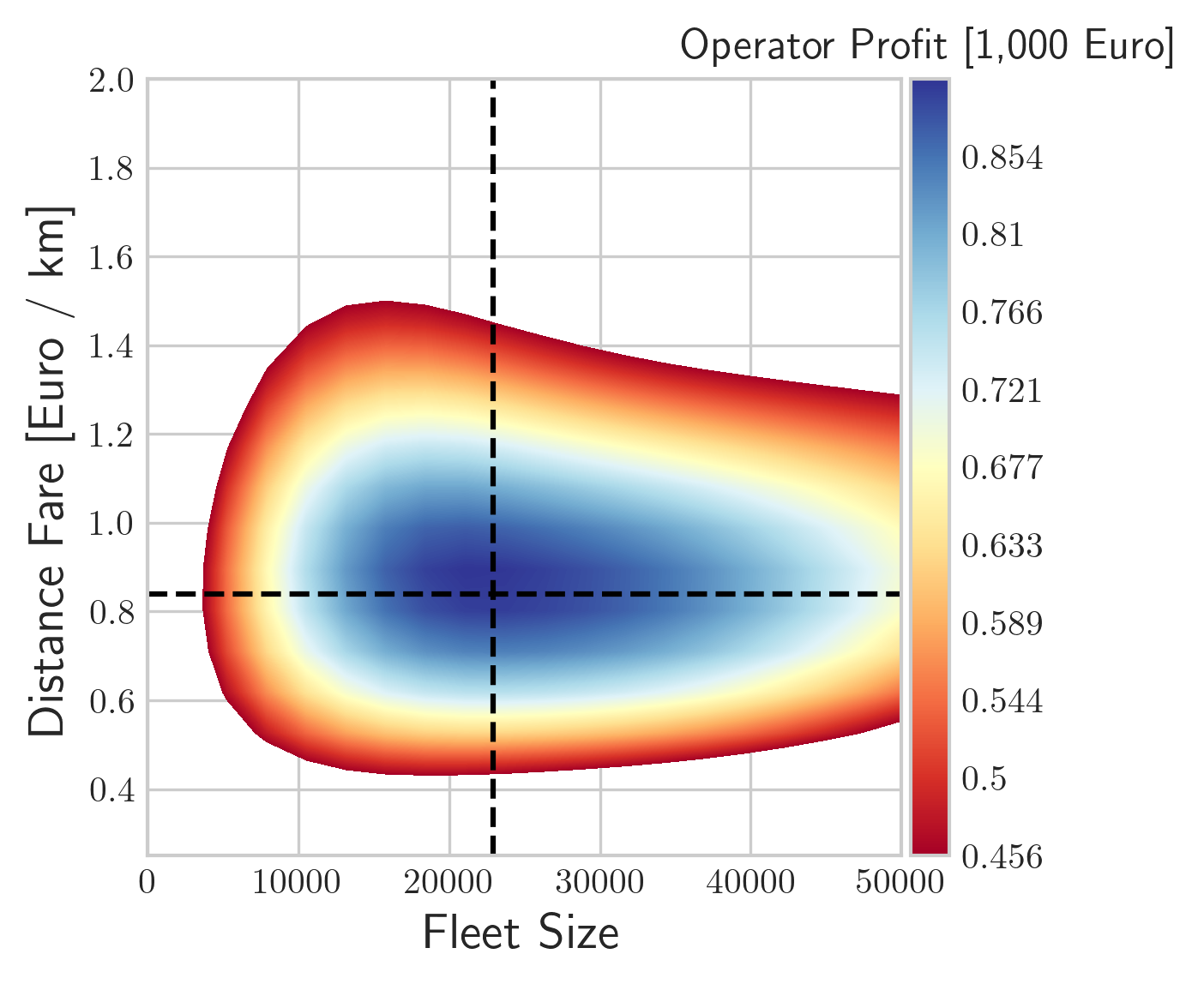}
    \caption{2D cuts of the AMOD profit surrogate function along AMOD operator variables in the hyper-planes defined by the described solutions for the Default Social Welfare Definition (left) and Pro-PT Social Welfare Definition (right). The white area shows the non-profitable area in the solution space and the dashed lines highlight the hyper-planes with the selected AMOD variables $y^{s,*}(x^r)$.}
    \label{fig:op_cuts}
\end{figure}

The AMOD profit surrogate function $P^S(x^r,y^s)$ is a 7-dimensional function. Since the AMOD service planner only has control over its own variables, the four regulator dimensions can be viewed as inaccessible. The regulator sets the hyper-planes in which the AMOD service planner can optimize. In the Default SW scenario, the regulator set a high toll of 1~\euro~per km and decreased the PT frequency. For this regulatory setting, the variable costs are increased by a factor of 4. As pooling allows a km-averaged occupancy of nearly 2, the AMOD operator can actually offer the service at fares of approximately 1 \euro~per km and be profitable. However, compared to the Pro-PT SW scenario, in which parking costs instead of toll costs were increased, the area of profitability is drastically reduced (see Fig.~\ref{fig:op_cuts}). 

\begin{figure}[th]
    \centering
    \includegraphics[width=0.4\textwidth]{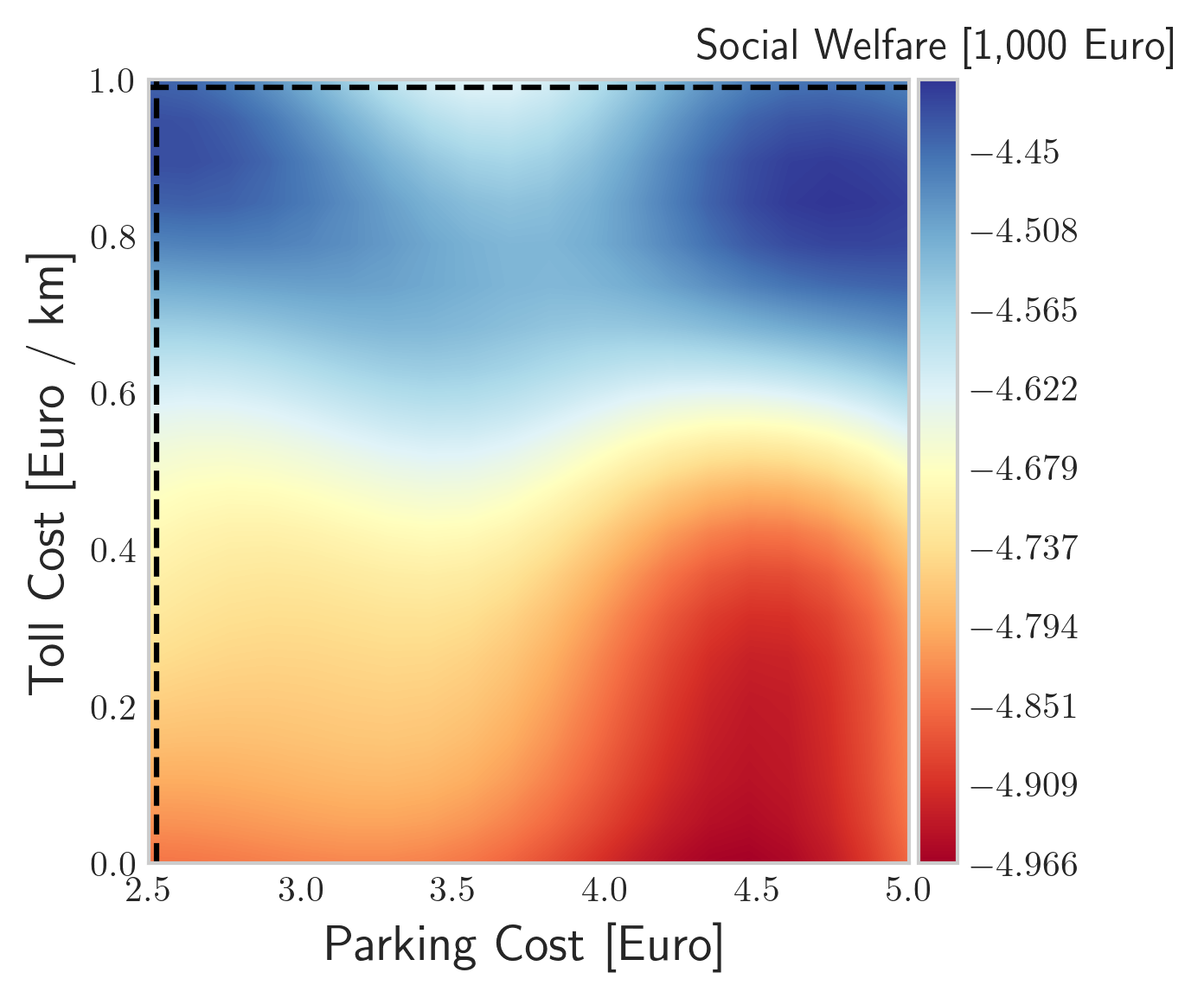}
    \includegraphics[width=0.4\textwidth]{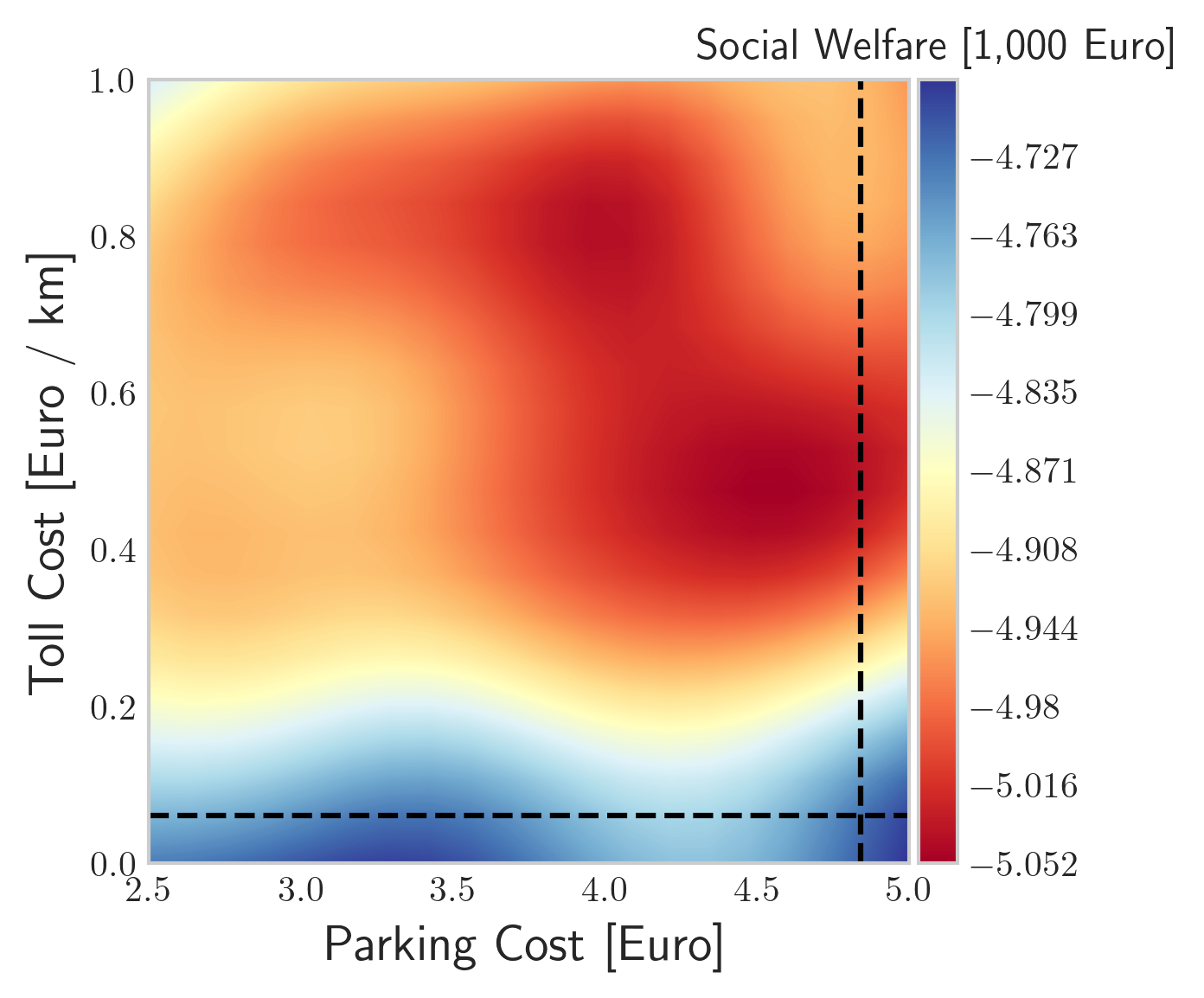} \\
    \includegraphics[width=0.4\textwidth]{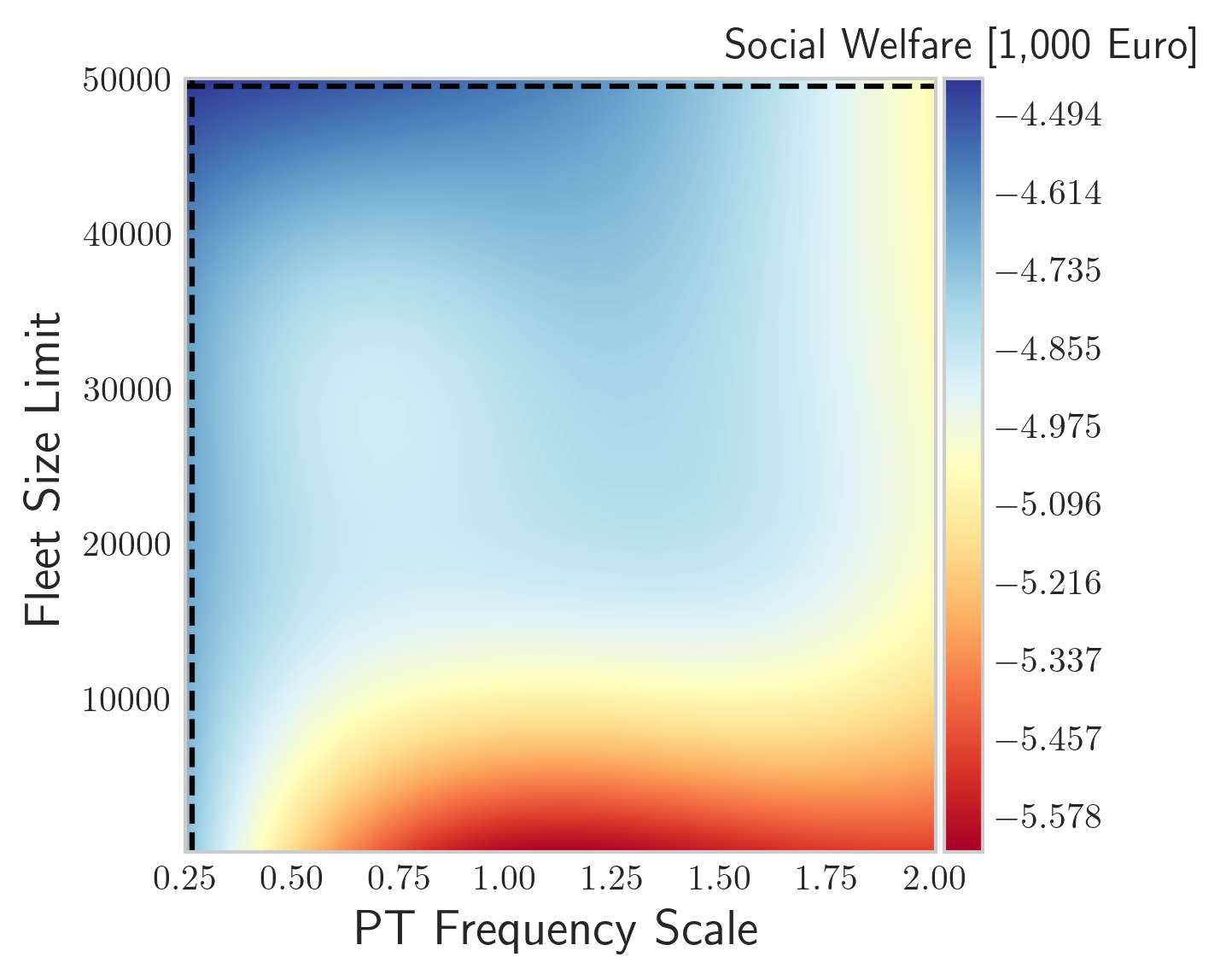}
    \includegraphics[width=0.4\textwidth]{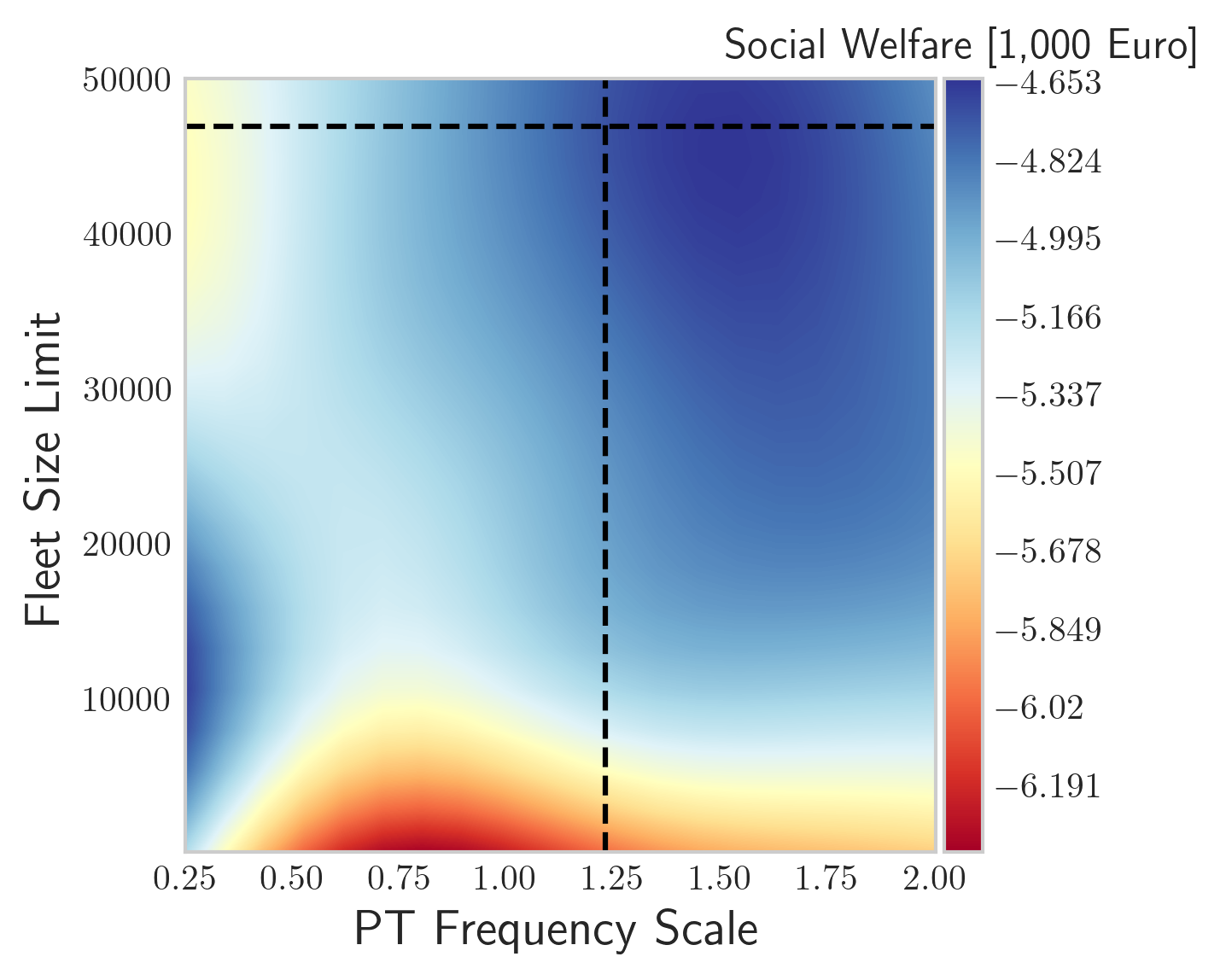}
    \caption{2D cuts of the social welfare surrogate functions along regulator variables in the hyper-planes defined by the described solutions for the Default Social Welfare Definition (left) and Pro-PT Social Welfare Definition (right). The dashed lines highlight the hyper-planes with the selected regulator variables $x^{r,*}$.}
    \label{fig:reg_cuts}
\end{figure}

The 4-dimensional surrogate functions $W^S(x^r)$ are based on less data, since only the simulations with the highest operator profit contribute data points -- one for each regulator setting. Figure~\ref{fig:reg_cuts} shows some two-dimensional cuts through the hyper-planes defined by the best solution for the two social welfare definition scenarios. The left and right sides clearly paint two very different pictures: while the Default SW function decreases with decreasing PT frequency, the best solution in the Pro-PT scenario is to increase PT frequency. Fleet size limitations below 10,000 vehicles have a very negative impact on social welfare and become minor up above that. As mentioned before, limitations above the profit-maximizing fleet size do not matter.
\par
The real advantage of the surrogate functions is the combined consideration of variables. Assuming a toll of 1~\euro~per km and parking costs of 2.50~\euro~per km, the worst decision according to this SW definition would actually be to forbid AMOD and keep the PT frequency as it is. As the ecological footprint is weighed highly in the Pro-PT case, both increases in PT frequency and AMOD fleet size (increasing fleet size limit from 0) are beneficial for social welfare as both modes attract travelers from PV users, which have the worst CO2 footprint.
\par
Parking and toll costs both reduce the travel utility of the PV thereby nudging more people to use PT. A combined increase of both toll and parking costs could be interesting for city administrations. The Pro-PT scenario shows that after a certain point, an increase is not really beneficial anymore because benefits from travelers changing mode do not compensate for the increased total travel costs anymore. At this point, city administrations will also have trouble justifying their measures and run into trouble with social equity.

\section{Discussion}
This study contained one implicit regulation: the city administration only allows a pooling-service. This does not mean that all AMOD customers have to be matched; they can also be alone in a vehicle. However, they would be pooled if it is beneficial according to the fleet control objective. The AMOD operator and city administration objectives are in line as both benefit from people sharing a ride: the operator has a better ratio of revenue to vehicle miles and the city benefits from fewer vehicles on the road.
\par
The AMOD ride-pooling system scales very well with increasing demand. AMOD customers only had minor waiting and detour times. On average, travelers had to endure 5 minutes for the sum of both. Considering the time and stress involved in finding a parking space in the city, the level-of-service is comparable to PV. This study used PV full costs (0.60~\euro~per km) in order to analyze long-term adoption. The AMOD operator adapted the prices according to its competition (0.85~\euro~per km in the scenarios without a toll), which was more than three times its variable cost. Combined with an average occupancy of nearly 2 passengers/vehicle, the profit margin of the AMOD system is very large. It is likely that an AMOD service has to compete with operating costs in the introductory phase, in which private vehicle ownership is rather high. The optimization of profit would motivate the AMOD operator to offer lower fares than in the studied scenarios. Therefore, the profit margin would be considerably lower. However, in the long term, travelers might be willing to abolish private vehicles if they have a better alternative.
\par
The introduction of an AMOD pooling service brings many social benefits. Travelers gain a new and very attractive mode, which can also help to reduce the number of vehicles on and off the road. Additionally, AMOD is environmentally more sustainable than PV as they are likely to be powered by electric motors and have a higher vehicle occupancy. They even compare favorably against today's PT system in Munich, where large buses drive rather often with only a few passengers.
\par
These benefits caused the policy optimization to not apply a fleet size limitation. This policy recommendation is likely general and transferable to other cities, but has to be analyzed in detail to be certain. Unless other factors not considered in this study give reason for a limitation, fixed vehicle costs limit the fleet size in the profit optimization. Therefore, policy makers should consider extending the PT portfolio to include an AMOD pooling service as it can be open to the public and does not exclude travelers.
In a PT integration process, AMOD should not only be considered as an intermodal feeder service. As mentioned in the beginning of this section, the large social benefits actually originate from its competitiveness with PV. Test simulations with an intermodal feeder service and the current PT system showed that only a minor share (less than 1\%) of travelers used AMOD to get to the next rail-based PT station. The bus stop network in the study area is very dense and frequencies rather high; travelers rarely chose to wait for an AMOD vehicle and pay extra for AMOD for a first/last mile trip. Due to its high computational burden and minor impact, the intermodal model was not included in the final version of the model. However, an intermodal feeder model should be included in an integrated PT-AMOD planning process, in which AMOD is considered to replace inefficient bus lines and off-peak hour bus services.
\par
It is difficult to derive other general policy recommendations. There will not be one set of regulations that are optimal for every city because measures of social welfare are subject to the preferences of each city and its residents. For example, some cities might weight emissions higher than traveler utilities. There is an effort to derive more general insights by classifying cities and simulating a set of archetype cities\footnote{http://web.mit.edu/afs/athena.mit.edu/org/i/its-lab/www/dashboard/new\%20dashboard/index.html}. However, the generalization still assumes that the social welfare definition and traveler behavior are similar in the same city classes. As shown with the two social welfare definition scenarios, the quantitative values of regulatory measures depend on the exact definition of social welfare. The framework in this study requires an explicit weighting of different social welfare components, the effect of which should be checked in order to set meaningful weights. The case study in Munich showed that the coefficient used in the Bundesverkehrswegeplan, a plan for future infrastructure projects by the German Federal Ministry from 2016, is so low that emissions had practically no impact on the policy optimization.
\par
In general, push measures are a useful tool to decrease the modal split of PV. Nevertheless, it should be ensured that the balance between additional traveler costs and social value remains in balance. A concern related to very high push factors is that they can increase social inequity. If the push factors are strong enough, only the wealthy travelers in the region will be able to use PVs. The framework presented in this study is a tool that will stop increasing the value of toll and parking costs when there is no more social gain. Further considerations, e.g. regarding social equity, can help to set additional boundaries on the variables.
\par
The application of two social welfare scenarios show that AMOD operators will react to policies and adopt their service, which in turn has an effect on travelers' choices. One contribution of this study is the development of a framework that allows regulators to consider this AMOD operator reaction in order to derive good policies for their respective social welfare definition. Moreover, the impact of weights in the social welfare definitions can be analyzed using the surrogate functions that are part of the BO procedure.
\par
The framework could integrate more complex mode-choice models to represent travel behavior more accurately. The acceptance of AMOD (especially that it is only offered as a pooling service) and travel behavior are likely the largest uncertainties in the system. More advanced behavioral models require more data to estimate a larger number of coefficients related to travel behavior. Generally, the solution approach to the tri-level problem can integrate more refined models (e.g. fleet control); however, there is direct trade-off between model resolution and computation time in the proposed tri-level model and simulation-optimization solution approach. Regarding AMOD profit, the cost structure of autonomous vehicles represents significant uncertainty. Nevertheless, as long as the AV costs remain below the fares, the prices could be market-driven and the modal split similar to the presented results.

\section{Conclusion}
\subsection{Summary}

It is well known that regulations can impact the behavior (e.g. mode choice) of travelers. With MSPs as a new intermediary interface between transportation planners/regulators and travelers, decision-making has become more challenging for city administrations. Usually, transportation planning, and in some cases transportation policy-making, are long-lasting processes that are often active for long time periods. Hence, the public sector's current regulatory decisions should anticipate how MSPs and travelers will be affected by various regulations. The proposed tri-level problem formulation in this study models a system, in which the administration's ultimate goal is to find regulations that maximize social welfare. The tri-level problem considers that an MSP will design their service in order to maximize profits according to the set regulations and, of course, the demand for their service.

This study employs a simulation-optimization solution approach that employs BO at the regulator and AMOD service planner levels and a transportation simulation model at the traveler level to solve the tri-level problem. After a large set of initial simulations, Gaussian Processes serve as surrogate models, which can be used to infer approximations for AMOD profit and social welfare. These approximations can help to understand and visualize a large-dimensional solution space and guide the optimization process, especially to derive good AMOD service planner variable values when jumping from one upper level iteration to the next. Therefore, the Bayesian framework with integrated demand-supply model can also serve as useful tool for AMOD service planners to optimize their system with respect to a changing regulations. 

This study presents a case study that includes the regulation of an AMOD pooling service in the city of Munich, Germany. The regulations include (1) increasing the costs of parking, (2) introducing a road toll, (3) increasing the PT budget by increasing the PT frequency and (4) a possible limitation of the AMOD fleet size. Each of these regulation can directly or indirectly decrease congestion and the number of PVs in the city center.

In this study, the regulator's objective was a social welfare measure containing the sum of generalized travel costs of all travelers, PT budget, revenues from parking fees and road tolls, as well as costs related to transport system emissions.

In general, the study shows that the proposed framework can serve as a very useful tool to model the tri-level problem of regulating MSP systems. With an initial set of simulations, regulators can even build surrogate functions for new definitions of social welfare, build a surrogate model for visualization, and use the optimization framework to investigate interesting areas of the solution space as long as the relevant components for the respective social welfare definition are saved explicitly for the simulations.

\subsection{Limitations}
The main shortcoming of the study and the modeling framework proposed within is the computational complexity associated with the tri-level model, and in particular the time needed to run each instance of the transportation simulation model. There is ultimately an inherent trade-off between computational complexity and (1) model resolution and (2) model sensitivity to various regulations. The current study includes a relatively large set of regulation variables and the components necessary for the model to be appropriately sensitive to these regulations. Hence, the study employs simulation-based methods to model the transportation system, as opposed to analytical models that tend to work well when one or two decision variables are present. In terms of resolution, the models employed in this study vary. The modeling framework spends significant computational effort modeling the real-time AMOD dispatching problem in a real road network; there would still be quite some potential for optimization of ride-pooling assignments, but the computational complexity would increase too much for the tri-level model. The detailed modeling framework considers street and public transport network parameters and demand patterns specific to the studied city to produce numeric results specific for this city. This implies that these results cannot be generalized directly to different cities, instead, the framework provides a tool that has to be applied for each city.
The profit margins for an AMOD operator seem extremely high when comparing it to the traditional modes (private vehicle and public transportation); competition between AMOD providers would likely lead to lower pooling efficiencies and even lower offered fares, thereby making this mode even more attractive to travelers. Furthermore, the mode choice model is quite straightforward, and the simplified model of network congestion across time and space significantly reduces computational time compared to a fully functional micro- or meso-scopic traffic simulation model. Finally, a more explicit public transport model is required: the general scaling of all PT frequencies should be extended to treat efficient and inefficient PT lines differently. Moreover, different PT frequency scaling for peak and off-peak hours with AMOD should be considered.

\section{Acknowledgements}
The authors gratefully acknowledge the Leibniz Supercomputing Centre for funding this project by providing computing time on its Linux-Cluster.
Otherwise this research did not receive any specific grant from funding agencies in the public, commercial, or not-for-profit sectors.

\appendix
\section{Glossary and Case Study Inputs}
\label{sec:appendix}
{\scriptsize  
\begin{center}
\begin{longtable}{llp{4.5cm}lll}
\textbf{Chapter}         &  \textbf{Symbol}      & \textbf{Description}                                         & \textbf{Unit}     & \textbf{Case Study Value} & \textbf{Source}                                             \\
\hline \hline
\endhead
\textbf{General}               & $x^r$                      & decision variable for regulator with regulation $r$                &                         &                                 &                                                                   \\
                               & $y^s$                      & decision variable for  AMOD service planner                                &                         &                                 &                                                                   \\
                               &                            & with service design parameter $s$                                  &                         &                                 &                                                                   \\
                               & $z_i^m$                    & mode-choice decision variable of traveler $i$ for mode $m$         &                         &                                 &                                                                   \\
                               & $W$                        & social welfare                                                     & \euro~   &                                 &                                                                   \\
                               & $P$                        & AMOD profit                                               & \euro~   &                                 &                                                                   \\
                               & $U_i^m$                    & Utility of traveler $i$ for mode $m$                               & \euro~   &                                 &                                                                   \\
                               & $Z_c$                      & Network Zone: $c=I$: inner Zone; $c=O$: outer Zone                 &                         &                                 &                                                                   \\ \hline
\textbf{Bayesian}               & $\mathcal{N}(\mu, \sigma)$ & Normal Distribution with mean $\mu$                                &                         &                                 &                                                                   \\
\textbf{Optimization}          &                            & and standard deviation $\sigma$                                    &                         &                                 &                                                                   \\
\textbf{}                      & $(x_p, f_p)$               & set of prior data points                                           &                         &                                 &                                                                   \\
                               & $K_{pq}$                   & covariance matrix of prior data points $k(x_p, x_q)$               &                         &                                 &                                                                   \\
                               & $\Gamma$                   & Gamma function                                                     &                         &                                 &                                                                   \\
                               & $H$                        & Bessel function                                                    &                         &                                 &                                                                   \\
                               & $\zeta$                    & Exponent of Bessel functions in Matern kernel                      &                         & $5/2$                   &                                                                   \\
                               & $A_n(x)$                   & Acquisition Function                                               &                         &                                 &                                                                   \\
                               & $d$                        & dimension of solution space                                        &                         &  &                                                                   \\
                               & $P(x^r, y^s)$              & Surrogate function for Baysian optimization                        &                         &                                 &                                                                   \\ \hline
\textbf{Traveler}              & $i$                        & index of traveler                                                  &                         &                                 &                                                                   \\
\textbf{Model}                 & $o_i$                      & origin of traveler $i$                                             &                         &                                 &                                                                   \\
                               & $d_i$                      & destination of traveler $i$                                        &                         &                                 &                                                                   \\
                               & $t_i$                      & time a traveler wants to start a trip                              & s                       &                                 &                                                                   \\
                               & $m$                        & mode choice index for each traveler                                &                         &                                 &                                                                   \\
                               & $P_i(m)$                   & probability of traveler $i$ choosing mode $m$                       &                         &                                 &                                                                   \\
                               & $U_m$                      & utility of mode $m$                                                & \euro~   &                                 &                                                                   \\
                               & $c^{VOT}$                  & value of time (same for each mode)                                 & \euro~/s  & -0.0045                         & \cite{Frei2017}                            \\
                               & $t_{od}$                   & travel time between origin $o$ and destination $d$                 & s                       &                                 &                                                                   \\
                               & $d_{od}$                   & travel distance between origin $o$ and destination $d$             & m                       &                                 &                                                                   \\
                               & $c_{D}^{PV}$               & distance dependent private vehicle cost                            & \euro~/km & 0.66                            & from\footnote{https://www.tcs.ch/de/testberichte-ratgeber/ratgeber/kontrollen-unterhalt/kilometerkosten.php}             \\
                               & $C_{od}^{toll}(t)$         & toll cost for travelling from origin $o$                           &                         &                                 &                                                                   \\
                               &                            & to destination $d$ at time $t$                                     & \euro~/m  &                                 &                                                                   \\
                               & $C^{park,PV}_{od}(t)$      & parking fee for travelling from origin $o$                        &                         &                                 &                                                                   \\
                               &                            & to destination $d$ at time $t$                                     & \euro~   &                                 &                                                                   \\
                               & $u^{PV}$                   & private vehicle mode choice intercept for modal calibration        & \euro~   & -4.70                           &                                                                   \\
                               & $f_{od}^{PT}$              & public transport fare for travelling                               &                         &                                 &                                                                   \\
                               &                            & from origin $o$ to destination $d$                                 & \euro~   & 1.00                            &                                                                   \\
                               & $d_i^{walk}$               & walking distance for walking                                       &                         &                                 &                                                                   \\
                               &                            & to/from am public transport stop                                   & m                       &                                 &                                                                   \\
                               & $v^{walk}$                 & walking speed                                                      & m/s                     & 1.33                            &                                                                   \\
                               & $N_T$                      & number of public transport transfers                               &                         &                                 &                                                                   \\
                               & $c^T$                      & transfer penalty                                                   & \euro~   & -0.8                            &                                                                   \\
                               & $\eta$                     & crowding of the public transport system                            &                         &                                 &                                                                   \\
                               & $f(\eta)$                  & function to weight crowding $\eta$ with travel time $t_{od}$       &                         &                                 &                                                                   \\
                               & $t_{wait}$                 & waiting time for AMOD                                              & s                       &                                 &                                                                   \\
                               & $S_i$                      & subsidy for traveler $i$ for an intermodal trip                    & \euro~   &                                 &                                                                   \\ \hline
\textbf{Street}                & $G=(N,E)$                  & graph network $G$ consisting of nodes $N$ and edges $E$            &                         &                                 &                                                                   \\
\textbf{Network}               & $\psi_c^0$                 & free flow cluster travel time factor of cluster $c$                &                         &                                 &                                                                   \\
\textbf{Model}                 & $n_o$                      & origin node $n_o \in N$                                            &                         &                                 &                                                                   \\
                               & $n_d$                      & destination node $n_d \in N$                                       &                         &                                 &                                                                   \\
                               & $t_e$                      & free flow travel time of edge $e \in E$                            & s                       &                                 &                                                                   \\
                               & $\psi_c^t$                 & cluster travel time factor of cluster $c$ at time $t$              &                         &                                 &                                                                   \\
                               & $t_e^t$                    & travel time of edge $e \in E$ at time \$t                          & s                       &                                 &                                                                   \\
                               & $k^t_c$                    & cluster average density of cluster $c$ at time $t$                 & $1/m$           &                                 &                                                                   \\
                               & $q^t_c$                    & cluster average flow of cluster $c$ at time $t$                    & $1/s$           &                                 &                                                                   \\
                               & $v_{1,i}$                  & first $\psi_i^t$fitting parameter for inner zone                   & $m/s$           & 5.87                            &                                                                   \\
                               & $v_{2,i}$                  & second $\psi_i^t$ fitting parameter for inner zone                 & $m/s$           & 7.37                            &                                                                   \\
                               & $v_{1,o}$                  & first $\psi_o^t$ fitting parameter for outer zone                  & $m/s$           & 10.32                           &                                                                   \\
                               & $v_{2,o}$                  & second $\psi_o^t$ fitting parameter for outer zone                 & $m/s$           & 15.35                           &                                                                   \\ \hline
\textbf{Public}                & $n_t^{PT}$                 & number of public transport travelers at time $t$                   &                         &                                 &                                                                   \\
\textbf{Transport}             & $\Omega_T^{PT}$            & total public transport capacity during time interval $T$           &                         &                                 &                                                                   \\
\textbf{Network}               & $\Omega_l^{PT}$            & capacity of public transport vehicle on line $l$                   &                         &                                 &                                                                   \\
\textbf{Model}            & $\vert\zeta_l^T\vert$      & number of trips on line $l$ during time interval $T$               &                         &                                 &                                                                   \\
                               & $C_T^{PT}$                 & costs of public transport system during time interval $T$          &                         &                                 &                                                                   \\
                               & $E_T^{PT}$                 & emissions of public transport system during time interval $T$      &                         &                                 &                                                                   \\
                               & $d_l$                      & mean length of a vehicle on line $l$                               &                         &                                 &                                                                   \\
                               & $c_l$                      & mean cost of a vehicle on line $l$                                 &                         &                                 &                                                                   \\
                               & $e_l$                      & mean emissions of a vehicle on line $l$                            &                         &                                 &                                                                   \\
                               & $\Omega^{Bus}$             & assumed capacity for bus                                         & pax                     & 100                             &   from\footnote{\label{fn:fzg}https://www.mvg.de/ueber/das-unternehmen/fahrzeuge.html}                                                                \\
                               & $\Omega^{Tram}$            & assumed capacity for tram                                        & pax                     & 216                             & from\footref{fn:fzg}                                         \\
                               & $\Omega^{Subway}$          & assumed capacity for subway train                                  & pax                     & 940                             & from\footref{fn:fzg}                                   \\
                               & $\Omega^{Urban Train}$     & assumed capacity for urban train                                & pax                     & 1088                            &                  from\footref{fn:fzg}                                                 \\
                               & $C^{Bus}$                  & cost for bus                                                       & \euro~/km & 3.00                            &         from\footnote{https://prof.beuth-hochschule.de/fileadmin/prof/jschlaich/200811\_Fr\_JS\_Kostenmodelle\_NAHVERKEHR.pdf\label{fn:cost}}                                                          \\
                               & $C^{Tram}$                 & cost for tram                                                      & \euro~/km & 5.00                            & from\footref{fn:cost}         \\
                               & $C^{Subway}$               & cost for subway train                                              & \euro~/km & 15.00                           & from\footref{fn:cost}                    \\
                               & $C^{Urban Train}$          & cost for urban train                                               & \euro~/km & 15.00                           &         from\footref{fn:cost}                                                          \\
                               & $w^{Bus}$                  & energy consumption for bus                                         & kWh/km                  & 34                              & from\footnote{https://www.ris-muenchen.de/RII/RII/DOK/ANTRAG/2337762.pdf\label{fn:energy}}                                                                  \\
                               & $w^{Tram}$                 & energy consumption for tram                                        & kWh/km                  & 26                              & from\footref{fn:energy}                                  \\
                               & $w^{Subway}$               & energy consumption for subway                                      & kWh/km                  & 113                             & from\footref{fn:energy}                                        \\
                               & $w^{Urban Train}$          & energy consumption for urban train                                 & kWh/km                  & 173                             & from\footref{fn:energy}                                                                  \\
                               & $e_w$                      & emission of CO2 per energy consumption                             & g/kWh                   & 112                             & from\footnote{https://www.swm.de/dam/swm/dokumente/geschaeftskunden/broschuere-strom-erdgas-gk.pdf}                             \\ \hline
\textbf{Fleet}                 & $R^s$                      & set of all served customers                                        &                         &                                 &                                                                   \\
\textbf{Control}               & $V$                        & set of all AMOD vehicles                                           &                         &                                 &                                                                   \\
\textbf{Model}                 & $v$                        & specific AMOD vehicle $v \in V$                                    &                         &                                 &                                                                   \\
                               & $d_v$                      & driven distance of vehicle $v$                                     & m                       &                                 &                                                                   \\
                               & $t_w^{max}$                & maximum customer waiting time constraint                           & s                       & 30                              &                                                                   \\
                               & $t^{d,rel}_{max}$          & maximum detour time constraint                                     &                         &                                 &                                                                   \\
                               &                            & relative to direct route travel time                               & \%                      & 40                              &                                                                   \\
                               & $\xi_v(R_v)$               & currently assigned route for vehicle $v$                           &                         &                                 &                                                                   \\
                               &                            & with assigned customer set $R_v$                                   &                         &                                 &                                                                   \\
                               & $B$                        & boarding duration                                                  & s                       & 30                              &                                                                   \\
                               & $\rho$                     & vehicle capacity                                                   & pax                     & 4                               &                                                                   \\
                               & $N^{i}_{v,max}$            & maximum number of vehicles considered                              &                         &                                 &                                                                   \\
                               &                            & for insertion of customer $i$                                      &                         & 10                              &                                                                   \\
\hline
\textbf{Modeling}              & $x^{P}$                    & regulator decision variable for parking regulations                & \euro~   & {[}2.50, 5.00{]}                &                                                                   \\
\textbf{the Impacts}           & $x^{RT}$                   & regulator decision variable for road toll                          & \euro~/km & {[}0.0, 1.0{]}                  &                                                                   \\
                               & $k_0$                      & threshold NFD traveler density to trigger road toll           
 & veh~/lane-km                & 5.0                        &                                                                   \\
\textbf{of Decision}           & $x^{PT}$                   & regulator decision variable for scaling public transport frequency &                         & {[}0.25, 2.0{]}                  &                                                                   \\
\textbf{Variables}             & $\eta^T_0$                 & baseline public transport crowding at time interval $T$            &                         &                                 &                                                                   \\
                               & $\eta^T$                   & scaled public transport crowding by increased PT frequency         &                         &                                 &                                                                   \\
                               & $c^T_0$                    & baseline public transport transfer penalty at time interval $T$   &                         &                                 &                                                                   \\
                               & $c^T$                      & scaled public transport transfer penalty at time interval $T$      &                         &                                 &                                                                   \\
                               & $C_{T,0}^{PT}$             & baseline global public transport capacity at time interval $T$     &                         &                                 &                                                                   \\
                               & $C_T^{PT}$                 & scaled global public transport capacity at time interval $T$       &                         &                                 &                                                                   \\
                               & $E_{T,0}^{PT}$             & baseline global public transport emissions at time interval $T$    &                         &                                 &                                                                   \\
                               & $E_T^{PT}$                 & scaled global public transport emission at time interval $T$       &                         &                                 &                                                                   \\
                               & $x^{F}$                    & regulator decision variable to limit the AMOD fleet size            & vehicles & {[}0, 50,000{]}                   &                                                                   \\
                               & $y^F$                      &  AMOD service planner decision variable for fleet size                     & vehicles                & {[}0,10000{]}                   &                                                                   \\
                               & $y^{PD}$                   &  AMOD service planner decision variable for distance-based fare            & \euro~/km & {[}0.25, 2.00{]}                &                                                                   \\
                               & $f^{AMOD}_{min}$           & minimum base fare for an AMOD trip                                 & \euro~   & 1.0                             &                                                                   \\
                               & $y^{PU}$                   &  AMOD service planner decision variable to scale fares if 75\% fleet utilization is exceeded                    &                         &           {[}1.0, 10.0{]}                       &                                                                   \\
\hline
\textbf{Social}        & $P$                        & AMOD profit                                               & \euro~   &                                 &                                                                   \\
\textbf{Welfare}      & $R^s$                      & set of served requests                                             &                         &                                 &                                                                   \\
\textbf{and Profit}            & $c_v^F$                    & (daily) fixed costs of an AMOD vehicle                             & \euro~   & 40                              & from\footnote{https://www.tcs.ch/de/testberichte-ratgeber/ratgeber/kotrollen-unterhalt/kilometerkosten.php\label{fn:fixcost}}             \\
\textbf{Model}                 & $c_v^D$                    & distance dependent operating cost of an AMOD vehicle               & \euro~/km & 0.25                            & from\footref{fn:fixcost}                           \\
                               & $d_v$                      & total driven distance of a vehicle                                 & m                       &                                 &                                                                   \\
                               & $C_t^{toll}$               & total toll paid by all AMOD vehicles                               & \euro~   &                                 &                                                                   \\
                               & $E_{PV}$             & avg private vehicle CO2 emissions                                  & g/km                    & 130                             & from\footnote{https://www.muenchen.de/rathaus/Stadtinfos/Statistik/Verkehr.html} \footnote{https://www.ris-muenchen.de/RII/RII/DOK/ANTRAG/2337762.pdf}     \\
                               & $C^{CO2}$                  & monetarization of CO2 emissions                   & \euro~/kg & 0.145                           & from\footnote{http://www.suedumfahrung-jetzt.de/wp-content/uploads/2016/04/bvwp-2030-methodenhandbuch.pdf}  \\                                                            
\caption{Collection of all variables and parameters within this study sorted by their first occurrence. Fixed parameter values within the case study are given with their source.}\\
\label{tab:collection}

\end{longtable}
\end{center}
}

\section*{References}
\bibliography{mybibfile}

\begin{thebibliography}{56}
\expandafter\ifx\csname natexlab\endcsname\relax\def\natexlab#1{#1}\fi
\providecommand{\url}[1]{\texttt{#1}}
\providecommand{\href}[2]{#2}
\providecommand{\path}[1]{#1}
\providecommand{\DOIprefix}{doi:}
\providecommand{\ArXivprefix}{arXiv:}
\providecommand{\URLprefix}{URL: }
\providecommand{\Pubmedprefix}{pmid:}
\providecommand{\doi}[1]{\href{http://dx.doi.org/#1}{\path{#1}}}
\providecommand{\Pubmed}[1]{\href{pmid:#1}{\path{#1}}}
\providecommand{\bibinfo}[2]{#2}
\ifx\xfnm\relax \def\xfnm[#1]{\unskip,\space#1}\fi
\bibitem[{Alonso-Mora et~al.(2017)Alonso-Mora, Samaranayake, Wallar, Frazzoli
  and Rus}]{AlonsoMora.2017}
\bibinfo{author}{Alonso-Mora, J.}, \bibinfo{author}{Samaranayake, S.},
  \bibinfo{author}{Wallar, A.}, \bibinfo{author}{Frazzoli, E.},
  \bibinfo{author}{Rus, D.}, \bibinfo{year}{2017}.
\newblock \bibinfo{title}{On-demand high-capacity ride-sharing via dynamic
  trip-vehicle assignment}.
\newblock \bibinfo{journal}{Proceedings of the National Academy of Sciences} ,
  \bibinfo{pages}{462--467}\DOIprefix\doi{10.1073/pnas.1611675114}.
\bibitem[{Beer et~al.(2017)Beer, Brakewood, Rahman and Viscardi}]{Beer.2017}
\bibinfo{author}{Beer, R.}, \bibinfo{author}{Brakewood, C.},
  \bibinfo{author}{Rahman, S.}, \bibinfo{author}{Viscardi, J.},
  \bibinfo{year}{2017}.
\newblock \bibinfo{title}{Qualitative analysis of ride-hailing regulations in
  major american cities}.
\newblock \bibinfo{journal}{Transportation Research Record: Journal of the
  Transportation Research Board} \bibinfo{volume}{2650},
  \bibinfo{pages}{84--91}.
\newblock \DOIprefix\doi{10.3141/2650-10}.
\bibitem[{Bellon(2019)}]{Bellon2019}
\bibinfo{author}{Bellon, T.}, \bibinfo{year}{2019}.
\newblock \bibinfo{title}{{Uber to limit drivers' app access to comply with NYC
  regulation}}.
\bibitem[{Bertsimas et~al.(2019)Bertsimas, Jaillet and Martin}]{Bertsimas2019}
\bibinfo{author}{Bertsimas, D.}, \bibinfo{author}{Jaillet, P.},
  \bibinfo{author}{Martin, S.}, \bibinfo{year}{2019}.
\newblock \bibinfo{title}{{Online Vehicle Routing: The Edge of Optimization in
  Large-Scale Applications}}.
\newblock \bibinfo{journal}{Operations Research} \bibinfo{volume}{67},
  \bibinfo{pages}{143--162}.
\newblock \DOIprefix\doi{10.1287/opre.2018.1763}.
\bibitem[{Bilali et~al.(2020)Bilali, Engelhardt, Dandl, Fastenrath and
  Bogenberger}]{Bilali2020}
\bibinfo{author}{Bilali, A.}, \bibinfo{author}{Engelhardt, R.},
  \bibinfo{author}{Dandl, F.}, \bibinfo{author}{Fastenrath, U.},
  \bibinfo{author}{Bogenberger, K.}, \bibinfo{year}{2020}.
\newblock \bibinfo{title}{{Analytical and Agent-Based Model to Evaluate
  Ride-Pooling Impact Factors}}.
\newblock \bibinfo{journal}{Transportation Research Record: Journal of the
  Transportation Research Board} \bibinfo{volume}{2674},
  \bibinfo{pages}{1--12}.
\newblock \DOIprefix\doi{10.1177/0361198120917666}.
\bibitem[{Bischoff and Maciejewski(2016)}]{Bischoff2016}
\bibinfo{author}{Bischoff, J.}, \bibinfo{author}{Maciejewski, M.},
  \bibinfo{year}{2016}.
\newblock \bibinfo{title}{{Simulation of City-wide Replacement of Private Cars
  with Autonomous Taxis in Berlin}}.
\newblock \bibinfo{journal}{Procedia Computer Science} \bibinfo{volume}{83},
  \bibinfo{pages}{237--244}.
\newblock \DOIprefix\doi{10.1016/j.procs.2016.04.121}.
\bibitem[{B{\"{o}}sch et~al.(2018)B{\"{o}}sch, Becker, Becker and
  Axhausen}]{Bosch2018}
\bibinfo{author}{B{\"{o}}sch, P.M.}, \bibinfo{author}{Becker, F.},
  \bibinfo{author}{Becker, H.}, \bibinfo{author}{Axhausen, K.W.},
  \bibinfo{year}{2018}.
\newblock \bibinfo{title}{{Cost-based analysis of autonomous mobility
  services}}.
\newblock \bibinfo{journal}{Transport Policy} \bibinfo{volume}{64},
  \bibinfo{pages}{76--91}.
\newblock \DOIprefix\doi{10.1016/J.TRANPOL.2017.09.005}.
\bibitem[{Bracher and Bogenberger(2017)}]{Bracher.2017}
\bibinfo{author}{Bracher, B.}, \bibinfo{author}{Bogenberger, K.},
  \bibinfo{year}{2017}.
\newblock \bibinfo{title}{A dynamic prizing scheme for a congestion charging
  zone based on a network fundamental diagram}, in: \bibinfo{booktitle}{5th
  IEEE International Conference on Models and Technologies for Intelligent
  Transportation Systems (MT-ITS)}, \bibinfo{publisher}{IEEE}. pp.
  \bibinfo{pages}{669--674}.
\newblock \DOIprefix\doi{10.1109/MTITS.2017.8005597}.
\bibitem[{Campbell(2019)}]{Campbell2019}
\bibinfo{author}{Campbell, A.F.}, \bibinfo{year}{2019}.
\newblock \bibinfo{title}{{California passes AB5 bill to regulate Uber, Lyft
  and the gig economy}}.
\bibitem[{Cetin and Deakin(2019)}]{Cetin.2019}
\bibinfo{author}{Cetin, T.}, \bibinfo{author}{Deakin, E.},
  \bibinfo{year}{2019}.
\newblock \bibinfo{title}{Regulation of taxis and the rise of ridesharing}.
\newblock \bibinfo{journal}{Transport Policy} \bibinfo{volume}{76},
  \bibinfo{pages}{149--158}.
\newblock \DOIprefix\doi{10.1016/j.tranpol.2017.09.002}.
\bibitem[{Chen and Kockelman(2016)}]{Chen2016}
\bibinfo{author}{Chen, T.D.}, \bibinfo{author}{Kockelman, K.M.},
  \bibinfo{year}{2016}.
\newblock \bibinfo{title}{{Management of a Shared Autonomous Electric Vehicle
  Fleet: Implications of Pricing Schemes}}.
\newblock \bibinfo{journal}{Transportation Research Record: Journal of the
  Transportation Research Board} \bibinfo{volume}{2572},
  \bibinfo{pages}{37--46}.
\newblock \DOIprefix\doi{10.3141/2572-05}.
\bibitem[{Conger(2020)}]{Conger.2020}
\bibinfo{author}{Conger, K.}, \bibinfo{year}{2020}.
\newblock \bibinfo{title}{{Uber and Lyft Drivers in California Will Remain
  Contractors}}.
\bibitem[{Daganzo(2007)}]{daganzo2007urban}
\bibinfo{author}{Daganzo, C.F.}, \bibinfo{year}{2007}.
\newblock \bibinfo{title}{Urban gridlock: Macroscopic modeling and mitigation
  approaches}.
\newblock \bibinfo{journal}{Transportation Research Part B: Methodological}
  \bibinfo{volume}{41}, \bibinfo{pages}{49--62}.
\newblock \DOIprefix\doi{10.1016/j.trb.2006.03.001}.
\bibitem[{Dandl and Bogenberger(2019)}]{Dandl2019}
\bibinfo{author}{Dandl, F.}, \bibinfo{author}{Bogenberger, K.},
  \bibinfo{year}{2019}.
\newblock \bibinfo{title}{{Comparing Future Autonomous Electric Taxis with an
  Existing Free-Floating Carsharing System}}.
\newblock \bibinfo{journal}{IEEE Transactions on Intelligent Transportation
  Systems} \bibinfo{volume}{20}.
\newblock \DOIprefix\doi{10.1109/TITS.2018.2857208}.
\bibitem[{Dandl et~al.(2017)Dandl, Bracher and Bogenberger}]{Dandl.2017}
\bibinfo{author}{Dandl, F.}, \bibinfo{author}{Bracher, B.},
  \bibinfo{author}{Bogenberger, K.}, \bibinfo{year}{2017}.
\newblock \bibinfo{title}{Microsimulation of an autonomous taxi-system in
  munich}, in: \bibinfo{booktitle}{2017 5th IEEE International Conference on
  Models and Technologies for Intelligent Transportation Systems (MT-ITS)}, pp.
  \bibinfo{pages}{833--838}.
\newblock \DOIprefix\doi{10.1109/MTITS.2017.8005628}.
\bibitem[{Dandl et~al.(2019)Dandl, Hyland, Bogenberger and
  Mahmassani}]{Dandl2019Mike}
\bibinfo{author}{Dandl, F.}, \bibinfo{author}{Hyland, M.},
  \bibinfo{author}{Bogenberger, K.}, \bibinfo{author}{Mahmassani, H.},
  \bibinfo{year}{2019}.
\newblock \bibinfo{title}{{Evaluating the impact of spatio-temporal demand
  forecast aggregation on the operational performance of shared autonomous
  mobility fleets}}.
\newblock \bibinfo{journal}{Transportation} \bibinfo{volume}{46}.
\newblock \DOIprefix\doi{10.1007/s11116-019-10007-9}.
\bibitem[{Dandl et~al.(2020)Dandl, Tilg, Rostami-Shahrbabaki and
  Bogenberger}]{Dandl.2020ITSC}
\bibinfo{author}{Dandl, F.}, \bibinfo{author}{Tilg, G.},
  \bibinfo{author}{Rostami-Shahrbabaki, M.}, \bibinfo{author}{Bogenberger, K.},
  \bibinfo{year}{2020}.
\newblock \bibinfo{title}{Network fundamental diagram based routing of vehicle
  fleets in dynamic traffic simulations}, in: \bibinfo{booktitle}{2020 IEEE
  23rd International Conference on Intelligent Transportation Systems (ITSC)},
  \bibinfo{publisher}{IEEE}. pp. \bibinfo{pages}{1--8}.
\newblock \DOIprefix\doi{10.1109/ITSC45102.2020.9294204}.
\bibitem[{Engelhardt et~al.(2019)Engelhardt, Dandl, Bilali and
  Bogenberger}]{Engelhardt.2019}
\bibinfo{author}{Engelhardt, R.}, \bibinfo{author}{Dandl, F.},
  \bibinfo{author}{Bilali, A.}, \bibinfo{author}{Bogenberger, K.},
  \bibinfo{year}{2019}.
\newblock \bibinfo{title}{Quantifying the benefits of autonomous on-demand
  ride-pooling: A simulation study for munich, germany}, in:
  \bibinfo{booktitle}{2019 22nd IEEE Intelligent Transportation Systems
  Conference (ITSC)}, \bibinfo{publisher}{IEEE}. pp.
  \bibinfo{pages}{2992--2997}.
\newblock \DOIprefix\doi{10.1109/ITSC.2019.8916955}.
\bibitem[{Erhardt et~al.(2019)Erhardt, Roy, Cooper, Sana, Chen and
  Castiglione}]{Erhardt2019}
\bibinfo{author}{Erhardt, G.D.}, \bibinfo{author}{Roy, S.},
  \bibinfo{author}{Cooper, D.}, \bibinfo{author}{Sana, B.},
  \bibinfo{author}{Chen, M.}, \bibinfo{author}{Castiglione, J.},
  \bibinfo{year}{2019}.
\newblock \bibinfo{title}{{Do transportation network companies decrease or
  increase congestion?}}
\newblock \bibinfo{journal}{Science Advances} \bibinfo{volume}{5},
  \bibinfo{pages}{eaau2670}.
\newblock \DOIprefix\doi{10.1126/sciadv.aau2670}.
\bibitem[{Fagnant and Kockelman(2014)}]{Fagnant2014}
\bibinfo{author}{Fagnant, D.J.}, \bibinfo{author}{Kockelman, K.M.},
  \bibinfo{year}{2014}.
\newblock \bibinfo{title}{{The travel and environmental implications of shared
  autonomous vehicles, using agent-based model scenarios}}.
\newblock \bibinfo{journal}{Transportation Research Part C: Emerging
  Technologies} \bibinfo{volume}{40}, \bibinfo{pages}{1--13}.
\newblock \DOIprefix\doi{10.1016/j.trc.2013.12.001}.
\bibitem[{Fagnant and Kockelman(2018)}]{Fagnant2018}
\bibinfo{author}{Fagnant, D.J.}, \bibinfo{author}{Kockelman, K.M.},
  \bibinfo{year}{2018}.
\newblock \bibinfo{title}{{Dynamic ride-sharing and fleet sizing for a system
  of shared autonomous vehicles in Austin, Texas}}.
\newblock \bibinfo{journal}{Transportation} \bibinfo{volume}{45},
  \bibinfo{pages}{143--158}.
\newblock \DOIprefix\doi{10.1007/s11116-016-9729-z},
  \href{http://arxiv.org/abs/arXiv:1011.1669v3}{\tt arXiv:arXiv:1011.1669v3}.
\bibitem[{Fagnant et~al.(2015)Fagnant, Kockelman and Bansal}]{Fagnant2015}
\bibinfo{author}{Fagnant, D.J.}, \bibinfo{author}{Kockelman, K.M.},
  \bibinfo{author}{Bansal, P.}, \bibinfo{year}{2015}.
\newblock \bibinfo{title}{{Operations of Shared Autonomous Vehicle Fleet for
  Austin, Texas, Market}}.
\newblock \bibinfo{journal}{Transportation Research Record: Journal of the
  Transportation Research Board} \bibinfo{volume}{2536},
  \bibinfo{pages}{98--106}.
\newblock \DOIprefix\doi{10.3141/2536-12}.
\bibitem[{Farvaresh and Sepehri(2013)}]{Farvaresh2013}
\bibinfo{author}{Farvaresh, H.}, \bibinfo{author}{Sepehri, M.M.},
  \bibinfo{year}{2013}.
\newblock \bibinfo{title}{{A Branch and Bound Algorithm for Bi-level Discrete
  Network Design Problem}}.
\newblock \bibinfo{journal}{Networks and Spatial Economics}
  \bibinfo{volume}{13}, \bibinfo{pages}{67--106}.
\newblock \DOIprefix\doi{10.1007/s11067-012-9173-3}.
\bibitem[{Fisk(1984)}]{Fisk1984}
\bibinfo{author}{Fisk, C.}, \bibinfo{year}{1984}.
\newblock \bibinfo{title}{{Game theory and transportation systems modelling}}.
\newblock \bibinfo{journal}{Transportation Research Part B: Methodological}
  \bibinfo{volume}{18}, \bibinfo{pages}{301--313}.
\newblock \DOIprefix\doi{10.1016/0191-2615(84)90013-4}.
\bibitem[{Frei et~al.(2017)Frei, Hyland and Mahmassani}]{Frei2017}
\bibinfo{author}{Frei, C.}, \bibinfo{author}{Hyland, M.},
  \bibinfo{author}{Mahmassani, H.S.}, \bibinfo{year}{2017}.
\newblock \bibinfo{title}{Flexing service schedules: Assessing the potential
  for demand-adaptive hybrid transit via a stated preference approach}.
\newblock \bibinfo{journal}{Transportation Research Part C: Emerging
  Technologies} \bibinfo{volume}{76}, \bibinfo{pages}{71 -- 89}.
\newblock \DOIprefix\doi{https://doi.org/10.1016/j.trc.2016.12.017}.
\bibitem[{Gao et~al.(2005)Gao, Wu and Sun}]{Gao2005}
\bibinfo{author}{Gao, Z.}, \bibinfo{author}{Wu, J.}, \bibinfo{author}{Sun, H.},
  \bibinfo{year}{2005}.
\newblock \bibinfo{title}{{Solution algorithm for the bi-level discrete network
  design problem}}.
\newblock \bibinfo{journal}{Transportation Research Part B: Methodological}
  \bibinfo{volume}{39}, \bibinfo{pages}{479--495}.
\newblock \DOIprefix\doi{10.1016/J.TRB.2004.06.004}.
\bibitem[{Geroliminis and Daganzo(2008)}]{Geroliminis.2008}
\bibinfo{author}{Geroliminis, N.}, \bibinfo{author}{Daganzo, C.F.},
  \bibinfo{year}{2008}.
\newblock \bibinfo{title}{Existence of urban-scale macroscopic fundamental
  diagrams: Some experimental findings}.
\newblock \bibinfo{journal}{Transportation Research Part B: Methodological}
  \bibinfo{volume}{42}, \bibinfo{pages}{759--770}.
\newblock \DOIprefix\doi{10.1016/j.trb.2008.02.002}.
\bibitem[{Gu et~al.(2019)Gu, Cai, Han and Wang}]{Gu.2019}
\bibinfo{author}{Gu, Y.}, \bibinfo{author}{Cai, X.}, \bibinfo{author}{Han, D.},
  \bibinfo{author}{Wang, D.Z.}, \bibinfo{year}{2019}.
\newblock \bibinfo{title}{A tri-level optimization model for a private road
  competition problem with traffic equilibrium constraints}.
\newblock \bibinfo{journal}{European Journal of Operational Research}
  \bibinfo{volume}{273}, \bibinfo{pages}{190--197}.
\newblock \DOIprefix\doi{10.1016/j.ejor.2018.07.041}.
\bibitem[{Gurumurthy and Kockelman(2018)}]{Gurumurthy2018}
\bibinfo{author}{Gurumurthy, K.M.}, \bibinfo{author}{Kockelman, K.M.},
  \bibinfo{year}{2018}.
\newblock \bibinfo{title}{{Analyzing the dynamic ride-sharing potential for
  shared autonomous vehicle fleets using cellphone data from Orlando,
  Florida}}.
\newblock \bibinfo{journal}{Computers, Environment and Urban Systems}
  \bibinfo{volume}{71}, \bibinfo{pages}{177--185}.
\newblock \DOIprefix\doi{10.1016/J.COMPENVURBSYS.2018.05.008}.
\bibitem[{Gurumurthy et~al.(2020)Gurumurthy, de~Souza, Enam and
  Auld}]{Gurumurthy2020}
\bibinfo{author}{Gurumurthy, K.M.}, \bibinfo{author}{de~Souza, F.},
  \bibinfo{author}{Enam, A.}, \bibinfo{author}{Auld, J.}, \bibinfo{year}{2020}.
\newblock \bibinfo{title}{{Integrating Supply and Demand Perspectives for a
  Large-Scale Simulation of Shared Autonomous Vehicles}}.
\newblock \bibinfo{journal}{Transportation Research Record: Journal of the
  Transportation Research Board} ,
  \bibinfo{pages}{036119812092115}\DOIprefix\doi{10.1177/0361198120921157}.
\bibitem[{Henao and Marshall(2018)}]{Henao2018}
\bibinfo{author}{Henao, A.}, \bibinfo{author}{Marshall, W.E.},
  \bibinfo{year}{2018}.
\newblock \bibinfo{title}{{The impact of ride-hailing on vehicle miles
  traveled}}.
\newblock \bibinfo{journal}{Transportation} ,
  \bibinfo{pages}{1--22}\DOIprefix\doi{10.1007/s11116-018-9923-2}.
\bibitem[{Holler et~al.(2019)Holler, Vuorio, Qin, Tang, Jiao, Jin, Singh, Wang
  and Ye}]{Holler2019}
\bibinfo{author}{Holler, J.}, \bibinfo{author}{Vuorio, R.},
  \bibinfo{author}{Qin, Z.}, \bibinfo{author}{Tang, X.}, \bibinfo{author}{Jiao,
  Y.}, \bibinfo{author}{Jin, T.}, \bibinfo{author}{Singh, S.},
  \bibinfo{author}{Wang, C.}, \bibinfo{author}{Ye, J.}, \bibinfo{year}{2019}.
\newblock \bibinfo{title}{{Deep Reinforcement Learning for Multi-driver Vehicle
  Dispatching and Repositioning Problem}}, in: \bibinfo{booktitle}{2019 IEEE
  International Conference on Data Mining (ICDM)}, \bibinfo{publisher}{IEEE}.
  pp. \bibinfo{pages}{1090--1095}.
\newblock \DOIprefix\doi{10.1109/ICDM.2019.00129}.
\bibitem[{H{\"{o}}rl et~al.()H{\"{o}}rl, Ruch, Becker, Frazzoli and
  Axhausen}]{Hoerl2018}
\bibinfo{author}{H{\"{o}}rl, S.}, \bibinfo{author}{Ruch, C.},
  \bibinfo{author}{Becker, F.}, \bibinfo{author}{Frazzoli, E.},
  \bibinfo{author}{Axhausen, K.W.}, .
\newblock \bibinfo{title}{{Fleet control algorithms for automated mobility : A
  simulation assessment for Zurich}}.
\newblock \bibinfo{journal}{TRB 2018 Annual Meeting}
  \DOIprefix\doi{10.3929/ETHZ-B-000175260}.
\bibitem[{Hu(2019)}]{Hu2019}
\bibinfo{author}{Hu, W.}, \bibinfo{year}{2019}.
\newblock \bibinfo{title}{{Confused About Congestion Pricing? Here's What We
  Know}}.
\bibitem[{Hyland et~al.(2019)Hyland, Dandl, Bogenberger and
  Mahmassani}]{Hyland2019}
\bibinfo{author}{Hyland, M.}, \bibinfo{author}{Dandl, F.},
  \bibinfo{author}{Bogenberger, K.}, \bibinfo{author}{Mahmassani, H.},
  \bibinfo{year}{2019}.
\newblock \bibinfo{title}{{Integrating demand forecasts into the operational
  strategies of shared automated vehicle mobility services: spatial resolution
  impacts}}.
\newblock \bibinfo{journal}{Transportation Letters} ,
  \bibinfo{pages}{1--6}\DOIprefix\doi{10.1080/19427867.2019.1691297}.
\bibitem[{Hyland and Mahmassani(2018)}]{Hyland2018}
\bibinfo{author}{Hyland, M.}, \bibinfo{author}{Mahmassani, H.S.},
  \bibinfo{year}{2018}.
\newblock \bibinfo{title}{{Dynamic autonomous vehicle fleet operations:
  Optimization-based strategies to assign AVs to immediate traveler demand
  requests}}.
\newblock \bibinfo{journal}{Transportation Research Part C: Emerging
  Technologies} \bibinfo{volume}{92}, \bibinfo{pages}{278--297}.
\newblock \DOIprefix\doi{10.1016/j.trc.2018.05.003}.
\bibitem[{Hyland and Mahmassani(2020)}]{Hyland2020}
\bibinfo{author}{Hyland, M.}, \bibinfo{author}{Mahmassani, H.S.},
  \bibinfo{year}{2020}.
\newblock \bibinfo{title}{{Operational benefits and challenges of shared-ride
  automated mobility-on-demand services}}.
\newblock \bibinfo{journal}{Transportation Research Part A: Policy and
  Practice} \bibinfo{volume}{134}, \bibinfo{pages}{251--270}.
\newblock \DOIprefix\doi{10.1016/J.TRA.2020.02.017}.
\bibitem[{Li et~al.(2019a)Li, Qin, Jiao, Yang, Wang, Wang, Wu and Ye}]{Li2019M}
\bibinfo{author}{Li, M.}, \bibinfo{author}{Qin, Z.}, \bibinfo{author}{Jiao,
  Y.}, \bibinfo{author}{Yang, Y.}, \bibinfo{author}{Wang, J.},
  \bibinfo{author}{Wang, C.}, \bibinfo{author}{Wu, G.}, \bibinfo{author}{Ye,
  J.}, \bibinfo{year}{2019}a.
\newblock \bibinfo{title}{{Efficient Ridesharing Order Dispatching with Mean
  Field Multi-Agent Reinforcement Learning}}, in: \bibinfo{booktitle}{The World
  Wide Web Conference on - WWW '19}, \bibinfo{publisher}{ACM Press},
  \bibinfo{address}{New York}. pp. \bibinfo{pages}{983--994}.
\newblock \DOIprefix\doi{10.1145/3308558.3313433}.
\bibitem[{Li et~al.(2019b)Li, Tavafoghi, Poolla and Varaiya}]{Li2019}
\bibinfo{author}{Li, S.}, \bibinfo{author}{Tavafoghi, H.},
  \bibinfo{author}{Poolla, K.}, \bibinfo{author}{Varaiya, P.},
  \bibinfo{year}{2019}b.
\newblock \bibinfo{title}{{Regulating TNCs: Should Uber and Lyft set their own
  rules?}}
\newblock \bibinfo{journal}{Transportation Research Part B: Methodological}
  \bibinfo{volume}{129}, \bibinfo{pages}{193--225}.
\newblock \DOIprefix\doi{10.1016/J.TRB.2019.09.008}.
\bibitem[{Liu et~al.(2019)Liu, Bansal, Daziano and Samaranayake}]{Liu2019}
\bibinfo{author}{Liu, Y.}, \bibinfo{author}{Bansal, P.},
  \bibinfo{author}{Daziano, R.}, \bibinfo{author}{Samaranayake, S.},
  \bibinfo{year}{2019}.
\newblock \bibinfo{title}{{A framework to integrate mode choice in the design
  of mobility-on-demand systems}}.
\newblock \bibinfo{journal}{Transportation Research Part C: Emerging
  Technologies} \bibinfo{volume}{105}, \bibinfo{pages}{648--665}.
\newblock \DOIprefix\doi{10.1016/J.TRC.2018.09.022}.
\bibitem[{Maciejewski et~al.(2016)Maciejewski, Bischoff and
  Nagel}]{Maciejewski2016}
\bibinfo{author}{Maciejewski, M.}, \bibinfo{author}{Bischoff, J.},
  \bibinfo{author}{Nagel, K.}, \bibinfo{year}{2016}.
\newblock \bibinfo{title}{{An Assignment-Based Approach to Efficient Real-Time
  City-Scale Taxi Dispatching}}.
\newblock \bibinfo{journal}{IEEE Intelligent Systems} \bibinfo{volume}{31},
  \bibinfo{pages}{68--77}.
\newblock \DOIprefix\doi{10.1109/MIS.2016.2}.
\bibitem[{Maget et~al.(2019)Maget, Pillat and Wa{\ss}muth}]{Maget.2019}
\bibinfo{author}{Maget, C.}, \bibinfo{author}{Pillat, J.},
  \bibinfo{author}{Wa{\ss}muth, V.}, \bibinfo{year}{2019}.
\newblock \bibinfo{title}{Transport demand model for the free state of bavaria
  -- basis for local transport planning}.
\newblock \bibinfo{journal}{Transportation Research Procedia}
  \bibinfo{volume}{41}, \bibinfo{pages}{219--228}.
\newblock \DOIprefix\doi{10.1016/j.trpro.2019.09.040}.
\bibitem[{Oda and Joe-Wong(2018)}]{Oda2018}
\bibinfo{author}{Oda, T.}, \bibinfo{author}{Joe-Wong, C.},
  \bibinfo{year}{2018}.
\newblock \bibinfo{title}{{MOVI: A Model-Free Approach to Dynamic Fleet
  Management}}, in: \bibinfo{booktitle}{IEEE INFOCOM 2018 - IEEE Conference on
  Computer Communications}, \bibinfo{publisher}{IEEE}. pp.
  \bibinfo{pages}{2708--2716}.
\newblock \DOIprefix\doi{10.1109/INFOCOM.2018.8485988}.
\bibitem[{Pinto et~al.(2019)Pinto, Hyland, Mahmassani and Verbas}]{Pinto2019}
\bibinfo{author}{Pinto, H.K.}, \bibinfo{author}{Hyland, M.F.},
  \bibinfo{author}{Mahmassani, H.S.}, \bibinfo{author}{Verbas, I.{\"{O}}.},
  \bibinfo{year}{2019}.
\newblock \bibinfo{title}{{Joint design of multimodal transit networks and
  shared autonomous mobility fleets}}.
\newblock \bibinfo{journal}{Transportation Research Part C: Emerging
  Technologies} \DOIprefix\doi{10.1016/j.trc.2019.06.010}.
\bibitem[{Saedi et~al.(2020)Saedi, Saeedmanesh, Zockaie, Saberi, Geroliminis
  and Mahmassani}]{Saedi.2020}
\bibinfo{author}{Saedi, R.}, \bibinfo{author}{Saeedmanesh, M.},
  \bibinfo{author}{Zockaie, A.}, \bibinfo{author}{Saberi, M.},
  \bibinfo{author}{Geroliminis, N.}, \bibinfo{author}{Mahmassani, H.S.},
  \bibinfo{year}{2020}.
\newblock \bibinfo{title}{Estimating network travel time reliability with
  network partitioning}.
\newblock \bibinfo{journal}{Transportation Research Part C: Emerging
  Technologies} \bibinfo{volume}{112}, \bibinfo{pages}{46--61}.
\newblock \DOIprefix\doi{10.1016/j.trc.2020.01.013}.
\bibitem[{Santi et~al.(2014)Santi, Resta, Szell, Sobolevsky, Strogatz and
  Ratti}]{Santi.2014}
\bibinfo{author}{Santi, P.}, \bibinfo{author}{Resta, G.},
  \bibinfo{author}{Szell, M.}, \bibinfo{author}{Sobolevsky, S.},
  \bibinfo{author}{Strogatz, S.H.}, \bibinfo{author}{Ratti, C.},
  \bibinfo{year}{2014}.
\newblock \bibinfo{title}{Quantifying the benefits of vehicle pooling with
  shareability networks}.
\newblock \bibinfo{journal}{Proceedings of the National Academy of Sciences of
  the United States of America} \bibinfo{volume}{111},
  \bibinfo{pages}{13290--13294}.
\newblock \DOIprefix\doi{10.1073/pnas.1403657111}.
\bibitem[{Simonetto et~al.(2019)Simonetto, Monteil and
  Gambella}]{Simonetto2019}
\bibinfo{author}{Simonetto, A.}, \bibinfo{author}{Monteil, J.},
  \bibinfo{author}{Gambella, C.}, \bibinfo{year}{2019}.
\newblock \bibinfo{title}{{Real-time city-scale ridesharing via linear
  assignment problems}}.
\newblock \bibinfo{journal}{Transportation Research Part C: Emerging
  Technologies} \bibinfo{volume}{101}, \bibinfo{pages}{208--232}.
\newblock \DOIprefix\doi{10.1016/J.TRC.2019.01.019}.
\bibitem[{Simoni et~al.(2019)Simoni, Kockelman, Gurumurthy and
  Bischoff}]{Simoni2019}
\bibinfo{author}{Simoni, M.D.}, \bibinfo{author}{Kockelman, K.M.},
  \bibinfo{author}{Gurumurthy, K.M.}, \bibinfo{author}{Bischoff, J.},
  \bibinfo{year}{2019}.
\newblock \bibinfo{title}{{Congestion pricing in a world of self-driving
  vehicles: An analysis of different strategies in alternative future
  scenarios}}.
\newblock \bibinfo{journal}{Transportation Research Part C: Emerging
  Technologies} \bibinfo{volume}{98}, \bibinfo{pages}{167--185}.
\newblock \DOIprefix\doi{10.1016/J.TRC.2018.11.002}.
\bibitem[{Srinivas et~al.(2012)Srinivas, Krause, Kakade and
  Seeger}]{Srinivas.2012}
\bibinfo{author}{Srinivas, N.}, \bibinfo{author}{Krause, A.},
  \bibinfo{author}{Kakade, S.M.}, \bibinfo{author}{Seeger, M.W.},
  \bibinfo{year}{2012}.
\newblock \bibinfo{title}{Information-theoretic regret bounds for gaussian
  process optimization in the bandit setting}.
\newblock \bibinfo{journal}{IEEE Transactions on Information Theory}
  \bibinfo{volume}{58}, \bibinfo{pages}{3250--3265}.
\newblock \DOIprefix\doi{10.1109/TIT.2011.2182033}.
\bibitem[{Tirachini et~al.(2017)Tirachini, Hurtubia, Dekker and
  Daziano}]{tirachini2017estimation}
\bibinfo{author}{Tirachini, A.}, \bibinfo{author}{Hurtubia, R.},
  \bibinfo{author}{Dekker, T.}, \bibinfo{author}{Daziano, R.A.},
  \bibinfo{year}{2017}.
\newblock \bibinfo{title}{Estimation of crowding discomfort in public
  transport: Results from santiago de chile}.
\newblock \bibinfo{journal}{Transportation Research Part A: Policy and
  Practice} \bibinfo{volume}{103}, \bibinfo{pages}{311 -- 326}.
\newblock \DOIprefix\doi{https://doi.org/10.1016/j.tra.2017.06.008}.
\bibitem[{Washbrook et~al.(2006)Washbrook, Haider and Jaccard}]{Washbrook.2006}
\bibinfo{author}{Washbrook, K.}, \bibinfo{author}{Haider, W.},
  \bibinfo{author}{Jaccard, M.}, \bibinfo{year}{2006}.
\newblock \bibinfo{title}{Estimating commuter mode choice: A discrete choice
  analysis of the impact of road pricing and parking charges}.
\newblock \bibinfo{journal}{Transportation} \bibinfo{volume}{33},
  \bibinfo{pages}{621--639}.
\newblock \DOIprefix\doi{10.1007/s11116-005-5711-x}.
\bibitem[{Xu et~al.(2016)Xu, Chen and Yang}]{Xu2016}
\bibinfo{author}{Xu, X.}, \bibinfo{author}{Chen, A.}, \bibinfo{author}{Yang,
  C.}, \bibinfo{year}{2016}.
\newblock \bibinfo{title}{{A review of sustainable network design for road
  networks}}.
\newblock \bibinfo{journal}{KSCE Journal of Civil Engineering}
  \bibinfo{volume}{20}, \bibinfo{pages}{1084--1098}.
\newblock \DOIprefix\doi{10.1007/s12205-016-1729-1}.
\bibitem[{Xu et~al.(2018)Xu, Li, Guan, Zhang, Li, Nan, Liu, Bian and
  Ye}]{Xu2018}
\bibinfo{author}{Xu, Z.}, \bibinfo{author}{Li, Z.}, \bibinfo{author}{Guan, Q.},
  \bibinfo{author}{Zhang, D.}, \bibinfo{author}{Li, Q.}, \bibinfo{author}{Nan,
  J.}, \bibinfo{author}{Liu, C.}, \bibinfo{author}{Bian, W.},
  \bibinfo{author}{Ye, J.}, \bibinfo{year}{2018}.
\newblock \bibinfo{title}{{Large-Scale Order Dispatch in On-Demand Ride-Hailing
  Platforms}}, in: \bibinfo{booktitle}{Proceedings of the 24th ACM SIGKDD
  International Conference on Knowledge Discovery {\&} Data Mining},
  \bibinfo{publisher}{ACM}, \bibinfo{address}{New York}. pp.
  \bibinfo{pages}{905--913}.
\newblock \DOIprefix\doi{10.1145/3219819.3219824}.
\bibitem[{Yang and {H. Bell}(1998)}]{Yang1998}
\bibinfo{author}{Yang, H.}, \bibinfo{author}{{H. Bell}, M.G.},
  \bibinfo{year}{1998}.
\newblock \bibinfo{title}{{Models and algorithms for road network design: a
  review and some new developments}}.
\newblock \bibinfo{journal}{Transport Reviews} \bibinfo{volume}{18},
  \bibinfo{pages}{257--278}.
\newblock \DOIprefix\doi{10.1080/01441649808717016}.
\bibitem[{Yin(2000)}]{Yin2000}
\bibinfo{author}{Yin, Y.}, \bibinfo{year}{2000}.
\newblock \bibinfo{title}{{Genetic-Algorithms-Based Approach for Bilevel
  Programming Models}}.
\newblock \bibinfo{journal}{Journal of Transportation Engineering}
  \bibinfo{volume}{126}, \bibinfo{pages}{115--120}.
\newblock \DOIprefix\doi{10.1061/(ASCE)0733-947X(2000)126:2(115)}.
\bibitem[{Zhang and Nie(2019)}]{Zhang2019}
\bibinfo{author}{Zhang, K.}, \bibinfo{author}{Nie, M.}, \bibinfo{year}{2019}.
\newblock \bibinfo{title}{{To Pool or Not to Pool: Equilibrium, Pricing and
  Regulation}}.
\newblock \bibinfo{journal}{SSRN Electronic Journal}
  \DOIprefix\doi{10.2139/ssrn.3497808}.

\end{thebibliography}

\end{document}